\documentclass[10pt,aps,prl,reprint,superscriptaddress,nobibnotes,nofootinbib,longbibliography]{revtex4-2}

\usepackage{multirow}
\usepackage{xcolor}

\usepackage{overpic}
\usepackage{graphicx}
\usepackage{epsfig}
\usepackage{epstopdf}
\usepackage{float}
\usepackage{subfigure}

\usepackage[version=4]{mhchem} 

\usepackage{dcolumn}
\newcolumntype{d}[1]{D{.}{.}{#1}}  
\usepackage{latexsym}
\usepackage{amsmath,amsfonts,amsthm,bm}
\usepackage{wasysym}
\usepackage{bm}

\usepackage[colorlinks,bookmarks=false,citecolor=blue,linkcolor=red,urlcolor=blue]{hyperref}
\usepackage{verbatim}

\usepackage{booktabs}

\usepackage[normalem]{ulem} 

\usepackage{indentfirst}
\usepackage{listings}
\usepackage{bbold}

\usepackage{array}
\usepackage{color}
\usepackage{siunitx}

\usepackage{tikz}
\definecolor{orcidlogocol}{HTML}{A6CE39}

\newcommand{\orcidicon}{%
  \begin{tikzpicture}[baseline=-0.65ex]
    \fill[orcidlogocol] (0,0) circle (0.9ex);

    \def\ybottom{-0.40ex}
    \def\ytop{0.36ex}

    \fill[white]
      (-0.43ex,\ybottom) rectangle (-0.31ex,\ytop);

    \fill[white] (-0.37ex,0.53ex) circle (0.08ex);

    \fill[white, even odd rule]
      (-0.13ex,\ybottom)
      -- (0.18ex,\ybottom)
      .. controls (0.72ex,\ybottom) and (0.72ex,\ytop) ..
      (0.18ex,\ytop)
      -- (-0.13ex,\ytop)
      -- cycle

      (0.01ex,-0.26ex)
      -- (0.18ex,-0.26ex)
      .. controls (0.51ex,-0.26ex) and (0.51ex,0.22ex) ..
      (0.18ex,0.22ex)
      -- (0.01ex,0.22ex)
      -- cycle;
  \end{tikzpicture}%
}

\newcommand{\orcid}[1]{\href{https://orcid.org/#1}{\orcidicon}}

\usepackage{xcolor}
\definecolor{apslinkblue}{RGB}{0,0,180}
\hypersetup{
  colorlinks=true,
  linkcolor=apslinkblue,
  citecolor=apslinkblue,
  urlcolor=apslinkblue
}

\AtBeginDocument{%
    {\usefont{U}{wasy}{m}{n}}%
    \DeclareFontShape{U}{wasy}{b}{n}{<-> wasyb10}{}%
}

\usepackage{docmute}
\usepackage{import}

\let\standardsection\section

\let\savedmaketitle\maketitle
\let\savedtitle\title
\let\savedauthor\author
\let\savedaffiliation\affiliation
\let\savedemail\email
\let\savedhomepage\homepage
\let\savedthanks\thanks
\let\saveddate\date

\usepackage{xparse}
\makeatletter

\def\prl@sectionstyle#1{%
  \normalfont\itshape #1---%
}
\renewcommand\section{%
  \@startsection
    {section}
    {1}
    {\z@}
    {1.0\baselineskip}
    {0pt}
    {\prl@sectionstyle}
}
\makeatother

\makeatletter
\renewcommand\appendix{%
  \par
  \addvspace{1.5\baselineskip}%
  \begin{center}
    \normalfont\bfseries End Matter
  \end{center}%
  \par
  \setcounter{section}{0}%
}
\makeatother

\begin{document}




\title{Bobkingite, a new coupled sawtooth chain platform}

\author{P. Peter Stavropoulos\orcid{0000-0002-4193-5844}}
\email{panagiotis@itp.uni-frankfurt.de}
\affiliation{Institut f\"ur Theoretische Physik, Goethe-Universit\"at Frankfurt, 60438 Frankfurt am Main, Germany}

\author{Aleksandar Razpopov\orcid{0009-0009-6935-3297}}
\affiliation{Institut f\"ur Theoretische Physik, Goethe-Universit\"at Frankfurt, 60438 Frankfurt am Main, Germany}

\author{Harrison LaBollita\orcid{0000-0002-6699-8577}}
\affiliation{Center for Computational Quantum Physics, Flatiron Institute, New York, New York 10010, USA}

\author{David~A.~S.~Kaib\orcid{0000-0002-1448-0063}}
\affiliation{Institut f\"ur Theoretische Physik, Goethe-Universit\"at Frankfurt, 60438 Frankfurt am Main, Germany}

\author{Michael R. Norman\orcid{0000-0002-9459-078X}}
\affiliation{Materials Science Division, Argonne National Laboratory, Lemont, IL 60439, USA}

\author{Antia S. Botana\orcid{0000-0001-5973-3039}}
\affiliation{Department of Physics, Arizona State University, Tempe, AZ 85287, USA}

\author{Roser Valent\'{\i}\orcid{0000-0003-0497-1165}}
\email{valenti@itp.uni-frankfurt.de}
\affiliation{Institut f\"ur Theoretische Physik, Goethe-Universit\"at Frankfurt, 60438 Frankfurt am Main, Germany}

\date{\today}

\begin{abstract}

We investigate the mineral bobkingite, \ce{Cu5(OH)8Cl2(H2O)2}, as a potential realization of the sawtooth chain. Using \textit{ab initio} methods, we estimate the magnetic exchange couplings and find that bobkingite hosts quasi-one-dimensional sawtooth chains, with residual three-dimensional interactions strongly suppressed by the crystal geometry. Examining the full exchange network, we find that the classical model exhibits an extensive manifold of nearly degenerate states with emergent two-dimensional character, which spin-wave theory shows to persist as Ising degrees of freedom. Unlike other sawtooth candidates, bobkingite has negligible vertical interchain coupling, preserving a one-dimensional degeneracy even in the presence of ordering, suggesting that any long-range order is weak. Thermal fluctuations stabilize a finite-temperature regime of multiplanar fluctuations rooted in the classical chain degeneracy, whose spin correlations resemble those of the quantum sawtooth chain, suggesting the persistence of quantum features in the full model. This makes bobkingite a compelling platform for exploring sawtooth chain physics.

\end{abstract}

\maketitle

%

\begin{figure*}[t]
	\centering
    \begin{overpic}[width=1.0\textwidth,percent,grid=false,tics=2,angle=0]{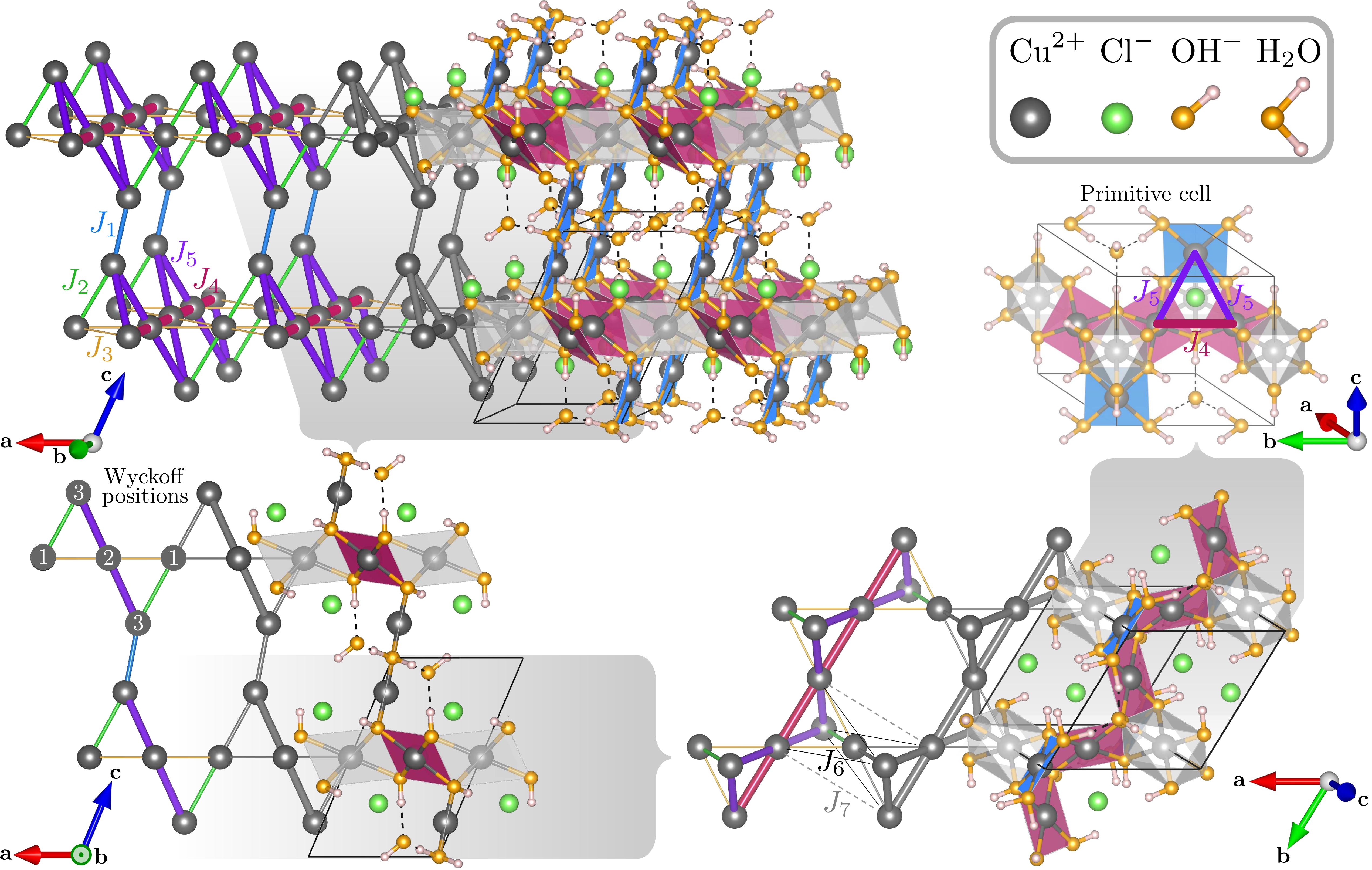}
        \put(0,61.5){(a)}
        \put(0,28){(b)}
        \put(54,28){(c)}
        \put(71,48.5){(d)}
    \end{overpic} 
    \caption{Crystal structure and bond exchange model definitions of bobkingite, from multiple perspectives. A general perspective is shown in (a), where the emerging sawtooth appears as purple ($J_5\!=\!J_{\text{bv}}$) and burgundy ($J_4\!=\!J_{\text{bb}}$) bonds. The weaker blue ($J_1$), green ($J_2$), and orange ($J_3$) are also indicated. (b) View of the $\mathbf{ac}$-plane, showing the distinct Wyckoff locations of the magnetic copper sites. (c) a top view of the $\mathbf{ab}$ plane, showing the kagome-like layered structure, and the weaker bonds $J_6$ and $J_7$. (d) A close-up look at the microscopic environment that sets the dominant  $J_4$ and $J_5$ couplings.  The various perspectives of bobkingite's structure were prepared using the software VESTA~\cite{VESTA}.}\label{fig:master_figure}
\end{figure*}

\section{Introduction}

One-dimensional quantum systems provide a natural setting for strong quantum fluctuations, which often preclude conventional symmetry-breaking orders and give rise to collective many-body phenomena~\cite{GiamarchiBook2003}. In the presence of frustration, arising from competing interactions that cannot be simultaneously minimized, these fluctuations are further enhanced and can stabilize highly entangled quantum ground states. A hallmark of such states is fractionalization, whereby the elementary excitations carry quantum numbers that differ from those of the microscopic degrees of freedom~\cite{balents2010spin,savary2017quantum}. Fractionalized phases have attracted a lot of interest both as paradigmatic examples of strongly correlated quantum matter and as potential platforms for topological quantum information processing~\cite{NayakRMP2008,FieldQST2018}. While fractionalization occurs in a variety of dimensions, one-dimensional systems provide some of its most direct realizations owing to the combined effects of strong quantum fluctuations and frustration~\cite{bethe1935statistical,faddeev1981spin,GiamarchiBook2003}.

A prototypical model in this context is the sawtooth ($\Delta$) chain~\cite{NakamuraPRB1996,ShastryPRB1996,hao2009fermionic}, 
forming a spin-$1/2$ antiferromagnetic (AFM) Heisenberg model on a lattice of corner-sharing triangles with base-base and base-vertex exchange couplings, $J_{\mathrm{bb}}$ and $J_{\mathrm{bv}}$ respectively. 
At the symmetric point $J_{\mathrm{bb}}\!=\!J_{\mathrm{bv}}$, the 
AFM model possesses an exactly dimerized twofold-degenerate ground state~\cite{MontiPLA1991}, while its elementary excitations correspond to domain-wall defects separating the two dimer coverings~\cite{NakamuraPRB1996}. The resulting fractionalized kink and antikink excitations make the sawtooth chain a canonical platform for studying the interplay of frustration, quantum fluctuations, and fractionalization in one dimension.

Realizing the ideal sawtooth chain in real materials remains challenging. Although candidate materials have been identified in a wide range of inorganic and metal-organic compounds~\cite{AvdeevSR2021,GiesterZfK1996,SobolevJAC2020,GnezdilovPRB2019,GarleaPRB2014,LauPRB2006,CavaJSSC1993,VanTendelooJSSC2001,LeBacqPRB2005,BaniodehNPJQM2018,InagakiJPSJ2005,RuizPerezInorgChem2000,KikuchiJPConfSeries2011,HeinzePRL2021,ZhuCM2023}, most of them deviate from the ideal model through longer-range exchange interactions, competing ferromagnetic (FM) and AFM couplings, or interchain interactions. In light of these realizations, numerous extensions of the sawtooth chain model have been investigated, including unequal base-base and base-vertex couplings ($J_{\mathrm{bb}}\!\neq\! J_{\mathrm{bv}}$)~\cite{blundell2004,Jiang2015,Rausch2025}, additional exchange interactions such as $J_{\mathrm{vv}}$~\cite{ChenPRL2001,ChenPRB2003}, Dzyaloshinskii--Moriya (DM) terms~\cite{HaoPRB2011}, and mixed-spin variants~\cite{ChandraPRB204}. Nevertheless, the robustness of sawtooth chain physics against residual long-range and interchain couplings in material realizations remains an open question.

In this work, we investigate the mineral bobkingite, \ce{Cu5(OH)8Cl2(H2O)2}~\cite{HawthorneMineralMag2002}, as a potential realization of the sawtooth chain. Performing \textit{ab initio} calculations, we find that its magnetic lattice consists of weakly coupled sawtooth chains, with residual interactions strongly suppressed by the crystal geometry. Motivated by this proximity to the ideal one-dimensional limit, we examine the effects of the remaining three-dimensional couplings. We show that the resulting classical model exhibits an extensive manifold of nearly degenerate states with emergent two-dimensional character. Leading order quantum corrections, treated within spin-wave theory, partially lift this degeneracy while preserving its low-dimensional nature, yielding an emergent two-dimensional effective description. 
Using thermal Monte Carlo simulations, we identify a finite-temperature regime in which multiplanar fluctuations rooted in the classical chain degeneracy are stabilized. We further establish a connection between this degeneracy and the quantum correlations of isolated sawtooth chains~\cite{Samarakoon2017PRB,Samarakoon2018PRB}, suggesting that quantum features may persist at finite temperatures in the full three-dimensional model \cite{LakeNatPhys2005,Banerjee2016,Rousochatzakis2019}. 
Thus, bobkingite may be viewed as an exciting theoretical and experimental platform to explore quantum phenomena in low dimensions.

\section{Crystal structure}

While bobkingite has not been synthesized in a laboratory setting, there is a crystallographic report of its discovery in mineral deposits~\cite{HawthorneMineralMag2002}.  The crystal admits the relatively low symmetry space group $C2/m$, similar to some related copper hydroxyhalide minerals~\cite{NormanRMP2016}. In the original discovery, only the positions of \ce{Cu}, \ce{O}, and \ce{Cl} were reported, with the approximate positions of the \ce{OH} and \ce{H2O} groups quoted without resolving the \ce{H} positions. To obtain the atomic positions, we fully relaxed the structure using density functional theory (DFT), as implemented in VASP~\cite{kresse1993,kresse1996,kresse1999}, see Supplemental Material (SM)~\cite{SI} for methods and a full crystallographic report, as well as references therein. 

The relaxed structure, in the primitive unit cell, is shown in Fig.~\ref{fig:master_figure} from multiple perspectives.
We find only minor changes in the lattice parameters and volume, while preserving the $C2/m$ space-group symmetry. 
The structure consists of distorted kagome layers, formed across the $\mathbf{ab}$ plane (panel (c)), connected by copper dimers along the $\mathbf{c}$ lattice vector (panel (b)). 
In the primitive cell, there is a total of 5 copper sites, occupying 3 distinct Wyckoff positions, which in turn define 3 distinct local environments (panel (d)). 
\ce{Cu}(1) is six-fold coordinated with \ce{OH} (gray octahedra). Together with \ce{Cu}(2), which is four-fold coordinated with \ce{OH} (burgundy squares), they form a distorted kagome lattice. Above and below this sits \ce{Cu}(3), which is also four-fold coordinated (blue squares), but with two \ce{OH} towards the kagome layer, and two $\ce{H2O}$ away from the kagome layer.
The interlayer coupling between the resulting pyrochlore slabs occurs between neighboring \ce{Cu}(3) sites.

\section{Magnetic exchanges}

The copper ions \ce{Cu^{2+}}, with a $3d^9$ configuration, have strong covalency with the nearby ligands. Looking at \ce{Cu}(2) or \ce{Cu}(3), consider the local coordinate $\tilde{x}\tilde{y}\tilde{z}$ system, with $\tilde{z}$ perpendicular to the square like local environment and $\tilde{x}$ and $\tilde{y}$ aligned towards the \ce{O} of the squares. The crystal-field splitting places the copper $d_{\tilde{x}^2-\tilde{y}^2}$ orbital highest in energy, leaving it half filled. The orientation of this orbital towards the surrounding ligands maximizes Cu--O $\sigma$ hybridization. An analogous local basis can be defined for the \ce{Cu}(1) sites, where the half-filled $d_{\tilde{x}^2-\tilde{y}^2}$ orbital likewise points toward neighboring oxygens. Thus, every copper site has a half-filled electronic level, and the addition of electron-electron interactions will drive the system into an insulating state with effective spin-$1/2$ moments. See SM~\cite{SI} for a discussion of the density of states from DFT, and references therein.

\begin{figure}[t]
	\centering
    \begin{overpic}[width=1\columnwidth,percent,angle=0,grid=false,tics=2]{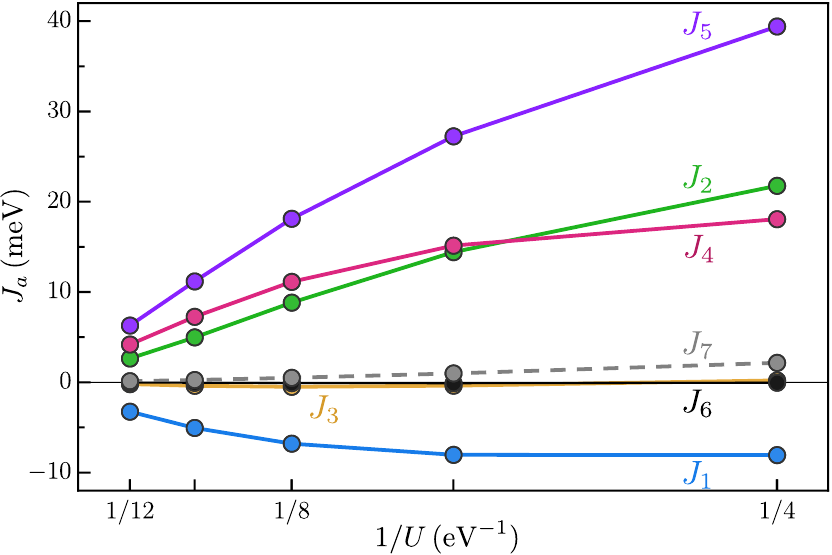}
    \end{overpic} 
    \caption{Heisenberg couplings extracted using TEMA up to the 7\textsuperscript{th}NN, using DFT+U, plotted as a function of $1/U$.}\label{fig:dft_res}
\end{figure}

\begin{table}
\caption{\label{tab:heis_mod} $a^{\text{th}}$NN Heisenberg couplings $J_a$, with bond lengths $d_a$, extracted from TEMA using $U\!=\!8\,\mathrm{eV}$ in the DFT+U calculations.}
\begin{ruledtabular}
\begin{tabular}{cd{3.5}|*{2}{d{3.5}}}
\multicolumn{1}{c}{$a$} &
\multicolumn{1}{c|}{$d_a$ (\AA)} &
\multicolumn{1}{c}{$J_a\,\mathrm{(meV)}$} &
\multicolumn{1}{c}{$J_a/|J_5|$} \\
\hline
 1 &    2.919 &    -6.79 &   -0.375 \\
 2 &    3.066 &     8.83 &    0.487 \\
 3 &    3.114 &    -0.49 &   -0.027 \\
 4 &    3.257 &    11.13 &    0.614 \\
 5 &    3.379 &    18.12 &    1.000 \\
 6 &    5.174 &    -0.08 &   -0.004 \\
 7 &    5.308 &     0.50 &    0.028 \\
\end{tabular}
\end{ruledtabular}
\end{table}

The strong Cu--O covalency mediates, via superexchange, interactions between the spin $1/2$ moments. Magnetocrystalline anisotropy in bobkingite is found to be of the order of $0.05\,\text{meV}$ per primitive cell, and a full microscopic derivation shows that the DM exchange~\cite{DzyaloshinskyJPCS1958,MoriyaPRL1960,MoriyaPR1960} is  an order of magnitude smaller than the Heisenberg exchange,
in agreement with other copper minerals~\cite{HerbAniso2008,HerbAniso2012,HerbAniso2025}, see SM~\cite{SI} for details. We thus only consider an isotropic Heisenberg model 
\begin{equation}\label{eq:spinmodel}
    \mathcal{H} = \sum\limits_{a=1}^7 \sum\limits_{\langle i,j \rangle^{(a)}} J_a \mathbf{S}_i \!\cdot \mathbf{S}_j,
\end{equation}
where $\langle i,j \rangle^{(a)}$ indicates summation of $i$, $j$ sites belonging to the $a^{\text{th}}$ nearest neighbor (NN). The respective couplings $J_{a}$ are depicted in Fig.~\ref{fig:master_figure} (a) and (c), up to the 7\textsuperscript{th}NN.
We extract the model couplings by the total energy mapping analysis (TEMA)~\cite{GlasbrennerNature2015,razpopov2023j,garcia2026microscopic}, which amounts to fitting the model to a set of spin-polarized DFT+U total-energy calculations, including an on-site electron-electron interaction $U$, see SM~\cite{SI} for details.
We plot the resulting couplings as a function of $1/U$ in Fig.~\ref{fig:dft_res}. 
The AFM exchanges follow the expected $1/U$ scaling. 
The dominant AFM exchange couplings, base-to-base $J_4\!=\!J_{\rm bb}$ and base-to-vertex $J_5\!=\!J_{\rm bv}$, form sawtooth chains as seen in Fig.~\ref{fig:master_figure}.
The leading interchain interactions are given by FM $J_1$ and AFM $J_2$. 
Since the first five exchange paths span similar Cu--Cu distances (2.9--3.4\,\AA), while the sixth-neighbor bond is much longer in legth (5.17\,\AA), $J_6$ is orders of magnitude smaller than the leading exchanges, justifying the truncation of the model. 
However, we also keep $J_7$, due to its special role in the model, as the first interaction to couple interchain base sites along the $\mathbf{a}$-axis. 
Its significance will be apparent below when we analyze the classical ground states.

Although $J_4$ and $J_5$ are symmetry inequivalent, their overall scale equivalence in bobkingite follows from the crystal structure. 
They correspond to \ce{Cu}(2)--\ce{Cu}(2) and \ce{Cu}(2)--\ce{Cu}(3) paths within similar corner-sharing \ce{Cu}--\ce{OH} units, with comparable bond angles ($113.4^\circ$ and $115.9^\circ$), distances (1.94--2.00\,\AA), and hydroxyl orientations. 
As a result, both exchanges proceed via similar superexchange paths, consistent with the Goodenough--Kanamori--Anderson rules, yielding comparable in magnitude AFM couplings. Nevertheless, the small differences in the bonds angles, and the exact orientation of the hydroxyl, can result in significantly different superexchange values ~\cite{LiPRL2018}, resulting in  $J_4 \!\neq\! J_5$. 
As a caveat, 
we note that some of the bond angles obtained after the full structural relaxation differ from those obtained from X-ray measurements in Ref.~\cite{HawthorneMineralMag2002}.
This provides additional impetus for attempting to grow bobkingite crystals and performing a more detailed crystallographic study.

The other three similar-length bond couplings, $J_1$, $J_2$ and $J_3$, involve edge-sharing geometries, where exchange is more sensitive to microscopic details and can involve competing FM and AFM contributions. Their superexchange pathways rely on mixed $\sigma$ and ligand-mediated $\pi$ processes, rather than the predominantly $\sigma$--$\sigma$ paths of the corner-sharing bonds. Consequently, $J_4$ and $J_5$ dominate, although $J_1$ and $J_2$ are sizable.



The lack of experimental data prevents us from refining the value of $U$ by comparison with measured quantities, such as the Curie--Weiss temperature extracted from magnetic susceptibility measurements. We therefore choose a reasonable value of $U\!=\!8\,\text{eV}$ for Cu and adopt the corresponding calculated exchange couplings, listed in Table~\ref{tab:heis_mod}, as our representative parameter set.
The ratio $J_{\rm bb}/J_{\rm bv}\!=\!J_4/J_5\!\simeq\!0.61$ places bobkingite in the dimerized gapped phase of the quantum sawtooth chain~\cite{blundell2004}, where fractionalized kinks and antikinks form a spin-triplet excitation gap at $k\!=\!0$. 
Given that the crystal consists of coupled chains, with some interchain couplings being of non-negligible magnitude, we focus in what follows on understanding the three-dimensional model at the classical level.

\section{Weakly coupled sawtooth chain model}
 
\begin{figure}[t]
	\centering
    \begin{overpic}[width=1.0\columnwidth,percent,angle=0,grid=false,tics=2]{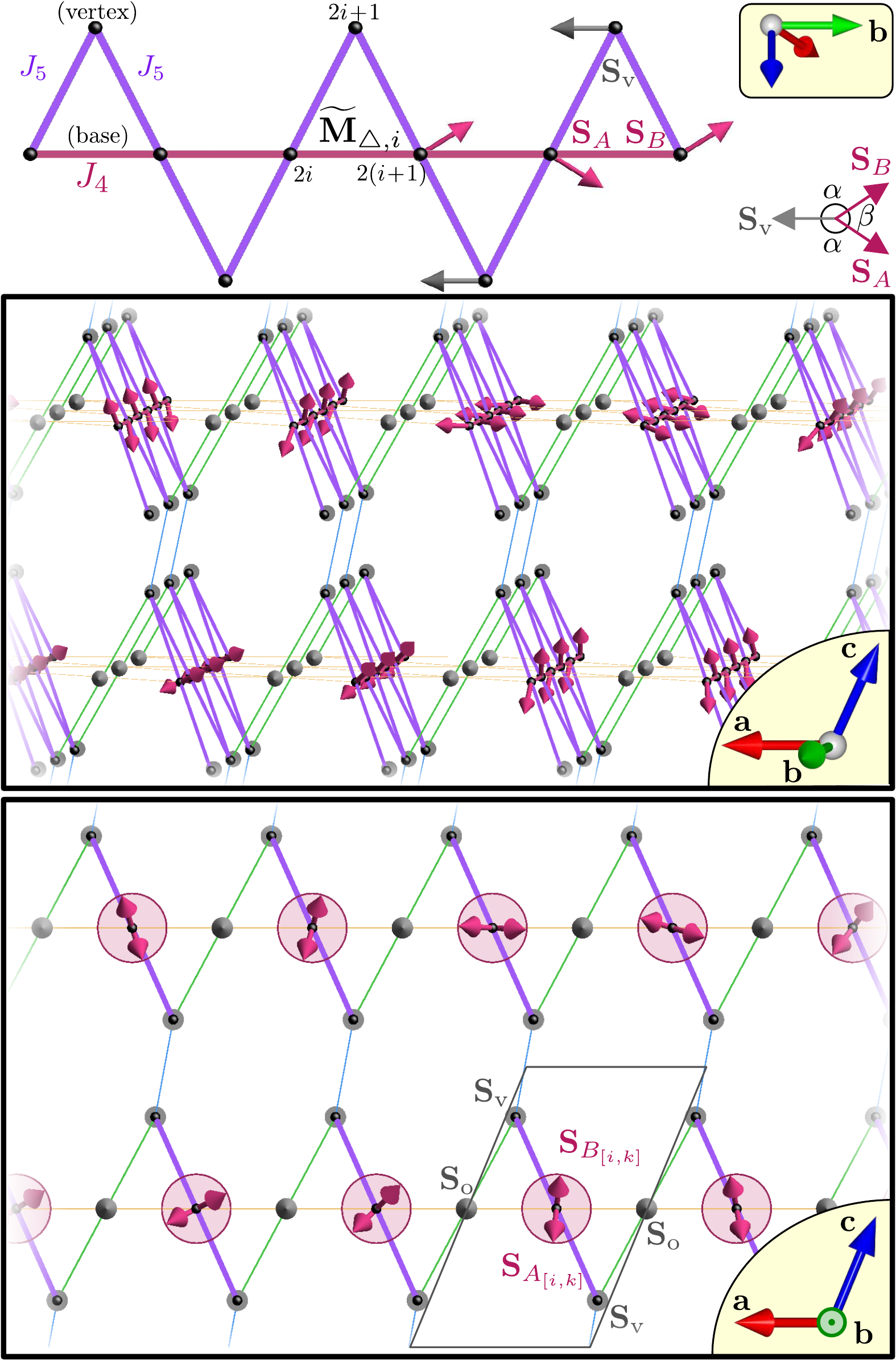}
        \put(0,98.05){(a)}
        \put(0.5,75.8){(b)}
        \put(0.5,38.8){(c)}
    \end{overpic} 
    \caption{(a) Primary definitions of the isolated sawtooth chain, and an example of coplanar order, with vertex spins $\mathbf{S}_{\text{v}}$ and base spins $\mathbf{S}_{A(B)}$. (b) and (c) show the simplest order formed in the three-dimensional model from two perspectives. Each sawtooth chain, running along the $\mathbf{b}$ axis, retains an overall rotational freedom $\psi_{[i,k]}$, highlighted by the red discs, where $i$ and $k$ denote the integer chain positions along $\mathbf{a}$ and $\mathbf{c}$, respectively.}\label{fig:bob_states}
\end{figure}

We investigate the full three-dimensional structure of coupled sawtooth chains in bobkingite via Monte Carlo simulations on clusters built out of the primitive cell, ranging from $1\!\times\!1\!\times\!1$ up to $12\!\times\!12\!\times\!12$ clusters, truncating the model at the 6\textsuperscript{th}NN, see SM~\cite{SI} for methods, and references therein. The $T=0$ converged classical ground-state energy is size independent and given by $\varepsilon_{\text{cl}}\!=\!\frac{1}{5}[J_1 - 2 J_2 +2J_4\cos(2\alpha)+ 4(J_5-J_3+J_6)\cos\alpha]$ per site, with $\alpha\!=\!\arccos[-(J_5-J_3+J_6)/(2 J_4)]$ the angle between base spins and vertex spins, see Fig.~\ref{fig:bob_states}(a).

To interpret this, we briefly recall the one-dimensional sawtooth chain, corresponding to the sawtooth $J_4-J_5$ model, Fig~\ref{fig:bob_states}(a), running along the $\mathbf{b}$ lattice vector.  In the ideal $J_4=J_5$ limit, the ground state exhibits a vast system-size scaling degeneracy; the simplest representatives are coplanar states in which spins form $120^\circ$ angles,  enforced by the vanishing of the triangle magnetization  $\mathbf{M}_{\triangle,i}\!=\! \mathbf{S}_{2i} \!+\!\mathbf{S}_{2i + 1} \!+\! \mathbf{S}_{2(i+1)}$  on every triangle. Away from the ideal limit, $J_4 \neq J_5$, this structure persists with a generalized conserved quantity 
 $\widetilde{\mathbf{M}}_{\triangle,i}\!=\! \mathbf{S}_{2i} \!-\! 2\cos\alpha\mathbf{S}_{2i + 1} \!+\! \mathbf{S}_{2(i+1)}$ where $\alpha\!=\!\arccos[-J_5/(2 J_4)]$, 
whose vanishing fixes the relative angles $\angle (\mathbf{S}_{A}, \mathbf{S}_{\text{v}}) \!=\! \angle (\mathbf{S}_{B}, \mathbf{S}_{\text{v}}) \!=\! \alpha$ and $\angle (\mathbf{S}_{A}, \mathbf{S}_{B}) \!=\! 2\pi-2\alpha=\beta$.
A detailed analysis of the classical one-dimensional sawtooth chain is presented in the \hyperref[EndMatter]{End Matter}.

Returning to the full three-dimensional structure, each sawtooth chain running along $\mathbf{b}$ realizes such a coplanar state, with $\alpha$ minimally renormalized by the interchain couplings, see Fig~\ref{fig:bob_states}(b). Chains couple vertically along $\mathbf{c}$ via FM $J_1$, promoting alignment of the vertex spins $\mathbf{S}_{\rm v}$ across stacked chains, and horizontally along $\mathbf{a}$ via AFM $J_2$ through an intermediate site $\mathbf{S}_{\text{o}}\!=\!-\mathbf{S}_{\rm v}$, which likewise promotes $\mathbf{S}_{\rm v}$ alignment across neighboring chains. The smaller couplings $J_3$ and $J_6$, do not qualitatively alter this picture, and being so small they leave $\mathbf{S}_{\text{o}}$ almost as a dangling site, which makes the sizable $J_2$ unfrustrated. See SM~\cite{SI} for additional details.

Although the vertex spins are ordered, a degeneracy remains: each chain $[i,k]$, labeled by its position in the $\mathbf{ac}$-plane, retains an independent rotational freedom $\psi_{i,k}$ about the $\mathbf{b}$ axis, see Fig~\ref{fig:bob_states}(c). With $N_a N_c$ chains in the system, this yields a ground-state degeneracy scaling as $SO(2)^{N_a N_c}$, revealing the emergent two-dimensional character of the classical manifold. Order by disorder will lift this classical degeneracy, leaving an Ising degree of freedom per chain,  see SM~\cite{SI} for details.

\begin{figure}[t]
	\centering
    \begin{overpic}[width=1.0\columnwidth,percent,angle=0,grid=false,tics=2]{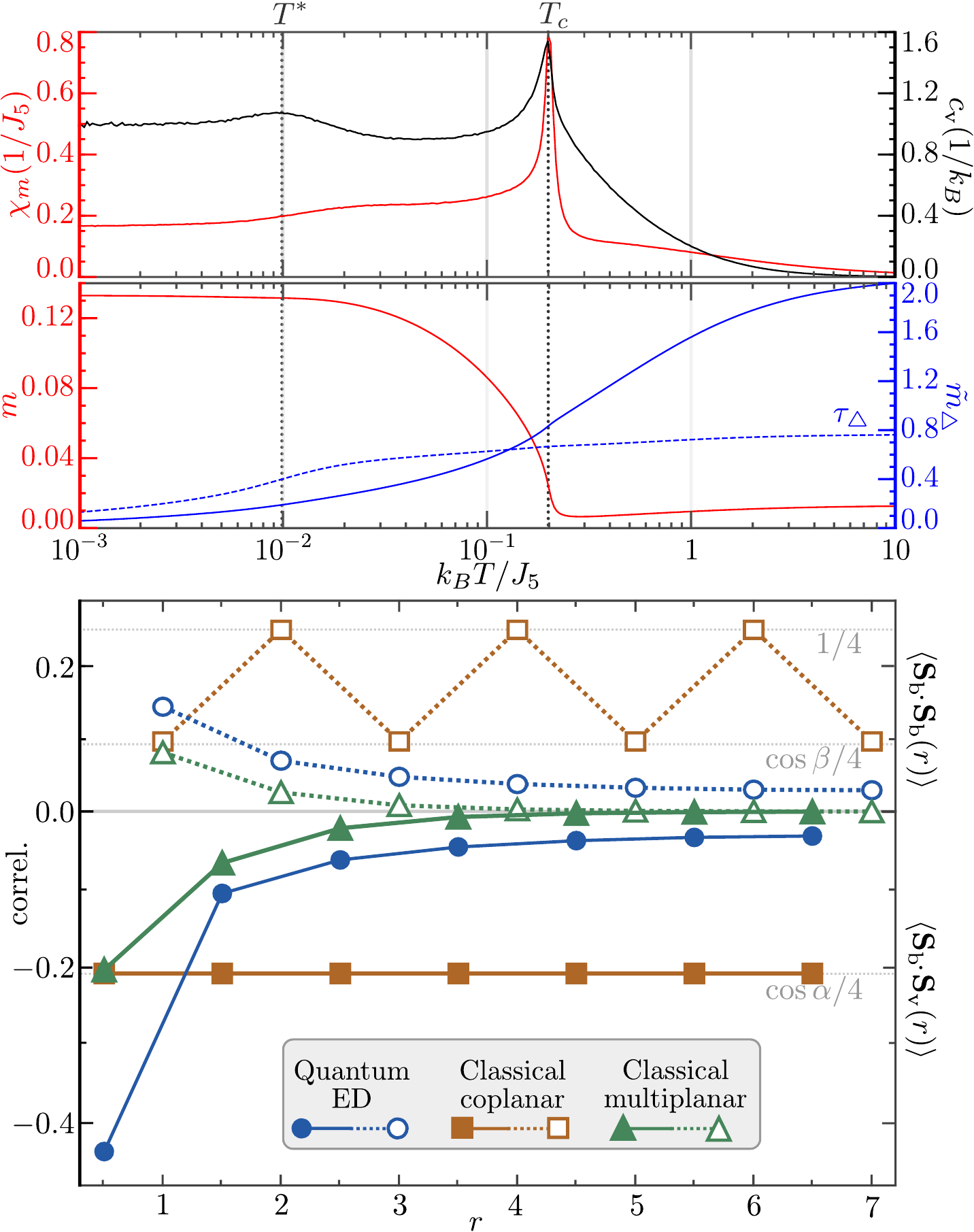}
        \put(0,97.5){(a)}
        \put(0,49.5){(b)}
    \end{overpic} 
    \caption{(a) Thermal Monte Carlo results for the $J_1 - J_6$ spin model for $L=10$, plotted against a logarithmic temperature scale. The heat capacity $c_v$, magnetic susceptibility $\chi_m$, magnetization $m$, average generalized triangle magnetization $m_{\triangle}$, and average NN torsion $\tau_{\triangle}$ are plotted. The second order magnetic transition $T_c$ and crossover from multiplanar fluctuations to coplanar order $T^*$ are also shown. (b) The base-to-base (dotted) and base-to-vertex (solid) quantum correlators (blue) as a function of distance, obtained from exact diagonalization of the $S=1/2$ sawtooth chain. As a reference the coplanar (brown) and average multiplanar (green) classical correlators are also plotted, with the coplanar correlators (canonically repeating) bound by the special values $1/4$, $\cos\beta/4$, and $\cos\alpha/4$. The average multiplanar correlators shows a strikingly similar behavior to the quantum correlators.}\label{fig:MCEDres}
\end{figure}

We perform parallel tempering Monte-Carlo simulations over a wide range of temperatures on symmetric $L\!\!\times\!\! L\!\!\times\!\! L$ clusters, with the full simulation details in the SM~\cite{SI}. In Fig.~\ref{fig:MCEDres}(a) we plot the most significant thermal Monte-Carlo results for $L\!=\!10$. The heat capacity per site $c_v$ shows two temperature scales. At $k_BT_c \!=\! 0.194 J_5$ there is a sharp peak that grows with system size, characteristic of a second order phase transition. This is corroborated by a strong peak in the magnetic susceptibility $\chi_m$, and the canonical second order behavior of the magnetization $m$, saturating at $(\sqrt{2\!+\!2\cos\beta}-1)/5 \!=\! 0.1332$ due to the net ferromagnetic moment along the base-base direction. More interestingly, there is a very low temperature $k_BT^* \!=\! 0.0095 J_5$ heat capacity bump, for which the magnetic susceptibility only shows  a very weak shoulder behavior. 

To understand the $T^*$ feature, we turn to the sawtooth chain. As discussed above, $\widetilde{\mathbf{M}}_{\triangle,i}$ must vanish in the classical ground state. Consistently, we observe a sharp drop in the average $m_{\triangle}$ at $T_c$, indicating the onset of local correlations that enforce the local constraints. To distinguish coplanar from multiplanar fluctuations on the sawtooth lattice, we measure the nearest-neighbor (NN) triangle torsion, $\tau_{\triangle}$, defined as the average of $\|\mathbf{n}_{l}\!\times\!\mathbf{n}_{l'}\|$ over neighboring sawtooth triangles $l$ and $l'$, where $\mathbf{n}_{l}$ denotes the unit normal to the spin plane of triangle $l$ (i.e., the vector chirality).
In the coplanar sawtooth state, the torsion must vanish identically, whereas it remains finite in multiplanar configurations. We find that the torsion remains finite and largely unaffected across $T_c$, before dropping sharply towards zero at $T^*$, signaling the onset of coplanar order. Thus, the intermediate temperature regime $T^* < T < T_c$ is only partially ordered, with multiplanar fluctuations remaining prevalent throughout the system.

The above result is particularly intriguing when juxtaposed with exact diagonalization results on the $S\!=\!1/2$ isolated one-dimensional sawtooth chain, with the quantum correlators shown in Fig.~\ref{fig:MCEDres}(b), along with the classical coplanar and averaged multiplanar correlators, see SM~\cite{SI} for details.
The quantum and classical multiplanar correlators show a qualitative similarity in both signs and distance decay, reminiscent of classical-to-quantum correspondences observed in other frustrated magnets with extensive classical degeneracies \cite{Samarakoon2017PRB,Samarakoon2018PRB}.
Given the finite-temperature window with multiplanar fluctuations in the three-dimensional classical model, this suggests that features of the quantum sawtooth dimerized phase will survive in the full quantum model, analogous to coupled-chain and proximate spin-liquid regimes \cite{LakeNatPhys2005,Banerjee2016,Rousochatzakis2019}.

We now look at the effect of the AFM $J_7$,  which will inevitably induce order within the $\mathbf{ab}$ plane at a temperature of $k_BT \!\approx\! 0.028 J_5$, which is already between $T^*$ and $T_c$, letting the intermediate multiplanar regime survive. Note that additional interchain interactions are also found to be AFM, see SM~\cite{SI}, which will compete with $J_7$ and further suppress this scale. Couplings along the more widely spaced $\mathbf{c}$ axis remain negligible, so that neighboring planes retain an $SO(2)$ rotational freedom relative to one another. Consequently, even in the presence of $J_7$, a one-dimensional system-size scaling degeneracy persists vertically (along $\mathbf{c}$), although this would be distinct and unrelated to sawtooth chain physics.

\section{Discussion}

Summarizing, bobkingite emerges as a near-ideal platform for sawtooth chain physics. Although not precisely at the ideal $J_4\!=\!J_5$ limit, it lies close enough to host massive system-size scaling degeneracies. The remaining interchain couplings are sufficiently weak that the system is well described as weakly coupled sawtooth chains, from which emergent two-dimensional degeneracies arise in the classical model and survive to leading-order quantum fluctuations as Ising degrees of freedom. Anisotropic terms are negligible, confirming this as the correct leading-order picture.

Strikingly, unlike other sawtooth candidates where horizontal and vertical interchain couplings are comparable, bobkingite possesses only a small horizontal coupling $J_7$ and essentially vanishing vertical ones. This asymmetry has profound consequences: even when $J_7$ drives ordering within the $\mathbf{ab}$ plane, a one-dimensional system-size scaling degeneracy persists along $\mathbf{c}$. The resulting in-plane order is therefore weak. Thermal fluctuations are shown to stabilize a finite-temperature multiplanar fluctuating phase.  This points to a cascade of transitions, first into a fluctuating multiplanar state, and then into an emergent two-dimensional classical state. 

The prospect of synthesizing bobkingite samples of sufficient quality to probe these phenomena experimentally makes it a particularly exciting target. It is further compelling to consider the fate of these emergent degeneracies in the full quantum model, where fractionalized excitations and beyond-linear spin-wave effects may give rise to yet richer behavior. Given that the quantum sawtooth chain exhibits a relationship to the classical multiplanar degeneracy~\cite{Samarakoon2017PRB,Samarakoon2018PRB}, quantum aspects of the sawtooth are expected to survive in the full three-dimensional model~\cite{LakeNatPhys2005,Banerjee2016,Rousochatzakis2019}.

\section{Acknowledgments}

P.P.S., A.R., D.A.S.K, and R.V. thank the Deutsche Forschungsgemeinschaft (DFG, German
Research Foundation) for funding through FOR5249 - 449872909
(Project P4) and TRR 288 —422213477 (Projects A05, B05). M.R.N. was supported by the Materials Sciences and Engineering Division, Basic Energy Sciences, Office of Science, US Dept.~of Energy.  A.S.B. and H.L. acknowledge support from NSF Grant No. DMR-2323971 and the ASU Research Computing Center for HPC resources.

\section{Data availability}

The data are available from the authors upon reasonable request.
\nocite{perdew1996,liechtenstein1995,PustogowPRB2017,PBEsol,Fazekas1999,WinterPRB2016,RiedlPRB2016,KoniecznaNPJQM2025,Rodrigues1840,fploWF,fplocite1,fplocite2,fploRel,CarterPRB2012,YildirimPRB1995,SwendsenPRL1986,HukushimaJPSJ1996,LakshmananLLGeq2011,CreutzPRD1987,KankiPRD2005,lehoucq1998arpack}
\IfFileExists{round.2/bibliography/thy_refs.bib}{%
    \bibliography{
        round.2/bibliography/jour_name_abbreviation,
        round.2/bibliography/thy_refs
    }%
}{%
    \IfFileExists{../bibliography/thy_refs.bib}{%
        \bibliography{
            ../bibliography/jour_name_abbreviation,
            ../bibliography/thy_refs
        }%
    }{%
        \bibliography{
            bibliography/jour_name_abbreviation,
            bibliography/thy_refs
        }%

\begin{thebibliography}{76}%
\makeatletter
\providecommand \@ifxundefined [1]{%
 \@ifx{#1\undefined}
}%
\providecommand \@ifnum [1]{%
 \ifnum #1\expandafter \@firstoftwo
 \else \expandafter \@secondoftwo
 \fi
}%
\providecommand \@ifx [1]{%
 \ifx #1\expandafter \@firstoftwo
 \else \expandafter \@secondoftwo
 \fi
}%
\providecommand \natexlab [1]{#1}%
\providecommand \enquote  [1]{``#1''}%
\providecommand \bibnamefont  [1]{#1}%
\providecommand \bibfnamefont [1]{#1}%
\providecommand \citenamefont [1]{#1}%
\providecommand \href@noop [0]{\@secondoftwo}%
\providecommand \href [0]{\begingroup \@sanitize@url \@href}%
\providecommand \@href[1]{\@@startlink{#1}\@@href}%
\providecommand \@@href[1]{\endgroup#1\@@endlink}%
\providecommand \@sanitize@url [0]{\catcode `\\12\catcode `\$12\catcode
  `\&12\catcode `\#12\catcode `\^12\catcode `\_12\catcode `\%12\relax}%
\providecommand \@@startlink[1]{}%
\providecommand \@@endlink[0]{}%
\providecommand \url  [0]{\begingroup\@sanitize@url \@url }%
\providecommand \@url [1]{\endgroup\@href {#1}{\urlprefix }}%
\providecommand \urlprefix  [0]{URL }%
\providecommand \Eprint [0]{\href }%
\providecommand \doibase [0]{https://doi.org/}%
\providecommand \selectlanguage [0]{\@gobble}%
\providecommand \bibinfo  [0]{\@secondoftwo}%
\providecommand \bibfield  [0]{\@secondoftwo}%
\providecommand \translation [1]{[#1]}%
\providecommand \BibitemOpen [0]{}%
\providecommand \bibitemStop [0]{}%
\providecommand \bibitemNoStop [0]{.\EOS\space}%
\providecommand \EOS [0]{\spacefactor3000\relax}%
\providecommand \BibitemShut  [1]{\csname bibitem#1\endcsname}%
\let\auto@bib@innerbib\@empty
\bibitem [{\citenamefont {Momma}\ and\ \citenamefont {Izumi}(2011)}]{VESTA}%
  \BibitemOpen
  \bibfield  {author} {\bibinfo {author} {\bibfnamefont {K.}~\bibnamefont
  {Momma}}\ and\ \bibinfo {author} {\bibfnamefont {F.}~\bibnamefont {Izumi}},\
  }\bibfield  {title} {\bibinfo {title} {{{\it VESTA3} for three-dimensional
  visualization of crystal, volumetric and morphology data}},\ }\href
  {https://doi.org/10.1107/S0021889811038970} {\bibfield  {journal} {\bibinfo
  {journal} {J. Appl. Crystallogr.}\ }\textbf {\bibinfo {volume} {44}},\
  \bibinfo {pages} {1272} (\bibinfo {year} {2011})}\BibitemShut {NoStop}%
\bibitem [{\citenamefont {Giamarchi}(2003)}]{GiamarchiBook2003}%
  \BibitemOpen
  \bibfield  {author} {\bibinfo {author} {\bibfnamefont {T.}~\bibnamefont
  {Giamarchi}},\ }\href
  {https://doi.org/10.1093/acprof:oso/9780198525004.001.0001} {\emph {\bibinfo
  {title} {Quantum Physics in One Dimension}}}\ (\bibinfo  {publisher} {Oxford
  University Press},\ \bibinfo {year} {2003})\BibitemShut {NoStop}%
\bibitem [{\citenamefont {Balents}(2010)}]{balents2010spin}%
  \BibitemOpen
  \bibfield  {author} {\bibinfo {author} {\bibfnamefont {L.}~\bibnamefont
  {Balents}},\ }\bibfield  {title} {\bibinfo {title} {Spin liquids in
  frustrated magnets},\ }\href {https://doi.org/10.1038/nature08917} {\bibfield
   {journal} {\bibinfo  {journal} {Nature}\ }\textbf {\bibinfo {volume}
  {464}},\ \bibinfo {pages} {199} (\bibinfo {year} {2010})}\BibitemShut
  {NoStop}%
\bibitem [{\citenamefont {Savary}\ and\ \citenamefont
  {Balents}(2017)}]{savary2017quantum}%
  \BibitemOpen
  \bibfield  {author} {\bibinfo {author} {\bibfnamefont {L.}~\bibnamefont
  {Savary}}\ and\ \bibinfo {author} {\bibfnamefont {L.}~\bibnamefont
  {Balents}},\ }\bibfield  {title} {\bibinfo {title} {Quantum spin liquids: a
  review},\ }\href {https://doi.org/10.1088/0034-4885/80/1/016502} {\bibfield
  {journal} {\bibinfo  {journal} {Rep. Prog. Phys.}\ }\textbf {\bibinfo
  {volume} {80}},\ \bibinfo {pages} {016502} (\bibinfo {year}
  {2017})}\BibitemShut {NoStop}%
\bibitem [{\citenamefont {Nayak}\ \emph {et~al.}(2008)\citenamefont {Nayak},
  \citenamefont {Simon}, \citenamefont {Stern}, \citenamefont {Freedman},\ and\
  \citenamefont {Das~Sarma}}]{NayakRMP2008}%
  \BibitemOpen
  \bibfield  {author} {\bibinfo {author} {\bibfnamefont {C.}~\bibnamefont
  {Nayak}}, \bibinfo {author} {\bibfnamefont {S.~H.}\ \bibnamefont {Simon}},
  \bibinfo {author} {\bibfnamefont {A.}~\bibnamefont {Stern}}, \bibinfo
  {author} {\bibfnamefont {M.}~\bibnamefont {Freedman}},\ and\ \bibinfo
  {author} {\bibfnamefont {S.}~\bibnamefont {Das~Sarma}},\ }\bibfield  {title}
  {\bibinfo {title} {Non-abelian anyons and topological quantum computation},\
  }\href {https://doi.org/10.1103/RevModPhys.80.1083} {\bibfield  {journal}
  {\bibinfo  {journal} {Rev. Mod. Phys.}\ }\textbf {\bibinfo {volume} {80}},\
  \bibinfo {pages} {1083} (\bibinfo {year} {2008})}\BibitemShut {NoStop}%
\bibitem [{\citenamefont {Field}\ and\ \citenamefont
  {Simula}(2018)}]{FieldQST2018}%
  \BibitemOpen
  \bibfield  {author} {\bibinfo {author} {\bibfnamefont {B.}~\bibnamefont
  {Field}}\ and\ \bibinfo {author} {\bibfnamefont {T.}~\bibnamefont {Simula}},\
  }\bibfield  {title} {\bibinfo {title} {Introduction to topological quantum
  computation with non-abelian anyons},\ }\href
  {https://doi.org/10.1088/2058-9565/aacad2} {\bibfield  {journal} {\bibinfo
  {journal} {Quantum Sci. Technol.}\ }\textbf {\bibinfo {volume} {3}},\
  \bibinfo {pages} {045004} (\bibinfo {year} {2018})}\BibitemShut {NoStop}%
\bibitem [{\citenamefont {Bethe}(1935)}]{bethe1935statistical}%
  \BibitemOpen
  \bibfield  {author} {\bibinfo {author} {\bibfnamefont {H.~A.}\ \bibnamefont
  {Bethe}},\ }\bibfield  {title} {\bibinfo {title} {Statistical theory of
  superlattices},\ }\href {https://doi.org/10.1098/rspa.1935.0122} {\bibfield
  {journal} {\bibinfo  {journal} {Proc. R. Soc. A}\ }\textbf {\bibinfo {volume}
  {150}},\ \bibinfo {pages} {552} (\bibinfo {year} {1935})}\BibitemShut
  {NoStop}%
\bibitem [{\citenamefont {Faddeev}\ and\ \citenamefont
  {Takhtajan}(1981)}]{faddeev1981spin}%
  \BibitemOpen
  \bibfield  {author} {\bibinfo {author} {\bibfnamefont {L.}~\bibnamefont
  {Faddeev}}\ and\ \bibinfo {author} {\bibfnamefont {L.}~\bibnamefont
  {Takhtajan}},\ }\bibfield  {title} {\bibinfo {title} {What is the spin of a
  spin wave?},\ }\href
  {https://doi.org/https://doi.org/10.1016/0375-9601(81)90335-2} {\bibfield
  {journal} {\bibinfo  {journal} {Phys. Lett. A}\ }\textbf {\bibinfo {volume}
  {85}},\ \bibinfo {pages} {375} (\bibinfo {year} {1981})}\BibitemShut
  {NoStop}%
\bibitem [{\citenamefont {Nakamura}\ and\ \citenamefont
  {Kubo}(1996)}]{NakamuraPRB1996}%
  \BibitemOpen
  \bibfield  {author} {\bibinfo {author} {\bibfnamefont {T.}~\bibnamefont
  {Nakamura}}\ and\ \bibinfo {author} {\bibfnamefont {K.}~\bibnamefont
  {Kubo}},\ }\bibfield  {title} {\bibinfo {title} {Elementary excitations in
  the \ensuremath{\Delta} chain},\ }\href
  {https://doi.org/10.1103/PhysRevB.53.6393} {\bibfield  {journal} {\bibinfo
  {journal} {Phys. Rev. B}\ }\textbf {\bibinfo {volume} {53}},\ \bibinfo
  {pages} {6393} (\bibinfo {year} {1996})}\BibitemShut {NoStop}%
\bibitem [{\citenamefont {Sen}\ \emph {et~al.}(1996)\citenamefont {Sen},
  \citenamefont {Shastry}, \citenamefont {Walstedt},\ and\ \citenamefont
  {Cava}}]{ShastryPRB1996}%
  \BibitemOpen
  \bibfield  {author} {\bibinfo {author} {\bibfnamefont {D.}~\bibnamefont
  {Sen}}, \bibinfo {author} {\bibfnamefont {B.~S.}\ \bibnamefont {Shastry}},
  \bibinfo {author} {\bibfnamefont {R.~E.}\ \bibnamefont {Walstedt}},\ and\
  \bibinfo {author} {\bibfnamefont {R.}~\bibnamefont {Cava}},\ }\bibfield
  {title} {\bibinfo {title} {Quantum solitons in the sawtooth lattice},\ }\href
  {https://doi.org/10.1103/PhysRevB.53.6401} {\bibfield  {journal} {\bibinfo
  {journal} {Phys. Rev. B}\ }\textbf {\bibinfo {volume} {53}},\ \bibinfo
  {pages} {6401} (\bibinfo {year} {1996})}\BibitemShut {NoStop}%
\bibitem [{\citenamefont {Hao}\ and\ \citenamefont
  {Tchernyshyov}(2009)}]{hao2009fermionic}%
  \BibitemOpen
  \bibfield  {author} {\bibinfo {author} {\bibfnamefont {Z.}~\bibnamefont
  {Hao}}\ and\ \bibinfo {author} {\bibfnamefont {O.}~\bibnamefont
  {Tchernyshyov}},\ }\bibfield  {title} {\bibinfo {title} {Fermionic spin
  excitations in two- and three-dimensional antiferromagnets},\ }\href
  {https://doi.org/10.1103/PhysRevLett.103.187203} {\bibfield  {journal}
  {\bibinfo  {journal} {Phys. Rev. Lett.}\ }\textbf {\bibinfo {volume} {103}},\
  \bibinfo {pages} {187203} (\bibinfo {year} {2009})}\BibitemShut {NoStop}%
\bibitem [{\citenamefont {Monti}\ and\ \citenamefont
  {S\"ut\H{o}}(1991)}]{MontiPLA1991}%
  \BibitemOpen
  \bibfield  {author} {\bibinfo {author} {\bibfnamefont {F.}~\bibnamefont
  {Monti}}\ and\ \bibinfo {author} {\bibfnamefont {A.}~\bibnamefont
  {S\"ut\H{o}}},\ }\bibfield  {title} {\bibinfo {title} {Spin-$1/2$
  {H}eisenberg model on $\delta$ trees},\ }\href
  {https://doi.org/https://doi.org/10.1016/0375-9601(91)90937-4} {\bibfield
  {journal} {\bibinfo  {journal} {Phys. Lett. A}\ }\textbf {\bibinfo {volume}
  {156}},\ \bibinfo {pages} {197} (\bibinfo {year} {1991})}\BibitemShut
  {NoStop}%
\bibitem [{\citenamefont {Nawa}\ \emph {et~al.}(2021)\citenamefont {Nawa},
  \citenamefont {Avdeev}, \citenamefont {Berdonosov}, \citenamefont {Sobolev},
  \citenamefont {Presniakov}, \citenamefont {Aslandukova}, \citenamefont
  {Kozlyakova}, \citenamefont {Vasiliev}, \citenamefont {Shchetinin},\ and\
  \citenamefont {Sato}}]{AvdeevSR2021}%
  \BibitemOpen
  \bibfield  {author} {\bibinfo {author} {\bibfnamefont {K.}~\bibnamefont
  {Nawa}}, \bibinfo {author} {\bibfnamefont {M.}~\bibnamefont {Avdeev}},
  \bibinfo {author} {\bibfnamefont {P.}~\bibnamefont {Berdonosov}}, \bibinfo
  {author} {\bibfnamefont {A.}~\bibnamefont {Sobolev}}, \bibinfo {author}
  {\bibfnamefont {I.}~\bibnamefont {Presniakov}}, \bibinfo {author}
  {\bibfnamefont {A.}~\bibnamefont {Aslandukova}}, \bibinfo {author}
  {\bibfnamefont {E.}~\bibnamefont {Kozlyakova}}, \bibinfo {author}
  {\bibfnamefont {A.}~\bibnamefont {Vasiliev}}, \bibinfo {author}
  {\bibfnamefont {I.}~\bibnamefont {Shchetinin}},\ and\ \bibinfo {author}
  {\bibfnamefont {T.~J.}\ \bibnamefont {Sato}},\ }\bibfield  {title} {\bibinfo
  {title} {Magnetic structure study of the sawtooth chain antiferromagnet
  \ce{Fe2Se2O7}},\ }\href {https://doi.org/10.1038/s41598-021-03058-5}
  {\bibfield  {journal} {\bibinfo  {journal} {Sci. Rep.}\ }\textbf {\bibinfo
  {volume} {11}},\ \bibinfo {pages} {24049} (\bibinfo {year}
  {2021})}\BibitemShut {NoStop}%
\bibitem [{\citenamefont {Giester}(1996)}]{GiesterZfK1996}%
  \BibitemOpen
  \bibfield  {author} {\bibinfo {author} {\bibfnamefont {G.}~\bibnamefont
  {Giester}},\ }\bibfield  {title} {\bibinfo {title} {Crystal structure of
  \ce{Fe2O(SeO3)2}, a new oxoselenite compound with ferric iron in distorted
  tetrahedral coordination},\ }\href
  {https://doi.org/doi:10.1524/zkri.1996.211.9.603} {\bibfield  {journal}
  {\bibinfo  {journal} {Z. Kristallogr. Cryst. Mater.}\ }\textbf {\bibinfo
  {volume} {211}},\ \bibinfo {pages} {603} (\bibinfo {year}
  {1996})}\BibitemShut {NoStop}%
\bibitem [{\citenamefont {Sobolev}\ \emph {et~al.}(2020)\citenamefont
  {Sobolev}, \citenamefont {Aslandukova}, \citenamefont {Kozlyakova},
  \citenamefont {Kuznetsova}, \citenamefont {Akhrorov}, \citenamefont
  {Berdonosov}, \citenamefont {Glazkova}, \citenamefont {Volkova},
  \citenamefont {Vasiliev},\ and\ \citenamefont {Presniakov}}]{SobolevJAC2020}%
  \BibitemOpen
  \bibfield  {author} {\bibinfo {author} {\bibfnamefont {A.~V.}\ \bibnamefont
  {Sobolev}}, \bibinfo {author} {\bibfnamefont {A.~A.}\ \bibnamefont
  {Aslandukova}}, \bibinfo {author} {\bibfnamefont {E.~S.}\ \bibnamefont
  {Kozlyakova}}, \bibinfo {author} {\bibfnamefont {E.~S.}\ \bibnamefont
  {Kuznetsova}}, \bibinfo {author} {\bibfnamefont {A.~Y.}\ \bibnamefont
  {Akhrorov}}, \bibinfo {author} {\bibfnamefont {P.~S.}\ \bibnamefont
  {Berdonosov}}, \bibinfo {author} {\bibfnamefont {I.~S.}\ \bibnamefont
  {Glazkova}}, \bibinfo {author} {\bibfnamefont {O.~S.}\ \bibnamefont
  {Volkova}}, \bibinfo {author} {\bibfnamefont {A.~N.}\ \bibnamefont
  {Vasiliev}},\ and\ \bibinfo {author} {\bibfnamefont {I.~A.}\ \bibnamefont
  {Presniakov}},\ }\bibfield  {title} {\bibinfo {title} {Magnetic hyperfine
  interactions in a sawtooth chain iron oxoselenite \ce{Fe2O(SeO3)2}:
  Experimental and theoretical investigation},\ }\href
  {https://doi.org/https://doi.org/10.1016/j.jallcom.2019.153549} {\bibfield
  {journal} {\bibinfo  {journal} {J. Alloys Compd.}\ }\textbf {\bibinfo
  {volume} {822}},\ \bibinfo {pages} {153549} (\bibinfo {year}
  {2020})}\BibitemShut {NoStop}%
\bibitem [{\citenamefont {Gnezdilov}\ \emph {et~al.}(2019)\citenamefont
  {Gnezdilov}, \citenamefont {Pashkevich}, \citenamefont {Kurnosov},
  \citenamefont {Zhuravlev}, \citenamefont {Wulferding}, \citenamefont
  {Lemmens}, \citenamefont {Menzel}, \citenamefont {Kozlyakova}, \citenamefont
  {Akhrorov}, \citenamefont {Kuznetsova}, \citenamefont {Berdonosov},
  \citenamefont {Dolgikh}, \citenamefont {Volkova},\ and\ \citenamefont
  {Vasiliev}}]{GnezdilovPRB2019}%
  \BibitemOpen
  \bibfield  {author} {\bibinfo {author} {\bibfnamefont {V.~P.}\ \bibnamefont
  {Gnezdilov}}, \bibinfo {author} {\bibfnamefont {Y.~G.}\ \bibnamefont
  {Pashkevich}}, \bibinfo {author} {\bibfnamefont {V.~S.}\ \bibnamefont
  {Kurnosov}}, \bibinfo {author} {\bibfnamefont {O.~V.}\ \bibnamefont
  {Zhuravlev}}, \bibinfo {author} {\bibfnamefont {D.}~\bibnamefont
  {Wulferding}}, \bibinfo {author} {\bibfnamefont {P.}~\bibnamefont {Lemmens}},
  \bibinfo {author} {\bibfnamefont {D.}~\bibnamefont {Menzel}}, \bibinfo
  {author} {\bibfnamefont {E.~S.}\ \bibnamefont {Kozlyakova}}, \bibinfo
  {author} {\bibfnamefont {A.~Y.}\ \bibnamefont {Akhrorov}}, \bibinfo {author}
  {\bibfnamefont {E.~S.}\ \bibnamefont {Kuznetsova}}, \bibinfo {author}
  {\bibfnamefont {P.~S.}\ \bibnamefont {Berdonosov}}, \bibinfo {author}
  {\bibfnamefont {V.~A.}\ \bibnamefont {Dolgikh}}, \bibinfo {author}
  {\bibfnamefont {O.~S.}\ \bibnamefont {Volkova}},\ and\ \bibinfo {author}
  {\bibfnamefont {A.~N.}\ \bibnamefont {Vasiliev}},\ }\bibfield  {title}
  {\bibinfo {title} {Flat-band spin dynamics and phonon anomalies of the
  saw-tooth spin-chain system \ce{Fe2O(SeO3)2}},\ }\href
  {https://doi.org/10.1103/PhysRevB.99.064413} {\bibfield  {journal} {\bibinfo
  {journal} {Phys. Rev. B}\ }\textbf {\bibinfo {volume} {99}},\ \bibinfo
  {pages} {064413} (\bibinfo {year} {2019})}\BibitemShut {NoStop}%
\bibitem [{\citenamefont {Garlea}\ \emph {et~al.}(2014)\citenamefont {Garlea},
  \citenamefont {Sanjeewa}, \citenamefont {McGuire}, \citenamefont {Kumar},
  \citenamefont {Sulejmanovic}, \citenamefont {He},\ and\ \citenamefont
  {Hwu}}]{GarleaPRB2014}%
  \BibitemOpen
  \bibfield  {author} {\bibinfo {author} {\bibfnamefont {V.~O.}\ \bibnamefont
  {Garlea}}, \bibinfo {author} {\bibfnamefont {L.~D.}\ \bibnamefont
  {Sanjeewa}}, \bibinfo {author} {\bibfnamefont {M.~A.}\ \bibnamefont
  {McGuire}}, \bibinfo {author} {\bibfnamefont {P.}~\bibnamefont {Kumar}},
  \bibinfo {author} {\bibfnamefont {D.}~\bibnamefont {Sulejmanovic}}, \bibinfo
  {author} {\bibfnamefont {J.}~\bibnamefont {He}},\ and\ \bibinfo {author}
  {\bibfnamefont {S.-J.}\ \bibnamefont {Hwu}},\ }\bibfield  {title} {\bibinfo
  {title} {Complex magnetic behavior of the sawtooth fe chains in
  \ce{Rb2Fe2O(AsO4)2}},\ }\href {https://doi.org/10.1103/PhysRevB.89.014426}
  {\bibfield  {journal} {\bibinfo  {journal} {Phys. Rev. B}\ }\textbf {\bibinfo
  {volume} {89}},\ \bibinfo {pages} {014426} (\bibinfo {year}
  {2014})}\BibitemShut {NoStop}%
\bibitem [{\citenamefont {Lau}\ \emph {et~al.}(2006)\citenamefont {Lau},
  \citenamefont {Ueland}, \citenamefont {Freitas}, \citenamefont {Dahlberg},
  \citenamefont {Schiffer},\ and\ \citenamefont {Cava}}]{LauPRB2006}%
  \BibitemOpen
  \bibfield  {author} {\bibinfo {author} {\bibfnamefont {G.~C.}\ \bibnamefont
  {Lau}}, \bibinfo {author} {\bibfnamefont {B.~G.}\ \bibnamefont {Ueland}},
  \bibinfo {author} {\bibfnamefont {R.~S.}\ \bibnamefont {Freitas}}, \bibinfo
  {author} {\bibfnamefont {M.~L.}\ \bibnamefont {Dahlberg}}, \bibinfo {author}
  {\bibfnamefont {P.}~\bibnamefont {Schiffer}},\ and\ \bibinfo {author}
  {\bibfnamefont {R.~J.}\ \bibnamefont {Cava}},\ }\bibfield  {title} {\bibinfo
  {title} {Magnetic characterization of the sawtooth-lattice olivines
  \ce{ZnL2S4} $(\mathrm{L}=\mathrm{Er},\mathrm{Tm},\mathrm{Yb})$},\ }\href
  {https://doi.org/10.1103/PhysRevB.73.012413} {\bibfield  {journal} {\bibinfo
  {journal} {Phys. Rev. B}\ }\textbf {\bibinfo {volume} {73}},\ \bibinfo
  {pages} {012413} (\bibinfo {year} {2006})}\BibitemShut {NoStop}%
\bibitem [{\citenamefont {Cava}\ \emph {et~al.}(1993)\citenamefont {Cava},
  \citenamefont {Zandbergen}, \citenamefont {Ramirez}, \citenamefont {Takagi},
  \citenamefont {Chen}, \citenamefont {Krajewski}, \citenamefont {Peck},
  \citenamefont {Waszczak}, \citenamefont {Meigs}, \citenamefont {Roth},\ and\
  \citenamefont {Schneemeyer}}]{CavaJSSC1993}%
  \BibitemOpen
  \bibfield  {author} {\bibinfo {author} {\bibfnamefont {R.}~\bibnamefont
  {Cava}}, \bibinfo {author} {\bibfnamefont {H.}~\bibnamefont {Zandbergen}},
  \bibinfo {author} {\bibfnamefont {A.}~\bibnamefont {Ramirez}}, \bibinfo
  {author} {\bibfnamefont {H.}~\bibnamefont {Takagi}}, \bibinfo {author}
  {\bibfnamefont {C.}~\bibnamefont {Chen}}, \bibinfo {author} {\bibfnamefont
  {J.}~\bibnamefont {Krajewski}}, \bibinfo {author} {\bibfnamefont
  {W.}~\bibnamefont {Peck}}, \bibinfo {author} {\bibfnamefont {J.}~\bibnamefont
  {Waszczak}}, \bibinfo {author} {\bibfnamefont {G.}~\bibnamefont {Meigs}},
  \bibinfo {author} {\bibfnamefont {R.}~\bibnamefont {Roth}},\ and\ \bibinfo
  {author} {\bibfnamefont {L.}~\bibnamefont {Schneemeyer}},\ }\bibfield
  {title} {\bibinfo {title} {\ce{LaCuO_{25+x}} and \ce{YCuO_{2.5+x}}
  delafossites: Materials with triangular \ce{Cu^{2+\delta}} planes},\ }\href
  {https://doi.org/https://doi.org/10.1006/jssc.1993.1179} {\bibfield
  {journal} {\bibinfo  {journal} {J. Solid State Chem.}\ }\textbf {\bibinfo
  {volume} {104}},\ \bibinfo {pages} {437} (\bibinfo {year}
  {1993})}\BibitemShut {NoStop}%
\bibitem [{\citenamefont {{Van Tendeloo}}\ \emph {et~al.}(2001)\citenamefont
  {{Van Tendeloo}}, \citenamefont {Garlea}, \citenamefont {Darie},
  \citenamefont {Bougerol-Chaillout},\ and\ \citenamefont
  {Bordet}}]{VanTendelooJSSC2001}%
  \BibitemOpen
  \bibfield  {author} {\bibinfo {author} {\bibfnamefont {G.}~\bibnamefont {{Van
  Tendeloo}}}, \bibinfo {author} {\bibfnamefont {O.}~\bibnamefont {Garlea}},
  \bibinfo {author} {\bibfnamefont {C.}~\bibnamefont {Darie}}, \bibinfo
  {author} {\bibfnamefont {C.}~\bibnamefont {Bougerol-Chaillout}},\ and\
  \bibinfo {author} {\bibfnamefont {P.}~\bibnamefont {Bordet}},\ }\bibfield
  {title} {\bibinfo {title} {The fine structure of \ce{YCuO_{2+x}} delafossite
  determined by synchrotron powder diffraction and electron microscopy},\
  }\href {https://doi.org/https://doi.org/10.1006/jssc.2000.9018} {\bibfield
  {journal} {\bibinfo  {journal} {J. Solid State Chem.}\ }\textbf {\bibinfo
  {volume} {156}},\ \bibinfo {pages} {428} (\bibinfo {year}
  {2001})}\BibitemShut {NoStop}%
\bibitem [{\citenamefont {Le~Bacq}\ \emph {et~al.}(2005)\citenamefont
  {Le~Bacq}, \citenamefont {Pasturel}, \citenamefont {Lacroix},\ and\
  \citenamefont {N\'u\~nez Regueiro}}]{LeBacqPRB2005}%
  \BibitemOpen
  \bibfield  {author} {\bibinfo {author} {\bibfnamefont {O.}~\bibnamefont
  {Le~Bacq}}, \bibinfo {author} {\bibfnamefont {A.}~\bibnamefont {Pasturel}},
  \bibinfo {author} {\bibfnamefont {C.}~\bibnamefont {Lacroix}},\ and\ \bibinfo
  {author} {\bibfnamefont {M.~D.}\ \bibnamefont {N\'u\~nez Regueiro}},\
  }\bibfield  {title} {\bibinfo {title} {First-principles determination of
  exchange interactions in delafossite \ce{YCuO_{2.5}}},\ }\href
  {https://doi.org/10.1103/PhysRevB.71.014432} {\bibfield  {journal} {\bibinfo
  {journal} {Phys. Rev. B}\ }\textbf {\bibinfo {volume} {71}},\ \bibinfo
  {pages} {014432} (\bibinfo {year} {2005})}\BibitemShut {NoStop}%
\bibitem [{\citenamefont {Baniodeh}\ \emph {et~al.}(2018)\citenamefont
  {Baniodeh}, \citenamefont {Magnani}, \citenamefont {Lan}, \citenamefont
  {Buth}, \citenamefont {Anson}, \citenamefont {Richter}, \citenamefont
  {Affronte}, \citenamefont {Schnack},\ and\ \citenamefont
  {Powell}}]{BaniodehNPJQM2018}%
  \BibitemOpen
  \bibfield  {author} {\bibinfo {author} {\bibfnamefont {A.}~\bibnamefont
  {Baniodeh}}, \bibinfo {author} {\bibfnamefont {N.}~\bibnamefont {Magnani}},
  \bibinfo {author} {\bibfnamefont {Y.}~\bibnamefont {Lan}}, \bibinfo {author}
  {\bibfnamefont {G.}~\bibnamefont {Buth}}, \bibinfo {author} {\bibfnamefont
  {C.~E.}\ \bibnamefont {Anson}}, \bibinfo {author} {\bibfnamefont
  {J.}~\bibnamefont {Richter}}, \bibinfo {author} {\bibfnamefont
  {M.}~\bibnamefont {Affronte}}, \bibinfo {author} {\bibfnamefont
  {J.}~\bibnamefont {Schnack}},\ and\ \bibinfo {author} {\bibfnamefont {A.~K.}\
  \bibnamefont {Powell}},\ }\bibfield  {title} {\bibinfo {title} {High spin
  cycles: topping the spin record for a single molecule verging on quantum
  criticality},\ }\href {https://doi.org/10.1038/s41535-018-0082-7} {\bibfield
  {journal} {\bibinfo  {journal} {npj Quantum Mater.}\ }\textbf {\bibinfo
  {volume} {3}},\ \bibinfo {pages} {10} (\bibinfo {year} {2018})}\BibitemShut
  {NoStop}%
\bibitem [{\citenamefont {Inagaki}\ \emph {et~al.}(2005)\citenamefont
  {Inagaki}, \citenamefont {Narumi}, \citenamefont {Kindo}, \citenamefont
  {Kikuchi}, \citenamefont {Kamikawa}, \citenamefont {Kunimoto}, \citenamefont
  {Okubo}, \citenamefont {Ohta}, \citenamefont {Saito}, \citenamefont {Azuma},
  \citenamefont {Takano}, \citenamefont {Nojiri}, \citenamefont {Kaburagi},\
  and\ \citenamefont {Tonegawa}}]{InagakiJPSJ2005}%
  \BibitemOpen
  \bibfield  {author} {\bibinfo {author} {\bibfnamefont {Y.}~\bibnamefont
  {Inagaki}}, \bibinfo {author} {\bibfnamefont {Y.}~\bibnamefont {Narumi}},
  \bibinfo {author} {\bibfnamefont {K.}~\bibnamefont {Kindo}}, \bibinfo
  {author} {\bibfnamefont {H.}~\bibnamefont {Kikuchi}}, \bibinfo {author}
  {\bibfnamefont {T.}~\bibnamefont {Kamikawa}}, \bibinfo {author}
  {\bibfnamefont {T.}~\bibnamefont {Kunimoto}}, \bibinfo {author}
  {\bibfnamefont {S.}~\bibnamefont {Okubo}}, \bibinfo {author} {\bibfnamefont
  {H.}~\bibnamefont {Ohta}}, \bibinfo {author} {\bibfnamefont {T.}~\bibnamefont
  {Saito}}, \bibinfo {author} {\bibfnamefont {M.}~\bibnamefont {Azuma}},
  \bibinfo {author} {\bibfnamefont {M.}~\bibnamefont {Takano}}, \bibinfo
  {author} {\bibfnamefont {H.}~\bibnamefont {Nojiri}}, \bibinfo {author}
  {\bibfnamefont {M.}~\bibnamefont {Kaburagi}},\ and\ \bibinfo {author}
  {\bibfnamefont {T.}~\bibnamefont {Tonegawa}},\ }\bibfield  {title} {\bibinfo
  {title} {Ferro-antiferromagnetic delta-chain system studied by high field
  magnetization measurements},\ }\href
  {https://doi.org/https://doi.org/10.1143/JPSJ.74.2831} {\bibfield  {journal}
  {\bibinfo  {journal} {J. Phys. Soc. Jpn.}\ }\textbf {\bibinfo {volume}
  {74}},\ \bibinfo {pages} {2831} (\bibinfo {year} {2005})}\BibitemShut
  {NoStop}%
\bibitem [{\citenamefont {Ruiz-P{\'e}rez}\ \emph {et~al.}(2000)\citenamefont
  {Ruiz-P{\'e}rez}, \citenamefont {Hern{\'a}ndez-Molina}, \citenamefont
  {Lorenzo-Luis}, \citenamefont {Lloret}, \citenamefont {Cano},\ and\
  \citenamefont {Julve}}]{RuizPerezInorgChem2000}%
  \BibitemOpen
  \bibfield  {author} {\bibinfo {author} {\bibfnamefont {C.}~\bibnamefont
  {Ruiz-P{\'e}rez}}, \bibinfo {author} {\bibfnamefont {M.}~\bibnamefont
  {Hern{\'a}ndez-Molina}}, \bibinfo {author} {\bibfnamefont {P.}~\bibnamefont
  {Lorenzo-Luis}}, \bibinfo {author} {\bibfnamefont {F.}~\bibnamefont
  {Lloret}}, \bibinfo {author} {\bibfnamefont {J.}~\bibnamefont {Cano}},\ and\
  \bibinfo {author} {\bibfnamefont {M.}~\bibnamefont {Julve}},\ }\bibfield
  {title} {\bibinfo {title} {Magnetic coupling through the carbon skeleton of
  malonate in two polymorphs of
  \ce{{\{}[Cu(bpy)(H2O)][Cu(bpy)(mal)(H2O)]{\}}(ClO4)2} (\ce{H2mal} = malonic
  acid; bpy = 2,2'-bipyridine)},\ }\href {https://doi.org/10.1021/ic000314n}
  {\bibfield  {journal} {\bibinfo  {journal} {Inorg. Chem.}\ }\textbf {\bibinfo
  {volume} {39}},\ \bibinfo {pages} {3845} (\bibinfo {year}
  {2000})}\BibitemShut {NoStop}%
\bibitem [{\citenamefont {Kikuchi}\ \emph {et~al.}(2011)\citenamefont
  {Kikuchi}, \citenamefont {Fujii}, \citenamefont {Takahashi}, \citenamefont
  {Azuma}, \citenamefont {Shimakawa}, \citenamefont {Taniguchi}, \citenamefont
  {Matsuo},\ and\ \citenamefont {Kindo}}]{KikuchiJPConfSeries2011}%
  \BibitemOpen
  \bibfield  {author} {\bibinfo {author} {\bibfnamefont {H.}~\bibnamefont
  {Kikuchi}}, \bibinfo {author} {\bibfnamefont {Y.}~\bibnamefont {Fujii}},
  \bibinfo {author} {\bibfnamefont {D.}~\bibnamefont {Takahashi}}, \bibinfo
  {author} {\bibfnamefont {M.}~\bibnamefont {Azuma}}, \bibinfo {author}
  {\bibfnamefont {Y.}~\bibnamefont {Shimakawa}}, \bibinfo {author}
  {\bibfnamefont {T.}~\bibnamefont {Taniguchi}}, \bibinfo {author}
  {\bibfnamefont {A.}~\bibnamefont {Matsuo}},\ and\ \bibinfo {author}
  {\bibfnamefont {K.}~\bibnamefont {Kindo}},\ }\bibfield  {title} {\bibinfo
  {title} {Spin gapped behavior of a frustrated delta chain compound
  euchroite},\ }\href {https://doi.org/10.1088/1742-6596/320/1/012045}
  {\bibfield  {journal} {\bibinfo  {journal} {J. Phys. Conf. Ser.}\ }\textbf
  {\bibinfo {volume} {320}},\ \bibinfo {pages} {012045} (\bibinfo {year}
  {2011})}\BibitemShut {NoStop}%
\bibitem [{\citenamefont {Heinze}\ \emph {et~al.}(2021)\citenamefont {Heinze},
  \citenamefont {Jeschke}, \citenamefont {Mazin}, \citenamefont
  {Metavitsiadis}, \citenamefont {Reehuis}, \citenamefont {Feyerherm},
  \citenamefont {Hoffmann}, \citenamefont {Bartkowiak}, \citenamefont
  {Prokhnenko}, \citenamefont {Wolter}, \citenamefont {Ding}, \citenamefont
  {Zapf}, \citenamefont {Corval\'an~Moya}, \citenamefont {Weickert},
  \citenamefont {Jaime}, \citenamefont {Rule}, \citenamefont {Menzel},
  \citenamefont {Valent\'{\i}}, \citenamefont {Brenig},\ and\ \citenamefont
  {S\"ullow}}]{HeinzePRL2021}%
  \BibitemOpen
  \bibfield  {author} {\bibinfo {author} {\bibfnamefont {L.}~\bibnamefont
  {Heinze}}, \bibinfo {author} {\bibfnamefont {H.~O.}\ \bibnamefont {Jeschke}},
  \bibinfo {author} {\bibfnamefont {I.~I.}\ \bibnamefont {Mazin}}, \bibinfo
  {author} {\bibfnamefont {A.}~\bibnamefont {Metavitsiadis}}, \bibinfo {author}
  {\bibfnamefont {M.}~\bibnamefont {Reehuis}}, \bibinfo {author} {\bibfnamefont
  {R.}~\bibnamefont {Feyerherm}}, \bibinfo {author} {\bibfnamefont {J.-U.}\
  \bibnamefont {Hoffmann}}, \bibinfo {author} {\bibfnamefont {M.}~\bibnamefont
  {Bartkowiak}}, \bibinfo {author} {\bibfnamefont {O.}~\bibnamefont
  {Prokhnenko}}, \bibinfo {author} {\bibfnamefont {A.~U.~B.}\ \bibnamefont
  {Wolter}}, \bibinfo {author} {\bibfnamefont {X.}~\bibnamefont {Ding}},
  \bibinfo {author} {\bibfnamefont {V.~S.}\ \bibnamefont {Zapf}}, \bibinfo
  {author} {\bibfnamefont {C.}~\bibnamefont {Corval\'an~Moya}}, \bibinfo
  {author} {\bibfnamefont {F.}~\bibnamefont {Weickert}}, \bibinfo {author}
  {\bibfnamefont {M.}~\bibnamefont {Jaime}}, \bibinfo {author} {\bibfnamefont
  {K.~C.}\ \bibnamefont {Rule}}, \bibinfo {author} {\bibfnamefont
  {D.}~\bibnamefont {Menzel}}, \bibinfo {author} {\bibfnamefont
  {R.}~\bibnamefont {Valent\'{\i}}}, \bibinfo {author} {\bibfnamefont
  {W.}~\bibnamefont {Brenig}},\ and\ \bibinfo {author} {\bibfnamefont
  {S.}~\bibnamefont {S\"ullow}},\ }\bibfield  {title} {\bibinfo {title}
  {Magnetization process of atacamite: A case of weakly coupled $s=1/2$
  sawtooth chains},\ }\href {https://doi.org/10.1103/PhysRevLett.126.207201}
  {\bibfield  {journal} {\bibinfo  {journal} {Phys. Rev. Lett.}\ }\textbf
  {\bibinfo {volume} {126}},\ \bibinfo {pages} {207201} (\bibinfo {year}
  {2021})}\BibitemShut {NoStop}%
\bibitem [{\citenamefont {Zhu}\ \emph {et~al.}(2023)\citenamefont {Zhu},
  \citenamefont {Zhu}, \citenamefont {Mentr{\'e}}, \citenamefont {Lee},
  \citenamefont {Chen}, \citenamefont {Jin}, \citenamefont {Zhu}, \citenamefont
  {Ar{\'e}valo-L{\'o}pez}, \citenamefont {Minaud}, \citenamefont {Choi},\ and\
  \citenamefont {L{\"u}}}]{ZhuCM2023}%
  \BibitemOpen
  \bibfield  {author} {\bibinfo {author} {\bibfnamefont {T.}~\bibnamefont
  {Zhu}}, \bibinfo {author} {\bibfnamefont {B.}~\bibnamefont {Zhu}}, \bibinfo
  {author} {\bibfnamefont {O.}~\bibnamefont {Mentr{\'e}}}, \bibinfo {author}
  {\bibfnamefont {S.}~\bibnamefont {Lee}}, \bibinfo {author} {\bibfnamefont
  {D.}~\bibnamefont {Chen}}, \bibinfo {author} {\bibfnamefont {Y.}~\bibnamefont
  {Jin}}, \bibinfo {author} {\bibfnamefont {W.}~\bibnamefont {Zhu}}, \bibinfo
  {author} {\bibfnamefont {{\'A}.~M.}\ \bibnamefont {Ar{\'e}valo-L{\'o}pez}},
  \bibinfo {author} {\bibfnamefont {C.}~\bibnamefont {Minaud}}, \bibinfo
  {author} {\bibfnamefont {K.-Y.}\ \bibnamefont {Choi}},\ and\ \bibinfo
  {author} {\bibfnamefont {M.}~\bibnamefont {L{\"u}}},\ }\bibfield  {title}
  {\bibinfo {title} {\ce{Cu3Te2O5(OH)4}: A frustrated two-dimensional quantum
  ``magnetic raft'' as a possible pathway to a spin liquid},\ }\href
  {https://doi.org/10.1021/acs.chemmater.3c00177} {\bibfield  {journal}
  {\bibinfo  {journal} {Chem. Mat.}\ }\textbf {\bibinfo {volume} {35}},\
  \bibinfo {pages} {3951} (\bibinfo {year} {2023})}\BibitemShut {NoStop}%
\bibitem [{\citenamefont {Blundell}\ and\ \citenamefont
  {Núñez-Regueiro}(2004)}]{blundell2004}%
  \BibitemOpen
  \bibfield  {author} {\bibinfo {author} {\bibfnamefont {S.~A.}\ \bibnamefont
  {Blundell}}\ and\ \bibinfo {author} {\bibfnamefont {M.~D.}\ \bibnamefont
  {Núñez-Regueiro}},\ }\bibfield  {title} {\bibinfo {title} {The $\delta$
  chain with different base–base and base–vertex interactions},\ }\href
  {https://doi.org/10.1088/0953-8984/16/11/031} {\bibfield  {journal} {\bibinfo
   {journal} {J. Phys.: Condens. Matter}\ }\textbf {\bibinfo {volume} {16}},\
  \bibinfo {pages} {S791} (\bibinfo {year} {2004})}\BibitemShut {NoStop}%
\bibitem [{\citenamefont {Jiang}\ \emph {et~al.}(2015)\citenamefont {Jiang},
  \citenamefont {Liu}, \citenamefont {Tang}, \citenamefont {Yang},\ and\
  \citenamefont {Sheng}}]{Jiang2015}%
  \BibitemOpen
  \bibfield  {author} {\bibinfo {author} {\bibfnamefont {J.-J.}\ \bibnamefont
  {Jiang}}, \bibinfo {author} {\bibfnamefont {Y.-J.}\ \bibnamefont {Liu}},
  \bibinfo {author} {\bibfnamefont {F.}~\bibnamefont {Tang}}, \bibinfo {author}
  {\bibfnamefont {C.-H.}\ \bibnamefont {Yang}},\ and\ \bibinfo {author}
  {\bibfnamefont {Y.-B.}\ \bibnamefont {Sheng}},\ }\bibfield  {title} {\bibinfo
  {title} {Analytical and numerical studies of the one-dimensional sawtooth
  chain},\ }\href {https://doi.org/https://doi.org/10.1016/j.physb.2015.01.036}
  {\bibfield  {journal} {\bibinfo  {journal} {Phys. B: Condens. Matter}\
  }\textbf {\bibinfo {volume} {463}},\ \bibinfo {pages} {30} (\bibinfo {year}
  {2015})}\BibitemShut {NoStop}%
\bibitem [{\citenamefont {Rausch}\ and\ \citenamefont
  {Karrasch}(2025)}]{Rausch2025}%
  \BibitemOpen
  \bibfield  {author} {\bibinfo {author} {\bibfnamefont {R.}~\bibnamefont
  {Rausch}}\ and\ \bibinfo {author} {\bibfnamefont {C.}~\bibnamefont
  {Karrasch}},\ }\bibfield  {title} {\bibinfo {title} {Noncollinear phase of
  the antiferromagnetic sawtooth chain},\ }\href
  {https://doi.org/10.1103/PhysRevB.111.045154} {\bibfield  {journal} {\bibinfo
   {journal} {Phys. Rev. B}\ }\textbf {\bibinfo {volume} {111}},\ \bibinfo
  {pages} {045154} (\bibinfo {year} {2025})}\BibitemShut {NoStop}%
\bibitem [{\citenamefont {Chen}\ \emph {et~al.}(2001)\citenamefont {Chen},
  \citenamefont {B\"uttner},\ and\ \citenamefont {Voit}}]{ChenPRL2001}%
  \BibitemOpen
  \bibfield  {author} {\bibinfo {author} {\bibfnamefont {S.}~\bibnamefont
  {Chen}}, \bibinfo {author} {\bibfnamefont {H.}~\bibnamefont {B\"uttner}},\
  and\ \bibinfo {author} {\bibfnamefont {J.}~\bibnamefont {Voit}},\ }\bibfield
  {title} {\bibinfo {title} {Phase diagram of an asymmetric spin ladder},\
  }\href {https://doi.org/10.1103/PhysRevLett.87.087205} {\bibfield  {journal}
  {\bibinfo  {journal} {Phys. Rev. Lett.}\ }\textbf {\bibinfo {volume} {87}},\
  \bibinfo {pages} {087205} (\bibinfo {year} {2001})}\BibitemShut {NoStop}%
\bibitem [{\citenamefont {Chen}\ \emph {et~al.}(2003)\citenamefont {Chen},
  \citenamefont {B\"uttner},\ and\ \citenamefont {Voit}}]{ChenPRB2003}%
  \BibitemOpen
  \bibfield  {author} {\bibinfo {author} {\bibfnamefont {S.}~\bibnamefont
  {Chen}}, \bibinfo {author} {\bibfnamefont {H.}~\bibnamefont {B\"uttner}},\
  and\ \bibinfo {author} {\bibfnamefont {J.}~\bibnamefont {Voit}},\ }\bibfield
  {title} {\bibinfo {title} {Ground state and excitation of an asymmetric spin
  ladder model},\ }\href {https://doi.org/10.1103/PhysRevB.67.054412}
  {\bibfield  {journal} {\bibinfo  {journal} {Phys. Rev. B}\ }\textbf {\bibinfo
  {volume} {67}},\ \bibinfo {pages} {054412} (\bibinfo {year}
  {2003})}\BibitemShut {NoStop}%
\bibitem [{\citenamefont {Hao}\ \emph {et~al.}(2011)\citenamefont {Hao},
  \citenamefont {Wan}, \citenamefont {Rousochatzakis}, \citenamefont
  {Wildeboer}, \citenamefont {Seidel}, \citenamefont {Mila},\ and\
  \citenamefont {Tchernyshyov}}]{HaoPRB2011}%
  \BibitemOpen
  \bibfield  {author} {\bibinfo {author} {\bibfnamefont {Z.}~\bibnamefont
  {Hao}}, \bibinfo {author} {\bibfnamefont {Y.}~\bibnamefont {Wan}}, \bibinfo
  {author} {\bibfnamefont {I.}~\bibnamefont {Rousochatzakis}}, \bibinfo
  {author} {\bibfnamefont {J.}~\bibnamefont {Wildeboer}}, \bibinfo {author}
  {\bibfnamefont {A.}~\bibnamefont {Seidel}}, \bibinfo {author} {\bibfnamefont
  {F.}~\bibnamefont {Mila}},\ and\ \bibinfo {author} {\bibfnamefont
  {O.}~\bibnamefont {Tchernyshyov}},\ }\bibfield  {title} {\bibinfo {title}
  {Destruction of valence-bond order in a ${S}=\frac{1}{2}$ sawtooth chain with
  a {D}zyaloshinskii-{M}oriya term},\ }\href
  {https://doi.org/10.1103/PhysRevB.84.094452} {\bibfield  {journal} {\bibinfo
  {journal} {Phys. Rev. B}\ }\textbf {\bibinfo {volume} {84}},\ \bibinfo
  {pages} {094452} (\bibinfo {year} {2011})}\BibitemShut {NoStop}%
\bibitem [{\citenamefont {Chandra}\ \emph {et~al.}(2004)\citenamefont
  {Chandra}, \citenamefont {Sen}, \citenamefont {Ivanov},\ and\ \citenamefont
  {Richter}}]{ChandraPRB204}%
  \BibitemOpen
  \bibfield  {author} {\bibinfo {author} {\bibfnamefont {V.~R.}\ \bibnamefont
  {Chandra}}, \bibinfo {author} {\bibfnamefont {D.}~\bibnamefont {Sen}},
  \bibinfo {author} {\bibfnamefont {N.~B.}\ \bibnamefont {Ivanov}},\ and\
  \bibinfo {author} {\bibfnamefont {J.}~\bibnamefont {Richter}},\ }\bibfield
  {title} {\bibinfo {title} {Antiferromagnetic sawtooth chain with
  spin-$\frac{1}{2}$ and spin-$1$ sites},\ }\href
  {https://doi.org/10.1103/PhysRevB.69.214406} {\bibfield  {journal} {\bibinfo
  {journal} {Phys. Rev. B}\ }\textbf {\bibinfo {volume} {69}},\ \bibinfo
  {pages} {214406} (\bibinfo {year} {2004})}\BibitemShut {NoStop}%
\bibitem [{\citenamefont {Hawthorne}\ \emph {et~al.}(2002)\citenamefont
  {Hawthorne}, \citenamefont {Cooper}, \citenamefont {Grice}, \citenamefont
  {Roberts},\ and\ \citenamefont {Hubbard}}]{HawthorneMineralMag2002}%
  \BibitemOpen
  \bibfield  {author} {\bibinfo {author} {\bibfnamefont {F.~C.}\ \bibnamefont
  {Hawthorne}}, \bibinfo {author} {\bibfnamefont {M.~A.}\ \bibnamefont
  {Cooper}}, \bibinfo {author} {\bibfnamefont {J.~D.}\ \bibnamefont {Grice}},
  \bibinfo {author} {\bibfnamefont {A.~C.}\ \bibnamefont {Roberts}},\ and\
  \bibinfo {author} {\bibfnamefont {N.}~\bibnamefont {Hubbard}},\ }\bibfield
  {title} {\bibinfo {title} {Description and crystal structure of bobkingite, a
  new mineral from {N}ew {C}liffe {H}ill {Q}uarry, {S}tanton-under-{B}ardon,
  {L}eicestershire, {UK}},\ }\href {https://doi.org/10.1180/0026461026620030}
  {\bibfield  {journal} {\bibinfo  {journal} {Mineralogical Magazine}\ }\textbf
  {\bibinfo {volume} {66}},\ \bibinfo {pages} {301–311} (\bibinfo {year}
  {2002})}\BibitemShut {NoStop}%
\bibitem [{\citenamefont {Samarakoon}\ \emph {et~al.}(2017)\citenamefont
  {Samarakoon}, \citenamefont {Banerjee}, \citenamefont {Zhang}, \citenamefont
  {Kamiya}, \citenamefont {Nagler}, \citenamefont {Tennant}, \citenamefont
  {Lee},\ and\ \citenamefont {Batista}}]{Samarakoon2017PRB}%
  \BibitemOpen
  \bibfield  {author} {\bibinfo {author} {\bibfnamefont {A.~M.}\ \bibnamefont
  {Samarakoon}}, \bibinfo {author} {\bibfnamefont {A.}~\bibnamefont
  {Banerjee}}, \bibinfo {author} {\bibfnamefont {S.-S.}\ \bibnamefont {Zhang}},
  \bibinfo {author} {\bibfnamefont {Y.}~\bibnamefont {Kamiya}}, \bibinfo
  {author} {\bibfnamefont {S.~E.}\ \bibnamefont {Nagler}}, \bibinfo {author}
  {\bibfnamefont {D.~A.}\ \bibnamefont {Tennant}}, \bibinfo {author}
  {\bibfnamefont {S.-H.}\ \bibnamefont {Lee}},\ and\ \bibinfo {author}
  {\bibfnamefont {C.~D.}\ \bibnamefont {Batista}},\ }\bibfield  {title}
  {\bibinfo {title} {{Comprehensive study of the dynamics of a classical Kitaev
  spin liquid}},\ }\href {https://doi.org/10.1103/PhysRevB.96.134408}
  {\bibfield  {journal} {\bibinfo  {journal} {Phys. Rev. B}\ }\textbf {\bibinfo
  {volume} {96}},\ \bibinfo {pages} {134408} (\bibinfo {year}
  {2017})}\BibitemShut {NoStop}%
\bibitem [{\citenamefont {Samarakoon}\ \emph {et~al.}(2018)\citenamefont
  {Samarakoon}, \citenamefont {Wachtel}, \citenamefont {Yamaji}, \citenamefont
  {Tennant}, \citenamefont {Batista},\ and\ \citenamefont
  {Kim}}]{Samarakoon2018PRB}%
  \BibitemOpen
  \bibfield  {author} {\bibinfo {author} {\bibfnamefont {A.~M.}\ \bibnamefont
  {Samarakoon}}, \bibinfo {author} {\bibfnamefont {G.}~\bibnamefont {Wachtel}},
  \bibinfo {author} {\bibfnamefont {Y.}~\bibnamefont {Yamaji}}, \bibinfo
  {author} {\bibfnamefont {D.~A.}\ \bibnamefont {Tennant}}, \bibinfo {author}
  {\bibfnamefont {C.~D.}\ \bibnamefont {Batista}},\ and\ \bibinfo {author}
  {\bibfnamefont {Y.~B.}\ \bibnamefont {Kim}},\ }\bibfield  {title} {\bibinfo
  {title} {{Classical and quantum spin dynamics of the honeycomb
  $\mathrm{\ensuremath{\Gamma}}$ model}},\ }\href
  {https://doi.org/10.1103/PhysRevB.98.045121} {\bibfield  {journal} {\bibinfo
  {journal} {Phys. Rev. B}\ }\textbf {\bibinfo {volume} {98}},\ \bibinfo
  {pages} {045121} (\bibinfo {year} {2018})}\BibitemShut {NoStop}%
\bibitem [{\citenamefont {Lake}\ \emph {et~al.}(2005)\citenamefont {Lake},
  \citenamefont {Tennant}, \citenamefont {Frost},\ and\ \citenamefont
  {Nagler}}]{LakeNatPhys2005}%
  \BibitemOpen
  \bibfield  {author} {\bibinfo {author} {\bibfnamefont {B.}~\bibnamefont
  {Lake}}, \bibinfo {author} {\bibfnamefont {D.~A.}\ \bibnamefont {Tennant}},
  \bibinfo {author} {\bibfnamefont {C.~D.}\ \bibnamefont {Frost}},\ and\
  \bibinfo {author} {\bibfnamefont {S.~E.}\ \bibnamefont {Nagler}},\ }\bibfield
   {title} {\bibinfo {title} {Quantum criticality and universal scaling of a
  quantum antiferromagnet},\ }\href {https://doi.org/10.1038/nmat1327}
  {\bibfield  {journal} {\bibinfo  {journal} {Nat. Mater.}\ }\textbf {\bibinfo
  {volume} {4}},\ \bibinfo {pages} {329} (\bibinfo {year} {2005})}\BibitemShut
  {NoStop}%
\bibitem [{\citenamefont {Banerjee}\ \emph {et~al.}(2016)\citenamefont
  {Banerjee}, \citenamefont {Bridges}, \citenamefont {Yan}, \citenamefont
  {Aczel}, \citenamefont {Li}, \citenamefont {Stone}, \citenamefont {Granroth},
  \citenamefont {Lumsden}, \citenamefont {Yiu}, \citenamefont {Knolle},
  \citenamefont {Bhattacharjee}, \citenamefont {Kovrizhin}, \citenamefont
  {Moessner}, \citenamefont {Tennant}, \citenamefont {Mandrus},\ and\
  \citenamefont {Nagler}}]{Banerjee2016}%
  \BibitemOpen
  \bibfield  {author} {\bibinfo {author} {\bibfnamefont {A.}~\bibnamefont
  {Banerjee}}, \bibinfo {author} {\bibfnamefont {C.~A.}\ \bibnamefont
  {Bridges}}, \bibinfo {author} {\bibfnamefont {J.-Q.}\ \bibnamefont {Yan}},
  \bibinfo {author} {\bibfnamefont {A.~A.}\ \bibnamefont {Aczel}}, \bibinfo
  {author} {\bibfnamefont {L.}~\bibnamefont {Li}}, \bibinfo {author}
  {\bibfnamefont {M.~B.}\ \bibnamefont {Stone}}, \bibinfo {author}
  {\bibfnamefont {G.~E.}\ \bibnamefont {Granroth}}, \bibinfo {author}
  {\bibfnamefont {M.~D.}\ \bibnamefont {Lumsden}}, \bibinfo {author}
  {\bibfnamefont {Y.}~\bibnamefont {Yiu}}, \bibinfo {author} {\bibfnamefont
  {J.}~\bibnamefont {Knolle}}, \bibinfo {author} {\bibfnamefont
  {S.}~\bibnamefont {Bhattacharjee}}, \bibinfo {author} {\bibfnamefont {D.~L.}\
  \bibnamefont {Kovrizhin}}, \bibinfo {author} {\bibfnamefont {R.}~\bibnamefont
  {Moessner}}, \bibinfo {author} {\bibfnamefont {D.~A.}\ \bibnamefont
  {Tennant}}, \bibinfo {author} {\bibfnamefont {D.~G.}\ \bibnamefont
  {Mandrus}},\ and\ \bibinfo {author} {\bibfnamefont {S.~E.}\ \bibnamefont
  {Nagler}},\ }\bibfield  {title} {\bibinfo {title} {Proximate kitaev quantum
  spin liquid behaviour in a honeycomb magnet},\ }\href
  {https://doi.org/10.1038/nmat4604} {\bibfield  {journal} {\bibinfo  {journal}
  {Nat. Mater.}\ }\textbf {\bibinfo {volume} {15}},\ \bibinfo {pages} {733}
  (\bibinfo {year} {2016})}\BibitemShut {NoStop}%
\bibitem [{\citenamefont {Rousochatzakis}\ \emph {et~al.}(2019)\citenamefont
  {Rousochatzakis}, \citenamefont {Kourtis}, \citenamefont {Knolle},
  \citenamefont {Moessner},\ and\ \citenamefont
  {Perkins}}]{Rousochatzakis2019}%
  \BibitemOpen
  \bibfield  {author} {\bibinfo {author} {\bibfnamefont {I.}~\bibnamefont
  {Rousochatzakis}}, \bibinfo {author} {\bibfnamefont {S.}~\bibnamefont
  {Kourtis}}, \bibinfo {author} {\bibfnamefont {J.}~\bibnamefont {Knolle}},
  \bibinfo {author} {\bibfnamefont {R.}~\bibnamefont {Moessner}},\ and\
  \bibinfo {author} {\bibfnamefont {N.~B.}\ \bibnamefont {Perkins}},\
  }\bibfield  {title} {\bibinfo {title} {Quantum spin liquid at finite
  temperature: Proximate dynamics and persistent typicality},\ }\href
  {https://doi.org/10.1103/PhysRevB.100.045117} {\bibfield  {journal} {\bibinfo
   {journal} {Phys. Rev. B}\ }\textbf {\bibinfo {volume} {100}},\ \bibinfo
  {pages} {045117} (\bibinfo {year} {2019})}\BibitemShut {NoStop}%
\bibitem [{\citenamefont {Norman}(2016)}]{NormanRMP2016}%
  \BibitemOpen
  \bibfield  {author} {\bibinfo {author} {\bibfnamefont {M.~R.}\ \bibnamefont
  {Norman}},\ }\bibfield  {title} {\bibinfo {title} {Colloquium:
  {H}erbertsmithite and the search for the quantum spin liquid},\ }\href
  {https://doi.org/10.1103/RevModPhys.88.041002} {\bibfield  {journal}
  {\bibinfo  {journal} {Rev. Mod. Phys.}\ }\textbf {\bibinfo {volume} {88}},\
  \bibinfo {pages} {041002} (\bibinfo {year} {2016})}\BibitemShut {NoStop}%
\bibitem [{\citenamefont {Kresse}\ and\ \citenamefont
  {Hafner}(1993)}]{kresse1993}%
  \BibitemOpen
  \bibfield  {author} {\bibinfo {author} {\bibfnamefont {G.}~\bibnamefont
  {Kresse}}\ and\ \bibinfo {author} {\bibfnamefont {J.}~\bibnamefont
  {Hafner}},\ }\bibfield  {title} {\bibinfo {title} {Ab initio molecular
  dynamics for liquid metals},\ }\href
  {https://doi.org/10.1103/PhysRevB.47.558} {\bibfield  {journal} {\bibinfo
  {journal} {Phys. Rev. B}\ }\textbf {\bibinfo {volume} {47}},\ \bibinfo
  {pages} {558} (\bibinfo {year} {1993})}\BibitemShut {NoStop}%
\bibitem [{\citenamefont {Kresse}\ and\ \citenamefont
  {Furthm\"uller}(1996)}]{kresse1996}%
  \BibitemOpen
  \bibfield  {author} {\bibinfo {author} {\bibfnamefont {G.}~\bibnamefont
  {Kresse}}\ and\ \bibinfo {author} {\bibfnamefont {J.}~\bibnamefont
  {Furthm\"uller}},\ }\bibfield  {title} {\bibinfo {title} {Efficient iterative
  schemes for ab initio total-energy calculations using a plane-wave basis
  set},\ }\href {https://doi.org/10.1103/PhysRevB.54.11169} {\bibfield
  {journal} {\bibinfo  {journal} {Phys. Rev. B}\ }\textbf {\bibinfo {volume}
  {54}},\ \bibinfo {pages} {11169} (\bibinfo {year} {1996})}\BibitemShut
  {NoStop}%
\bibitem [{\citenamefont {Kresse}\ and\ \citenamefont
  {Joubert}(1999)}]{kresse1999}%
  \BibitemOpen
  \bibfield  {author} {\bibinfo {author} {\bibfnamefont {G.}~\bibnamefont
  {Kresse}}\ and\ \bibinfo {author} {\bibfnamefont {D.}~\bibnamefont
  {Joubert}},\ }\bibfield  {title} {\bibinfo {title} {From ultrasoft
  pseudopotentials to the projector augmented-wave method},\ }\href
  {https://doi.org/10.1103/PhysRevB.59.1758} {\bibfield  {journal} {\bibinfo
  {journal} {Phys. Rev. B}\ }\textbf {\bibinfo {volume} {59}},\ \bibinfo
  {pages} {1758} (\bibinfo {year} {1999})}\BibitemShut {NoStop}%
\bibitem [{SI()}]{SI}%
  \BibitemOpen
  \href@noop {} {}\bibinfo {note} {See Supplemental Material at \textit{[URL
  will be inserted by publisher]} for \textit{ab initio} simulation package
  VASP method details, electronic structure from DFT and DFT+U, the relaxed DFT
  crystal structure, magnetocrystalline anisotropy methods, microscopic
  derivation of anisotropic exchanges, TEMA method details with tabulated $J_a$
  values as a function of $U$, additional details on the classical state and
  its relation to the exchanges, classical T=0 minimization, order by disorder
  from linear spin-wave theory, classical thermal Monte Carlo details, and
  exact diagonalization details on the $S\!=\!1/2$ sawtooth chain,
  Refs.~\cite{kresse1993,kresse1996,kresse1999,perdew1996,liechtenstein1995,PustogowPRB2017,PBEsol,HerbAniso2008,HerbAniso2012,HerbAniso2025,GlasbrennerNature2015,razpopov2023j,garcia2026microscopic,DzyaloshinskyJPCS1958,MoriyaPRL1960,MoriyaPR1960,Fazekas1999,WinterPRB2016,RiedlPRB2016,KoniecznaNPJQM2025,Rodrigues1840,fploWF,fplocite1,fplocite2,fploRel,CarterPRB2012,YildirimPRB1995,SwendsenPRL1986,HukushimaJPSJ1996,LakshmananLLGeq2011,CreutzPRD1987,KankiPRD2005,lehoucq1998arpack,Samarakoon2017PRB,Samarakoon2018PRB,LakeNatPhys2005,Banerjee2016,Rousochatzakis2019}}\BibitemShut
  {NoStop}%
\bibitem [{\citenamefont {Dzyaloshinsky}(1958)}]{DzyaloshinskyJPCS1958}%
  \BibitemOpen
  \bibfield  {author} {\bibinfo {author} {\bibfnamefont {I.}~\bibnamefont
  {Dzyaloshinsky}},\ }\bibfield  {title} {\bibinfo {title} {A thermodynamic
  theory of “weak” ferromagnetism of antiferromagnetics},\ }\href
  {https://doi.org/https://doi.org/10.1016/0022-3697(58)90076-3} {\bibfield
  {journal} {\bibinfo  {journal} {J. Phys. Chem. Solids}\ }\textbf {\bibinfo
  {volume} {4}},\ \bibinfo {pages} {241} (\bibinfo {year} {1958})}\BibitemShut
  {NoStop}%
\bibitem [{\citenamefont {Moriya}(1960{\natexlab{a}})}]{MoriyaPRL1960}%
  \BibitemOpen
  \bibfield  {author} {\bibinfo {author} {\bibfnamefont {T.}~\bibnamefont
  {Moriya}},\ }\bibfield  {title} {\bibinfo {title} {New mechanism of
  anisotropic superexchange interaction},\ }\href
  {https://doi.org/10.1103/PhysRevLett.4.228} {\bibfield  {journal} {\bibinfo
  {journal} {Phys. Rev. Lett.}\ }\textbf {\bibinfo {volume} {4}},\ \bibinfo
  {pages} {228} (\bibinfo {year} {1960}{\natexlab{a}})}\BibitemShut {NoStop}%
\bibitem [{\citenamefont {Moriya}(1960{\natexlab{b}})}]{MoriyaPR1960}%
  \BibitemOpen
  \bibfield  {author} {\bibinfo {author} {\bibfnamefont {T.}~\bibnamefont
  {Moriya}},\ }\bibfield  {title} {\bibinfo {title} {Anisotropic superexchange
  interaction and weak ferromagnetism},\ }\href
  {https://doi.org/10.1103/PhysRev.120.91} {\bibfield  {journal} {\bibinfo
  {journal} {Phys. Rev.}\ }\textbf {\bibinfo {volume} {120}},\ \bibinfo {pages}
  {91} (\bibinfo {year} {1960}{\natexlab{b}})}\BibitemShut {NoStop}%
\bibitem [{\citenamefont {Zorko}\ \emph {et~al.}(2008)\citenamefont {Zorko},
  \citenamefont {Nellutla}, \citenamefont {van Tol}, \citenamefont {Brunel},
  \citenamefont {Bert}, \citenamefont {Duc}, \citenamefont {Trombe},
  \citenamefont {de~Vries}, \citenamefont {Harrison},\ and\ \citenamefont
  {Mendels}}]{HerbAniso2008}%
  \BibitemOpen
  \bibfield  {author} {\bibinfo {author} {\bibfnamefont {A.}~\bibnamefont
  {Zorko}}, \bibinfo {author} {\bibfnamefont {S.}~\bibnamefont {Nellutla}},
  \bibinfo {author} {\bibfnamefont {J.}~\bibnamefont {van Tol}}, \bibinfo
  {author} {\bibfnamefont {L.~C.}\ \bibnamefont {Brunel}}, \bibinfo {author}
  {\bibfnamefont {F.}~\bibnamefont {Bert}}, \bibinfo {author} {\bibfnamefont
  {F.}~\bibnamefont {Duc}}, \bibinfo {author} {\bibfnamefont {J.-C.}\
  \bibnamefont {Trombe}}, \bibinfo {author} {\bibfnamefont {M.~A.}\
  \bibnamefont {de~Vries}}, \bibinfo {author} {\bibfnamefont {A.}~\bibnamefont
  {Harrison}},\ and\ \bibinfo {author} {\bibfnamefont {P.}~\bibnamefont
  {Mendels}},\ }\bibfield  {title} {\bibinfo {title} {Dzyaloshinsky-{M}oriya
  anisotropy in the spin-1/2 kagome compound \ce{ZnCu3(OH)6Cl2}},\ }\href
  {https://doi.org/10.1103/PhysRevLett.101.026405} {\bibfield  {journal}
  {\bibinfo  {journal} {Phys. Rev. Lett.}\ }\textbf {\bibinfo {volume} {101}},\
  \bibinfo {pages} {026405} (\bibinfo {year} {2008})}\BibitemShut {NoStop}%
\bibitem [{\citenamefont {Han}\ \emph {et~al.}(2012)\citenamefont {Han},
  \citenamefont {Chu},\ and\ \citenamefont {Lee}}]{HerbAniso2012}%
  \BibitemOpen
  \bibfield  {author} {\bibinfo {author} {\bibfnamefont {T.}~\bibnamefont
  {Han}}, \bibinfo {author} {\bibfnamefont {S.}~\bibnamefont {Chu}},\ and\
  \bibinfo {author} {\bibfnamefont {Y.~S.}\ \bibnamefont {Lee}},\ }\bibfield
  {title} {\bibinfo {title} {Refining the spin {H}amiltonian in the
  spin-$\frac{1}{2}$ kagome lattice antiferromagnet \ce{ZnCu3(OH)6Cl2} using
  single crystals},\ }\href {https://doi.org/10.1103/PhysRevLett.108.157202}
  {\bibfield  {journal} {\bibinfo  {journal} {Phys. Rev. Lett.}\ }\textbf
  {\bibinfo {volume} {108}},\ \bibinfo {pages} {157202} (\bibinfo {year}
  {2012})}\BibitemShut {NoStop}%
\bibitem [{\citenamefont {Heully-Alary}\ \emph {et~al.}(2025)\citenamefont
  {Heully-Alary}, \citenamefont {Ben~Amor}, \citenamefont {Suaud},
  \citenamefont {Messio}, \citenamefont {de~Graaf},\ and\ \citenamefont
  {Guihéry}}]{HerbAniso2025}%
  \BibitemOpen
  \bibfield  {author} {\bibinfo {author} {\bibfnamefont {F.}~\bibnamefont
  {Heully-Alary}}, \bibinfo {author} {\bibfnamefont {N.}~\bibnamefont
  {Ben~Amor}}, \bibinfo {author} {\bibfnamefont {N.}~\bibnamefont {Suaud}},
  \bibinfo {author} {\bibfnamefont {L.}~\bibnamefont {Messio}}, \bibinfo
  {author} {\bibfnamefont {C.}~\bibnamefont {de~Graaf}},\ and\ \bibinfo
  {author} {\bibfnamefont {N.}~\bibnamefont {Guihéry}},\ }\bibfield  {title}
  {\bibinfo {title} {Is herbertsmithite far from an ideal antiferromagnet?
  {A}b-initio answer including in-plane {D}zyaloshinskii-{M}oriya interactions
  and coupling with extra-plane impurities},\ }\href
  {https://doi.org/10.21468/SciPostPhysCore.8.4.092} {\bibfield  {journal}
  {\bibinfo  {journal} {SciPost Phys. Core}\ }\textbf {\bibinfo {volume} {8}},\
  \bibinfo {pages} {092} (\bibinfo {year} {2025})}\BibitemShut {NoStop}%
\bibitem [{\citenamefont {Glasbrenner}\ \emph {et~al.}(2015)\citenamefont
  {Glasbrenner}, \citenamefont {Mazin}, \citenamefont {Jeschke}, \citenamefont
  {Hirschfeld}, \citenamefont {Fernandes},\ and\ \citenamefont
  {Valent{\'i}}}]{GlasbrennerNature2015}%
  \BibitemOpen
  \bibfield  {author} {\bibinfo {author} {\bibfnamefont {J.~K.}\ \bibnamefont
  {Glasbrenner}}, \bibinfo {author} {\bibfnamefont {I.~I.}\ \bibnamefont
  {Mazin}}, \bibinfo {author} {\bibfnamefont {H.~O.}\ \bibnamefont {Jeschke}},
  \bibinfo {author} {\bibfnamefont {P.~J.}\ \bibnamefont {Hirschfeld}},
  \bibinfo {author} {\bibfnamefont {R.~M.}\ \bibnamefont {Fernandes}},\ and\
  \bibinfo {author} {\bibfnamefont {R.}~\bibnamefont {Valent{\'i}}},\
  }\bibfield  {title} {\bibinfo {title} {Effect of magnetic frustration on
  nematicity and superconductivity in iron chalcogenides},\ }\href
  {https://doi.org/10.1038/nphys3434} {\bibfield  {journal} {\bibinfo
  {journal} {Nat. Phys.}\ }\textbf {\bibinfo {volume} {11}},\ \bibinfo {pages}
  {953} (\bibinfo {year} {2015})}\BibitemShut {NoStop}%
\bibitem [{\citenamefont {Razpopov}\ \emph {et~al.}(2023)\citenamefont
  {Razpopov}, \citenamefont {Kaib}, \citenamefont {Backes}, \citenamefont
  {Balents}, \citenamefont {Wilson}, \citenamefont {Ferrari}, \citenamefont
  {Riedl},\ and\ \citenamefont {Valent{\'\i}}}]{razpopov2023j}%
  \BibitemOpen
  \bibfield  {author} {\bibinfo {author} {\bibfnamefont {A.}~\bibnamefont
  {Razpopov}}, \bibinfo {author} {\bibfnamefont {D.~A.}\ \bibnamefont {Kaib}},
  \bibinfo {author} {\bibfnamefont {S.}~\bibnamefont {Backes}}, \bibinfo
  {author} {\bibfnamefont {L.}~\bibnamefont {Balents}}, \bibinfo {author}
  {\bibfnamefont {S.~D.}\ \bibnamefont {Wilson}}, \bibinfo {author}
  {\bibfnamefont {F.}~\bibnamefont {Ferrari}}, \bibinfo {author} {\bibfnamefont
  {K.}~\bibnamefont {Riedl}},\ and\ \bibinfo {author} {\bibfnamefont
  {R.}~\bibnamefont {Valent{\'\i}}},\ }\bibfield  {title} {\bibinfo {title} {A
  $j_{eff}$= $1/2$ {K}itaev material on the triangular lattice: the case of
  {N}a{R}u{O}$_2$},\ }\href {https://doi.org/10.1038/s41535-023-00567-6}
  {\bibfield  {journal} {\bibinfo  {journal} {npj Quantum Mater.}\ }\textbf
  {\bibinfo {volume} {8}},\ \bibinfo {pages} {36} (\bibinfo {year}
  {2023})}\BibitemShut {NoStop}%
\bibitem [{\citenamefont {Garcia-Gassull}\ \emph {et~al.}(2026)\citenamefont
  {Garcia-Gassull}, \citenamefont {Razpopov}, \citenamefont {Stavropoulos},
  \citenamefont {Mazin},\ and\ \citenamefont
  {Valent{\'\i}}}]{garcia2026microscopic}%
  \BibitemOpen
  \bibfield  {author} {\bibinfo {author} {\bibfnamefont {L.}~\bibnamefont
  {Garcia-Gassull}}, \bibinfo {author} {\bibfnamefont {A.}~\bibnamefont
  {Razpopov}}, \bibinfo {author} {\bibfnamefont {P.~P.}\ \bibnamefont
  {Stavropoulos}}, \bibinfo {author} {\bibfnamefont {I.~I.}\ \bibnamefont
  {Mazin}},\ and\ \bibinfo {author} {\bibfnamefont {R.}~\bibnamefont
  {Valent{\'\i}}},\ }\bibfield  {title} {\bibinfo {title} {Microscopic origin
  of the magnetic interactions and their experimental signatures in
  altermagnetic {L}a$_2${O}$_3${M}n$_2${S}e$_2$},\ }\href
  {https://doi.org/10.1038/s44306-025-00125-9} {\bibfield  {journal} {\bibinfo
  {journal} {npj Spintron.}\ }\textbf {\bibinfo {volume} {4}},\ \bibinfo
  {pages} {9} (\bibinfo {year} {2026})}\BibitemShut {NoStop}%
\bibitem [{\citenamefont {Li}\ \emph {et~al.}(2018)\citenamefont {Li},
  \citenamefont {Winter},\ and\ \citenamefont {Valent\'{\i}}}]{LiPRL2018}%
  \BibitemOpen
  \bibfield  {author} {\bibinfo {author} {\bibfnamefont {Y.}~\bibnamefont
  {Li}}, \bibinfo {author} {\bibfnamefont {S.~M.}\ \bibnamefont {Winter}},\
  and\ \bibinfo {author} {\bibfnamefont {R.}~\bibnamefont {Valent\'{\i}}},\
  }\bibfield  {title} {\bibinfo {title} {Role of hydrogen in the
  spin-orbital-entangled quantum liquid candidate \ce{H3LiIr2O6}},\ }\href
  {https://doi.org/10.1103/PhysRevLett.121.247202} {\bibfield  {journal}
  {\bibinfo  {journal} {Phys. Rev. Lett.}\ }\textbf {\bibinfo {volume} {121}},\
  \bibinfo {pages} {247202} (\bibinfo {year} {2018})}\BibitemShut {NoStop}%
\bibitem [{\citenamefont {Perdew}\ \emph {et~al.}(1996)\citenamefont {Perdew},
  \citenamefont {Burke},\ and\ \citenamefont {Ernzerhof}}]{perdew1996}%
  \BibitemOpen
  \bibfield  {author} {\bibinfo {author} {\bibfnamefont {J.~P.}\ \bibnamefont
  {Perdew}}, \bibinfo {author} {\bibfnamefont {K.}~\bibnamefont {Burke}},\ and\
  \bibinfo {author} {\bibfnamefont {M.}~\bibnamefont {Ernzerhof}},\ }\bibfield
  {title} {\bibinfo {title} {Generalized gradient approximation made simple},\
  }\href {https://doi.org/10.1103/PhysRevLett.77.3865} {\bibfield  {journal}
  {\bibinfo  {journal} {Phys. Rev. Lett.}\ }\textbf {\bibinfo {volume} {77}},\
  \bibinfo {pages} {3865} (\bibinfo {year} {1996})}\BibitemShut {NoStop}%
\bibitem [{\citenamefont {Liechtenstein}\ \emph {et~al.}(1995)\citenamefont
  {Liechtenstein}, \citenamefont {Anisimov},\ and\ \citenamefont
  {Zaanen}}]{liechtenstein1995}%
  \BibitemOpen
  \bibfield  {author} {\bibinfo {author} {\bibfnamefont {A.~I.}\ \bibnamefont
  {Liechtenstein}}, \bibinfo {author} {\bibfnamefont {V.~I.}\ \bibnamefont
  {Anisimov}},\ and\ \bibinfo {author} {\bibfnamefont {J.}~\bibnamefont
  {Zaanen}},\ }\bibfield  {title} {\bibinfo {title} {Density-functional theory
  and strong interactions: Orbital ordering in {M}ott-{H}ubbard insulators},\
  }\href {https://doi.org/10.1103/PhysRevB.52.R5467} {\bibfield  {journal}
  {\bibinfo  {journal} {Phys. Rev. B}\ }\textbf {\bibinfo {volume} {52}},\
  \bibinfo {pages} {R5467} (\bibinfo {year} {1995})}\BibitemShut {NoStop}%
\bibitem [{\citenamefont {Pustogow}\ \emph {et~al.}(2017)\citenamefont
  {Pustogow}, \citenamefont {Li}, \citenamefont {Voloshenko}, \citenamefont
  {Puphal}, \citenamefont {Krellner}, \citenamefont {Mazin}, \citenamefont
  {Dressel},\ and\ \citenamefont {Valent\'{\i}}}]{PustogowPRB2017}%
  \BibitemOpen
  \bibfield  {author} {\bibinfo {author} {\bibfnamefont {A.}~\bibnamefont
  {Pustogow}}, \bibinfo {author} {\bibfnamefont {Y.}~\bibnamefont {Li}},
  \bibinfo {author} {\bibfnamefont {I.}~\bibnamefont {Voloshenko}}, \bibinfo
  {author} {\bibfnamefont {P.}~\bibnamefont {Puphal}}, \bibinfo {author}
  {\bibfnamefont {C.}~\bibnamefont {Krellner}}, \bibinfo {author}
  {\bibfnamefont {I.~I.}\ \bibnamefont {Mazin}}, \bibinfo {author}
  {\bibfnamefont {M.}~\bibnamefont {Dressel}},\ and\ \bibinfo {author}
  {\bibfnamefont {R.}~\bibnamefont {Valent\'{\i}}},\ }\bibfield  {title}
  {\bibinfo {title} {Nature of optical excitations in the frustrated kagome
  compound herbertsmithite},\ }\href
  {https://doi.org/10.1103/PhysRevB.96.241114} {\bibfield  {journal} {\bibinfo
  {journal} {Phys. Rev. B}\ }\textbf {\bibinfo {volume} {96}},\ \bibinfo
  {pages} {241114(R)} (\bibinfo {year} {2017})}\BibitemShut {NoStop}%
\bibitem [{\citenamefont {Perdew}\ \emph {et~al.}(2008)\citenamefont {Perdew},
  \citenamefont {Ruzsinszky}, \citenamefont {Csonka}, \citenamefont {Vydrov},
  \citenamefont {Scuseria}, \citenamefont {Constantin}, \citenamefont {Zhou},\
  and\ \citenamefont {Burke}}]{PBEsol}%
  \BibitemOpen
  \bibfield  {author} {\bibinfo {author} {\bibfnamefont {J.~P.}\ \bibnamefont
  {Perdew}}, \bibinfo {author} {\bibfnamefont {A.}~\bibnamefont {Ruzsinszky}},
  \bibinfo {author} {\bibfnamefont {G.~I.}\ \bibnamefont {Csonka}}, \bibinfo
  {author} {\bibfnamefont {O.~A.}\ \bibnamefont {Vydrov}}, \bibinfo {author}
  {\bibfnamefont {G.~E.}\ \bibnamefont {Scuseria}}, \bibinfo {author}
  {\bibfnamefont {L.~A.}\ \bibnamefont {Constantin}}, \bibinfo {author}
  {\bibfnamefont {X.}~\bibnamefont {Zhou}},\ and\ \bibinfo {author}
  {\bibfnamefont {K.}~\bibnamefont {Burke}},\ }\bibfield  {title} {\bibinfo
  {title} {Restoring the density-gradient expansion for exchange in solids and
  surfaces},\ }\href {https://doi.org/10.1103/PhysRevLett.100.136406}
  {\bibfield  {journal} {\bibinfo  {journal} {Phys. Rev. Lett.}\ }\textbf
  {\bibinfo {volume} {100}},\ \bibinfo {pages} {136406} (\bibinfo {year}
  {2008})}\BibitemShut {NoStop}%
\bibitem [{\citenamefont {Fazekas}(1999)}]{Fazekas1999}%
  \BibitemOpen
  \bibfield  {author} {\bibinfo {author} {\bibfnamefont {P.}~\bibnamefont
  {Fazekas}},\ }\href@noop {} {\emph {\bibinfo {title} {Lecture Notes on
  Electron Correlation and Magnetism}}},\ \bibinfo {series} {Series in Modern
  Condensed Matter Physics}, Vol.~\bibinfo {volume} {5}\ (\bibinfo  {publisher}
  {World Scientific},\ \bibinfo {address} {Singapore},\ \bibinfo {year}
  {1999})\BibitemShut {NoStop}%
\bibitem [{\citenamefont {Winter}\ \emph {et~al.}(2016)\citenamefont {Winter},
  \citenamefont {Li}, \citenamefont {Jeschke},\ and\ \citenamefont
  {Valent{\'{\i}}}}]{WinterPRB2016}%
  \BibitemOpen
  \bibfield  {author} {\bibinfo {author} {\bibfnamefont {S.~M.}\ \bibnamefont
  {Winter}}, \bibinfo {author} {\bibfnamefont {Y.}~\bibnamefont {Li}}, \bibinfo
  {author} {\bibfnamefont {H.~O.}\ \bibnamefont {Jeschke}},\ and\ \bibinfo
  {author} {\bibfnamefont {R.}~\bibnamefont {Valent{\'{\i}}}},\ }\bibfield
  {title} {\bibinfo {title} {Challenges in design of kitaev materials: Magnetic
  interactions from competing energy scales},\ }\href
  {https://doi.org/10.1103/PhysRevB.93.214431} {\bibfield  {journal} {\bibinfo
  {journal} {Phys. Rev. B}\ }\textbf {\bibinfo {volume} {93}},\ \bibinfo
  {pages} {214431} (\bibinfo {year} {2016})}\BibitemShut {NoStop}%
\bibitem [{\citenamefont {Riedl}\ \emph {et~al.}(2016)\citenamefont {Riedl},
  \citenamefont {Guterding}, \citenamefont {Jeschke}, \citenamefont {Gingras},\
  and\ \citenamefont {Valent{\'{\i}}}}]{RiedlPRB2016}%
  \BibitemOpen
  \bibfield  {author} {\bibinfo {author} {\bibfnamefont {K.}~\bibnamefont
  {Riedl}}, \bibinfo {author} {\bibfnamefont {D.}~\bibnamefont {Guterding}},
  \bibinfo {author} {\bibfnamefont {H.~O.}\ \bibnamefont {Jeschke}}, \bibinfo
  {author} {\bibfnamefont {M.~J.~P.}\ \bibnamefont {Gingras}},\ and\ \bibinfo
  {author} {\bibfnamefont {R.}~\bibnamefont {Valent{\'{\i}}}},\ }\bibfield
  {title} {\bibinfo {title} {Ab initio determination of spin hamiltonians with
  anisotropic exchange interactions: The case of the pyrochlore ferromagnet
  \ce{Lu2V2O7}},\ }\href {https://doi.org/10.1103/PhysRevB.94.014410}
  {\bibfield  {journal} {\bibinfo  {journal} {Phys. Rev. B}\ }\textbf {\bibinfo
  {volume} {94}},\ \bibinfo {pages} {014410} (\bibinfo {year}
  {2016})}\BibitemShut {NoStop}%
\bibitem [{\citenamefont {Konieczna}\ \emph {et~al.}(2025)\citenamefont
  {Konieczna}, \citenamefont {Kaib}, \citenamefont {Winter},\ and\
  \citenamefont {Valent{\'i}}}]{KoniecznaNPJQM2025}%
  \BibitemOpen
  \bibfield  {author} {\bibinfo {author} {\bibfnamefont {A.~A.}\ \bibnamefont
  {Konieczna}}, \bibinfo {author} {\bibfnamefont {D.~A.~S.}\ \bibnamefont
  {Kaib}}, \bibinfo {author} {\bibfnamefont {S.~M.}\ \bibnamefont {Winter}},\
  and\ \bibinfo {author} {\bibfnamefont {R.}~\bibnamefont {Valent{\'i}}},\
  }\bibfield  {title} {\bibinfo {title} {Understanding the microscopic origin
  of the magnetic interactions in \ce{CoNb2O6}},\ }\href
  {https://doi.org/10.1038/s41535-024-00715-6} {\bibfield  {journal} {\bibinfo
  {journal} {npj Quantum Mater.}\ }\textbf {\bibinfo {volume} {10}},\ \bibinfo
  {pages} {8} (\bibinfo {year} {2025})}\BibitemShut {NoStop}%
\bibitem [{\citenamefont {Rodrigues}(1840)}]{Rodrigues1840}%
  \BibitemOpen
  \bibfield  {author} {\bibinfo {author} {\bibfnamefont {O.}~\bibnamefont
  {Rodrigues}},\ }\bibfield  {title} {\bibinfo {title} {Des lois
  g{\'e}om{\'e}triques qui r{\'e}gissent les d{\'e}placements d'un syst\`eme
  solide dans l'espace, et de la variation des coordonn{\'e}es provenant de ces
  d{\'e}placements consid{\'e}r{\'e}s ind{\'e}pendamment des causes qui peuvent
  les produire},\ }\href {https://www.numdam.org/item/JMPA_1840_1_5__380_0/}
  {\bibfield  {journal} {\bibinfo  {journal} {J. Math. Pures Appl.}\ }\textbf
  {\bibinfo {volume} {5}},\ \bibinfo {pages} {380} (\bibinfo {year}
  {1840})}\BibitemShut {NoStop}%
\bibitem [{\citenamefont {Koepernik}\ \emph {et~al.}(2023)\citenamefont
  {Koepernik}, \citenamefont {Janson}, \citenamefont {Sun},\ and\ \citenamefont
  {van~den Brink}}]{fploWF}%
  \BibitemOpen
  \bibfield  {author} {\bibinfo {author} {\bibfnamefont {K.}~\bibnamefont
  {Koepernik}}, \bibinfo {author} {\bibfnamefont {O.}~\bibnamefont {Janson}},
  \bibinfo {author} {\bibfnamefont {Y.}~\bibnamefont {Sun}},\ and\ \bibinfo
  {author} {\bibfnamefont {J.}~\bibnamefont {van~den Brink}},\ }\bibfield
  {title} {\bibinfo {title} {Symmetry-conserving maximally projected wannier
  functions},\ }\href {https://doi.org/10.1103/PhysRevB.107.235135} {\bibfield
  {journal} {\bibinfo  {journal} {Phys. Rev. B}\ }\textbf {\bibinfo {volume}
  {107}},\ \bibinfo {pages} {235135} (\bibinfo {year} {2023})}\BibitemShut
  {NoStop}%
\bibitem [{\citenamefont {Koepernik}\ and\ \citenamefont
  {Eschrig}(1999)}]{fplocite1}%
  \BibitemOpen
  \bibfield  {author} {\bibinfo {author} {\bibfnamefont {K.}~\bibnamefont
  {Koepernik}}\ and\ \bibinfo {author} {\bibfnamefont {H.}~\bibnamefont
  {Eschrig}},\ }\bibfield  {title} {\bibinfo {title} {Full-potential
  nonorthogonal local-orbital minimum-basis band-structure scheme},\ }\href
  {https://doi.org/10.1103/PhysRevB.59.1743} {\bibfield  {journal} {\bibinfo
  {journal} {Phys. Rev. B}\ }\textbf {\bibinfo {volume} {59}},\ \bibinfo
  {pages} {1743} (\bibinfo {year} {1999})}\BibitemShut {NoStop}%
\bibitem [{\citenamefont {Opahle}\ \emph {et~al.}(1999)\citenamefont {Opahle},
  \citenamefont {Koepernik},\ and\ \citenamefont {Eschrig}}]{fplocite2}%
  \BibitemOpen
  \bibfield  {author} {\bibinfo {author} {\bibfnamefont {I.}~\bibnamefont
  {Opahle}}, \bibinfo {author} {\bibfnamefont {K.}~\bibnamefont {Koepernik}},\
  and\ \bibinfo {author} {\bibfnamefont {H.}~\bibnamefont {Eschrig}},\
  }\bibfield  {title} {\bibinfo {title} {Full-potential band-structure
  calculation of iron pyrite},\ }\href
  {https://doi.org/10.1103/PhysRevB.60.14035} {\bibfield  {journal} {\bibinfo
  {journal} {Phys. Rev. B}\ }\textbf {\bibinfo {volume} {60}},\ \bibinfo
  {pages} {14035} (\bibinfo {year} {1999})}\BibitemShut {NoStop}%
\bibitem [{\citenamefont {Eschrig}\ \emph {et~al.}(2004)\citenamefont
  {Eschrig}, \citenamefont {Richter},\ and\ \citenamefont {Opahle}}]{fploRel}%
  \BibitemOpen
  \bibfield  {author} {\bibinfo {author} {\bibfnamefont {H.}~\bibnamefont
  {Eschrig}}, \bibinfo {author} {\bibfnamefont {M.}~\bibnamefont {Richter}},\
  and\ \bibinfo {author} {\bibfnamefont {I.}~\bibnamefont {Opahle}},\
  }\bibfield  {title} {\bibinfo {title} {Relativistic solid state
  calculations},\ }in\ \href {https://doi.org/10.1016/S1380-7323(04)80039-6}
  {\emph {\bibinfo {booktitle} {Relativistic Electronic Structure Theory, Part
  2: Applications}}},\ \bibinfo {series} {Theoretical and Computational
  Chemistry}, Vol.~\bibinfo {volume} {14},\ \bibinfo {editor} {edited by\
  \bibinfo {editor} {\bibfnamefont {P.}~\bibnamefont {Schwerdtfeger}}}\
  (\bibinfo  {publisher} {Elsevier},\ \bibinfo {address} {Amsterdam},\ \bibinfo
  {year} {2004})\ pp.\ \bibinfo {pages} {723--776}\BibitemShut {NoStop}%
\bibitem [{\citenamefont {Carter}\ \emph {et~al.}(2013)\citenamefont {Carter},
  \citenamefont {Shankar~V.},\ and\ \citenamefont {Kee}}]{CarterPRB2012}%
  \BibitemOpen
  \bibfield  {author} {\bibinfo {author} {\bibfnamefont {J.-M.}\ \bibnamefont
  {Carter}}, \bibinfo {author} {\bibfnamefont {V.}~\bibnamefont {Shankar~V.}},\
  and\ \bibinfo {author} {\bibfnamefont {H.-Y.}\ \bibnamefont {Kee}},\
  }\bibfield  {title} {\bibinfo {title} {Theory of metal-insulator transition
  in the family of perovskite iridium oxides},\ }\href
  {https://doi.org/10.1103/PhysRevB.88.035111} {\bibfield  {journal} {\bibinfo
  {journal} {Phys. Rev. B}\ }\textbf {\bibinfo {volume} {88}},\ \bibinfo
  {pages} {035111} (\bibinfo {year} {2013})}\BibitemShut {NoStop}%
\bibitem [{\citenamefont {Yildirim}\ \emph {et~al.}(1995)\citenamefont
  {Yildirim}, \citenamefont {Harris}, \citenamefont {Aharony},\ and\
  \citenamefont {Entin-Wohlman}}]{YildirimPRB1995}%
  \BibitemOpen
  \bibfield  {author} {\bibinfo {author} {\bibfnamefont {T.}~\bibnamefont
  {Yildirim}}, \bibinfo {author} {\bibfnamefont {A.~B.}\ \bibnamefont
  {Harris}}, \bibinfo {author} {\bibfnamefont {A.}~\bibnamefont {Aharony}},\
  and\ \bibinfo {author} {\bibfnamefont {O.}~\bibnamefont {Entin-Wohlman}},\
  }\bibfield  {title} {\bibinfo {title} {Anisotropic spin hamiltonians due to
  spin-orbit and coulomb exchange interactions},\ }\href
  {https://doi.org/10.1103/PhysRevB.52.10239} {\bibfield  {journal} {\bibinfo
  {journal} {Phys. Rev. B}\ }\textbf {\bibinfo {volume} {52}},\ \bibinfo
  {pages} {10239} (\bibinfo {year} {1995})}\BibitemShut {NoStop}%
\bibitem [{\citenamefont {Swendsen}\ and\ \citenamefont
  {Wang}(1986)}]{SwendsenPRL1986}%
  \BibitemOpen
  \bibfield  {author} {\bibinfo {author} {\bibfnamefont {R.~H.}\ \bibnamefont
  {Swendsen}}\ and\ \bibinfo {author} {\bibfnamefont {J.-S.}\ \bibnamefont
  {Wang}},\ }\bibfield  {title} {\bibinfo {title} {Replica monte carlo
  simulation of spin-glasses},\ }\href
  {https://doi.org/10.1103/PhysRevLett.57.2607} {\bibfield  {journal} {\bibinfo
   {journal} {Phys. Rev. Lett.}\ }\textbf {\bibinfo {volume} {57}},\ \bibinfo
  {pages} {2607} (\bibinfo {year} {1986})}\BibitemShut {NoStop}%
\bibitem [{\citenamefont {Hukushima}\ and\ \citenamefont
  {Nemoto}(1996)}]{HukushimaJPSJ1996}%
  \BibitemOpen
  \bibfield  {author} {\bibinfo {author} {\bibfnamefont {K.}~\bibnamefont
  {Hukushima}}\ and\ \bibinfo {author} {\bibfnamefont {K.}~\bibnamefont
  {Nemoto}},\ }\bibfield  {title} {\bibinfo {title} {Exchange monte carlo
  method and application to spin glass simulations},\ }\href
  {https://doi.org/10.1143/JPSJ.65.1604} {\bibfield  {journal} {\bibinfo
  {journal} {J. Phys. Soc. Jpn.}\ }\textbf {\bibinfo {volume} {65}},\ \bibinfo
  {pages} {1604} (\bibinfo {year} {1996})}\BibitemShut {NoStop}%
\bibitem [{\citenamefont {Lakshmanan}(2011)}]{LakshmananLLGeq2011}%
  \BibitemOpen
  \bibfield  {author} {\bibinfo {author} {\bibfnamefont {M.}~\bibnamefont
  {Lakshmanan}},\ }\bibfield  {title} {\bibinfo {title} {The fascinating world
  of the {L}andau-{L}ifshitz-{G}ilbert equation: an overview},\ }\href
  {https://doi.org/10.1098/rsta.2010.0319} {\bibfield  {journal} {\bibinfo
  {journal} {Philos. Trans. R. Soc. A}\ }\textbf {\bibinfo {volume} {369}},\
  \bibinfo {pages} {1280} (\bibinfo {year} {2011})}\BibitemShut {NoStop}%
\bibitem [{\citenamefont {Creutz}(1987)}]{CreutzPRD1987}%
  \BibitemOpen
  \bibfield  {author} {\bibinfo {author} {\bibfnamefont {M.}~\bibnamefont
  {Creutz}},\ }\bibfield  {title} {\bibinfo {title} {Overrelaxation and monte
  carlo simulation},\ }\href {https://doi.org/10.1103/PhysRevD.36.515}
  {\bibfield  {journal} {\bibinfo  {journal} {Phys. Rev. D}\ }\textbf {\bibinfo
  {volume} {36}},\ \bibinfo {pages} {515} (\bibinfo {year} {1987})}\BibitemShut
  {NoStop}%
\bibitem [{\citenamefont {Kanki}\ \emph {et~al.}(2005)\citenamefont {Kanki},
  \citenamefont {Loison},\ and\ \citenamefont {Schotte}}]{KankiPRD2005}%
  \BibitemOpen
  \bibfield  {author} {\bibinfo {author} {\bibfnamefont {K.}~\bibnamefont
  {Kanki}}, \bibinfo {author} {\bibfnamefont {D.}~\bibnamefont {Loison}},\ and\
  \bibinfo {author} {\bibfnamefont {K.~D.}\ \bibnamefont {Schotte}},\
  }\bibfield  {title} {\bibinfo {title} {Efficiency of the microcanonical
  over-relaxation algorithm for vector spins analyzing first and second order
  transitions},\ }\href {https://doi.org/10.1140/epjb/e2005-00130-7} {\bibfield
   {journal} {\bibinfo  {journal} {Eur. Phys. J. B}\ }\textbf {\bibinfo
  {volume} {44}},\ \bibinfo {pages} {309} (\bibinfo {year} {2005})}\BibitemShut
  {NoStop}%
\bibitem [{\citenamefont {Lehoucq}\ \emph {et~al.}(1998)\citenamefont
  {Lehoucq}, \citenamefont {Sorensen},\ and\ \citenamefont
  {Yang}}]{lehoucq1998arpack}%
  \BibitemOpen
  \bibfield  {author} {\bibinfo {author} {\bibfnamefont {R.~B.}\ \bibnamefont
  {Lehoucq}}, \bibinfo {author} {\bibfnamefont {D.~C.}\ \bibnamefont
  {Sorensen}},\ and\ \bibinfo {author} {\bibfnamefont {C.}~\bibnamefont
  {Yang}},\ }\href@noop {} {\emph {\bibinfo {title} {{ARPACK} Users' Guide:
  Solution of Large-scale Eigenvalue Problems with Implicitly Restarted Arnoldi
  Methods}}},\ Software, Environments, and Tools\ (\bibinfo  {publisher}
  {Society for Industrial and Applied Mathematics},\ \bibinfo {year}
  {1998})\BibitemShut {NoStop}%
\end{thebibliography}%
    }%
}

\phantomsection
\label{EndMatter}

\appendix

\renewcommand{\theequation}{A\arabic{equation}}
\setcounter{equation}{0}

In this End Matter, we present systematically the classical solutions to the one-dimensional AFM sawtooth chain. First we examine the origin of the ground state constraint in the general case where $J_{\text{bb}}\neq J_{\text{bv}}$. Then we present the simple coplanar state, which plays a key role in the three-dimensional model described in the main text. Finally we classify the multiplanar state, to reveal the full degeneracy of the one-dimensional chain.

\section{One-dimensional sawtooth chain}
\begin{figure}[t]
	\centering
    \begin{overpic}[width=1.0\columnwidth,percent,angle=0,grid=false,tics=2]{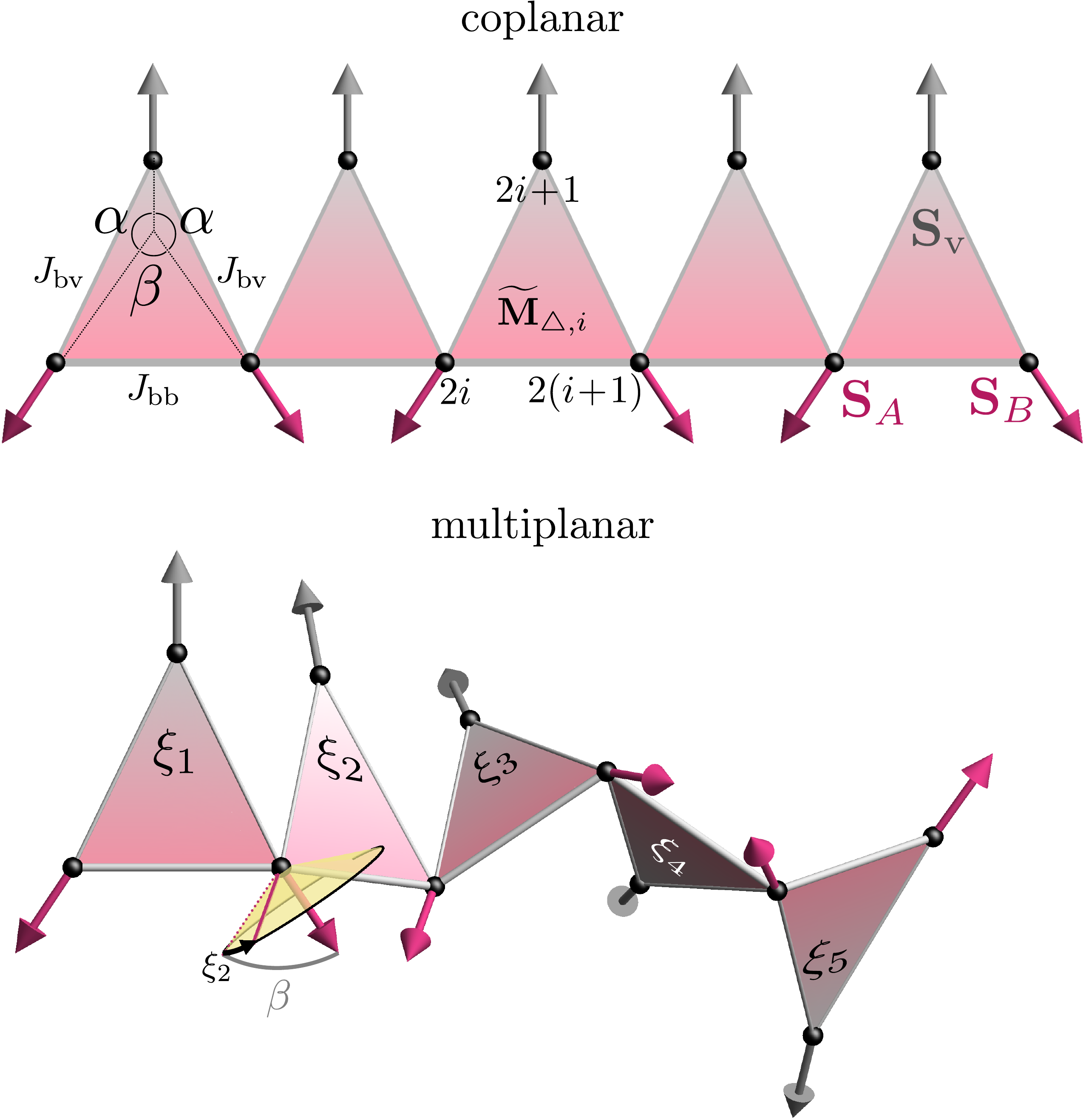}
        \put(0,97){(a)}
        \put(0,51.55){(b)}
    \end{overpic} 
    \caption{An example of the coplanar (a) and multiplanar (b) states are shown. The triangle sites have been rotated to highlight that every triangle must form a single plane of spins, due to the conserved generalized magnetization constraint. The multiplanar degeneracy is found constructively by consecutive cone rotations of each added triangle.}\label{fig_sup:sawtooth_states}
\end{figure}

Treating in isolation the largest $J_{\text{bb}}(=J_4)$ and $J_{\text{bv}}(=J_5)$ parameters leads to the sawtooth chain. The schematic of the sawtooth chain is seen in the main text, and we write the reduced chain model Hamiltonian
\begin{equation}
\begin{array}{rcl}
\mathcal{H}_{\text{chain}} &\!\!\!\!=\!\!\!\!& \displaystyle\sum\limits_{i}\big[J_{\text{bb}} \mathbf{S}_{2i}\!\cdot\!\mathbf{S}_{2(i+1)} \\[12pt]
& & \ \ \ \ \ + J_{\text{bv}} \left( \mathbf{S}_{2i}\!\cdot\!\mathbf{S}_{2i+1} + \mathbf{S}_{2(i+1)}\!\cdot\!\mathbf{S}_{2i+1}\right)\big],
\end{array}
\end{equation}
where $i$ sums over triangles, as defined in the main text, and seen here in Fig.~\ref{fig_sup:sawtooth_states}.
Collections of three sites $2 i$, $2 i+\!1$, and $2(i\!+\!1)$ form the $i^{\text{th}}$ triangle, which share one common base along $2 i$ and $2(i\!+\!1)$, and a vertex $2 i+\!1$. We can rewrite the chain model
\begin{equation}
\begin{array}{rcl}
\mathcal{H}_{\text{chain}} &\!\!\!\!=\!\!\!\!& \dfrac{J_{\text{bb}}}{2}\displaystyle\sum\limits_{i} \|\widetilde{\mathbf{M}}_{\triangle,i}\|^2 \\[12pt]
& & \ \ \ \ +\  N_{\triangle}J_{\text{bb}}S^2\left(
     2 + \cos(2\alpha)
    \right),
\end{array}
\end{equation}
where $\widetilde{\mathbf{M}}_{\triangle,i}= \mathbf{S}_{2i} - 2\cos\alpha\mathbf{S}_{2i + 1} + \mathbf{S}_{2(i+1)}$ is the generalized triangle magnetization and $\alpha\!=\!\arccos[-J_{\text{bv}}/(2 J_{\text{bb}})]$ as defined also in the main text, and $N_{\triangle}$ the number of triangles. Note that when $J_{\text{bb}} = J_{\text{bv}}$, the angle $\alpha$ will be $2\pi/3$, which reduces $\widetilde{\mathbf{M}}_{\triangle,i}$ to the familiar $\mathbf{M}_{\triangle,i}$ in the isotropic limit.
 
 The generalized triangle magnetization forms an individual local conserved quantity, one for every triangle. Being that the chain Hamiltonian is a sum of squares, the energy is minimized when $\widetilde{\mathbf{M}}_{\triangle,i}=\mathbf{0}$ which becomes a constraint for the ground state.

\section{Coplanar and multiplanar states}

The ground state needs to satisfy the condition $\widetilde{\mathbf{M}}_{\triangle,i}=\mathbf{0}$, which can be achieved in many ways. This condition is equivalent to saying that the three spins on the $i^{\text{th}}$ triangle need to live on the same plane. The simplest way to achieve this is for all spins to be on the same plane, which is the coplanar state, an example of which we saw in the main text. In Fig.~\ref{fig_sup:sawtooth_states}(a) we show an equivalent coplanar state which amounts to a global SO(3) rotation.

In addition to the fully coplanar order, there is a continuous family of degenerate multiplanar states. While each triangle individually needs to have spins living on a plane, the relative planes between triangles need not align, as shown constructively in Fig.~\ref{fig_sup:sawtooth_states}(b). To construct such a state, we start from the first triangle in the figure, which already has an arbitrary choice $\xi_1$ related to the global SO(3) rotational freedom. Once fixed, the next triangle will share one base spin with the previous triangle, and must itself form a new plane. This plane can be chosen out of rotations on a cone (see yellow cone in the figure), rotated by $\xi_2$. In this way, the other base spin is guaranteed to form the required $\beta=2\pi - 2\alpha$ angle. Upon choosing one value for $\xi_2$, this also uniquely determines the vertex spin of the second triangle, which needs to satisfy the $\alpha$ angle relative to the base. We can repeat this for every subsequent triangle, always adding a new $\xi_i$ redundancy, one for each added triangle. In this way, coupling $N_{\triangle}$ triangles reveals an $SO(2)^{N_{\triangle}}$ degeneracy in the ground state. The aforementioned degeneracies are an exact description in the limit of an infinite sawtooth chain, or a finite chain with open boundaries. If we wish to impose periodic boundary conditions, this creates one constraint from the periodic boundary condition.

We have verified in Monte Carlo, by simulating sawtooth chains of length up to $N_{\triangle}=200$, that a random multiplanar state is always selected with no canonical pattern to the chosen triangle planes, while always maintaining $\widetilde{\mathbf{M}}_{\triangle,i}=\bm{0}$, and $\alpha$ and $\beta$ angle relations.


\clearpage
\newpage
\setcounter{page}{1}

\let\section\standardsection
\let\maketitle\savedmaketitle
\let\title\savedtitle
\let\author\savedauthor
\let\affiliation\savedaffiliation
\let\email\savedemail
\let\homepage\savedhomepage
\let\thanks\savedthanks
\let\date\saveddate
\setcounter{affil}{0}
\setcounter{secnumdepth}{2}
\setcounter{section}{0}
\setcounter{subsection}{0}
\setcounter{figure}{0}
\setcounter{table}{0}
\setcounter{equation}{0}
\renewcommand{\thesection}{S\arabic{section}}
\renewcommand{\thesubsection}{\thesection.\arabic{subsection}}
\renewcommand{\thefigure}{S\arabic{figure}}
\renewcommand{\thetable}{S\arabic{table}}
\renewcommand{\theequation}{S\arabic{equation}}
\def\suppresssupbibliography{}

\title{SUPPLEMENTAL MATERIAL FOR:\\[3pt]Bobkingite, a new coupled sawtooth chain platform}

\author{P. Peter Stavropoulos\orcid{0000-0002-4193-5844}}
\email{panagiotis@itp.uni-frankfurt.de}
\affiliation{Institut f\"ur Theoretische Physik, Goethe-Universit\"at Frankfurt, 60438 Frankfurt am Main, Germany}

\author{Aleksandar Razpopov\orcid{0009-0009-6935-3297}}
\affiliation{Institut f\"ur Theoretische Physik, Goethe-Universit\"at Frankfurt, 60438 Frankfurt am Main, Germany}

\author{Harrison LaBollita\orcid{0000-0002-6699-8577}}
\affiliation{Center for Computational Quantum Physics, Flatiron Institute, New York, New York 10010, USA}

\author{David~A.~S.~Kaib\orcid{0000-0002-1448-0063}}
\affiliation{Institut f\"ur Theoretische Physik, Goethe-Universit\"at Frankfurt, 60438 Frankfurt am Main, Germany}

\author{Michael R. Norman\orcid{0000-0002-9459-078X}}
\affiliation{Materials Science Division, Argonne National Laboratory, Lemont, IL 60439, USA}

\author{Antia S. Botana\orcid{0000-0001-5973-3039}}
\affiliation{Department of Physics, Arizona State University, Tempe, AZ 85287, USA}

\author{Roser Valent\'{\i}\orcid{0000-0003-0497-1165}}
\email{valenti@itp.uni-frankfurt.de}
\affiliation{Institut f\"ur Theoretische Physik, Goethe-Universit\"at Frankfurt, 60438 Frankfurt am Main, Germany}

\date{\today}

\maketitle

\begin{table*}
\caption{\label{tab:comp_functionals} Conventional lattice parameter changes between PBE and PBEsol relaxations. Included is a comparison of the $a^{\text{th}}$ neighbor \ce{Cu}-\ce{Cu} distances, and \ce{Cu}(2)-\ce{O}-\ce{Cu}(w) superexchange angles, corresponding to $J_4$ for $w=2$ and $J_5$ for $w=3$.}
\begin{ruledtabular}
\begin{tabular}{l|*{1}{d{3.3}}*{1}{d{2.3}}*{1}{d{1.3}}*{1}{d{1.3}}*{1}{d{3.3}}|*{5}{d{1.3}}|*{2}{d{3.3}}}
\multicolumn{1}{c|}{\multirow[c]{2}{*}{GGA}} &
\multicolumn{5}{c|}{lattice parameters} &
\multicolumn{5}{c|}{$a^{\text{th}}$ \ce{Cu}-\ce{Cu} bond (\AA)} &
\multicolumn{2}{c}{\ce{Cu}(2)-O-\ce{Cu}($w$) angles (${}^\circ$)} \\
 \multicolumn{1}{c|}{}&
 \multicolumn{1}{c}{vol (\AA$^3$)} & 
 \multicolumn{1}{c}{$a^c$ (\AA)} &
 \multicolumn{1}{c}{$b^c$ (\AA)} &
 \multicolumn{1}{c}{$c^c$ (\AA)} &
 \multicolumn{1}{c|}{$\beta^c$ (${}^\circ$)} &
 \multicolumn{1}{c}{$a=1$} & 
 \multicolumn{1}{c}{$a=2$} & 
 \multicolumn{1}{c}{$a=3$} & 
 \multicolumn{1}{c}{$a=4$} & 
 \multicolumn{1}{c|}{$a=5$} & 
 \multicolumn{1}{c}{$w=2$} & 
 \multicolumn{1}{c}{$w=3$} \\
 \hline
PBE &   572.440 &    10.616 &     6.514 &     8.956 &   108.990 &     2.919 &     3.066 &     3.114 &     3.257 &     3.379 &   113.412 &   115.863 \\
PBEsol &   540.004 &    10.354 &     6.412 &     8.803 &   108.975 &     2.875 &     3.009 &     3.045 &     3.206 &     3.316 &   112.574 &   116.011 \\
\end{tabular}
\end{ruledtabular}
\end{table*}

\begin{table*}
\caption{\label{tab:heis_mod_all_data}  Heisenberg couplings $J_a$, of the $a^{\text{th}}$ neighbor, with length $d_a$, for a range of $U$.}
\begin{ruledtabular}
\begin{tabular}{cd{1.3}|*{5}{d{3.3}}}
\multicolumn{1}{c}{\multirow[c]{2}{*}{$a$}} &
\multicolumn{1}{c|}{\multirow[c|]{2}{*}{$d_a$ (\AA)}} & 
\multicolumn{5}{c}{$J_a\,\mathrm{(meV)}$}\\ 
 \multicolumn{1}{c}{} &
 \multicolumn{1}{c|}{} & 
 \multicolumn{1}{c}{$U\!=\!4\,\mathrm{eV}$} & 
 \multicolumn{1}{c}{$U\!=\!6\,\mathrm{eV}$} & 
 \multicolumn{1}{c}{$U\!=\!8\,\mathrm{eV}$} & 
 \multicolumn{1}{c}{$U\!=\!10\,\mathrm{eV}$} & 
 \multicolumn{1}{c}{$U\!=\!12\,\mathrm{eV}$}\\
\hline
1 &    2.919 &   -8.06 &   -8.03 &   -6.79 &   -5.05 &   -3.25 \\
2 &    3.066 &   21.77 &   14.42 &    8.83 &    4.98 &    2.64 \\
3 &    3.114 &    0.21 &   -0.37 &   -0.49 &   -0.38 &   -0.20 \\
4 &    3.257 &   18.07 &   15.14 &   11.13 &    7.26 &    4.19 \\
5 &    3.379 &   39.42 &   27.25 &   18.12 &   11.18 &    6.29 \\
6 &    5.174 &   -0.04 &   -0.10 &   -0.08 &   -0.05 &   -0.02 \\
7 &    5.308 &    2.16 &    0.99 &    0.50 &    0.26 &    0.13 \\
\hline
\multicolumn{2}{c|}{$J_4/J_5$} &  0.458  &   0.556 &   0.614 &   0.649 &   0.666
\end{tabular}
\end{ruledtabular}
\end{table*}

\begin{table}
\caption{\label{tab:relaxed_crystal_data} Crystallographic data, in the $\mathbf{a}^c$, $\mathbf{b}^c$, $\mathbf{c}^c$ conventional unit cell, of the fully relaxed structure.}
\begin{ruledtabular}
\begin{tabular}{l*{3}{d{3.5}}l}\\[-6pt]
\multicolumn{1}{l}{SG} & \multicolumn{4}{c}{\text{$C2/m$ (unique $b$)}} \\[2pt]
\multicolumn{1}{l}{\text{$a^{c}$ (\AA)}} & \multicolumn{4}{d{3.5}}{10.61633} \\
\multicolumn{1}{l}{\text{$b^{c}$ (\AA)}} & \multicolumn{4}{d{3.5}}{6.51396} \\
\multicolumn{1}{l}{\text{$c^{c}$ (\AA)}} & \multicolumn{4}{d{3.5}}{8.95612} \\
\multicolumn{1}{l}{\text{$\beta$ (${}^\circ$)}} & \multicolumn{4}{d{3.5}}{112.44407} \\[2pt]
\multicolumn{1}{l}{\text{vol (\AA\textsuperscript{$3$})}} & \multicolumn{4}{c}{572.44} \\[6pt]
\hline
\multicolumn{1}{l}{Atom}  & 
\multicolumn{1}{d{3.5}}{\text{$x$}} & 
\multicolumn{1}{d{3.5}}{\text{$y$}} & 
\multicolumn{1}{d{3.5}}{\text{$z$}} & 
\multicolumn{1}{l}{WP}  \\ 
\hline
\multicolumn{1}{l}{Cu(1)} &        0 &        0 &      0.5 & \multicolumn{1}{c}{2b} \\ 
\multicolumn{1}{l}{Cu(2)} &     0.25 &     0.25 &      0.5 & \multicolumn{1}{c}{4i} \\ 
\multicolumn{1}{l}{Cu(3)} &  0.96901 &        0 &  0.82658 & \multicolumn{1}{c}{4g} \\ 
\multicolumn{1}{l}{Cl   } &  0.68127 &        0 &  0.73175 & \multicolumn{1}{c}{4g} \\ 
\multicolumn{1}{l}{O(1) } &  0.44366 &  0.30575 &  0.64719 & \multicolumn{1}{c}{8j} \\ 
\multicolumn{1}{l}{O(2) } &  0.77629 &        0 &  0.39601 & \multicolumn{1}{c}{4g} \\ 
\multicolumn{1}{l}{O(3) } &        0 &  0.79112 &        0 & \multicolumn{1}{c}{4i} \\ 
\multicolumn{1}{l}{O(4) } &  0.65968 &        0 &  0.07382 & \multicolumn{1}{c}{4g} \\ 
\multicolumn{1}{l}{H(1) } &  0.50744 &  0.18634 &  0.68005 & \multicolumn{1}{c}{8j} \\ 
\multicolumn{1}{l}{H(2) } &  0.73335 &        0 &  0.27353 & \multicolumn{1}{c}{4g} \\ 
\multicolumn{1}{l}{H(3) } &   0.4154 &  0.16549 &  0.96464 & \multicolumn{1}{c}{8j} \\ 
\multicolumn{1}{l}{H(4) } &  0.29264 &        0 &  0.00116 & \multicolumn{1}{c}{4g}
\end{tabular}
\end{ruledtabular}
\end{table}

\section{Density functional theory}
Density functional theory (DFT) calculations are performed using the projector augmented wave method as implemented in the \textit{Vienna ab-initio simulation package} (VASP)~\cite{kresse1993,kresse1996,kresse1999}.
The Perdew-Burke-Ernzerhof (PBE) version of the generalized gradient approximation (GGA) is used as the exchange-correlation functional~\cite{perdew1996}.
For total energy calculations, a kinetic energy cutoff of 600 eV sets the size of the plane-wave basis.
An on-site Coulomb repulsion is introduced for the strongly localized Cu-3$d$ electrons.
The Brillouin zone (BZ) integration is performed on $8\!\times\!8\!\times\!6$ and $4\!\times\!7\!\times\!5$ Monkhorst-Pack $\mathbf{k}$-grids with a $0.1$ Gaussian smearing for the primitive and conventional unit cells, respectively, and an energy criteria of $10^{-8}$ eV is used for the self-consistent energy convergence.

\subsection{Electronic structure}
\begin{figure}[t]
	\centering
    \begin{overpic}[width=1\columnwidth,percent,angle=0]{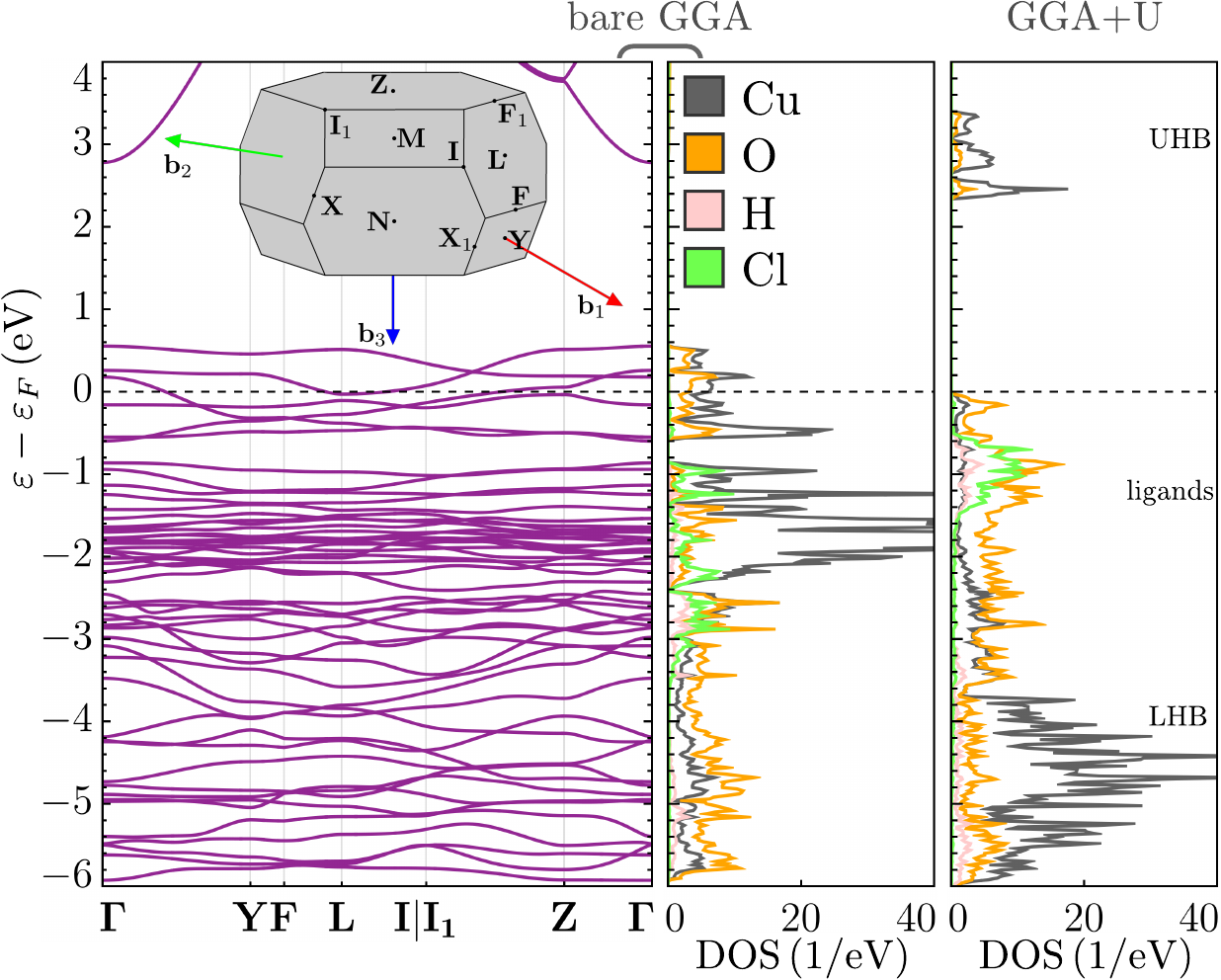}
    \end{overpic} 
    \caption{Bands and DOS from bare GGA, with the primitive BZ shown as an inset. Band counting reveals a $3d^9$ configuration, with significant oxygen character below the Fermi level that acts to mediate the superexchange. As a comparison, the DOS from GGA+U is also shown, demonstrating the gapping out of copper $d$ states near the Fermi level, leading to a Mott insulator.}\label{fig:gga_bands}
\end{figure}

In Fig.~\ref{fig:gga_bands} we show the nonmagnetic bare GGA band structure and species-resolved density of states of bobkingite, which serves as a guide to the leading degrees of freedom. The states around the Fermi level are predominantly of \ce{Cu} $d$ character, indicating that the low-energy degrees of freedom primarily reside on the copper sites. Consistent with the expected \ce{Cu^{2+}} ($3d^9$) configuration, we find approximately $2.5$ Kramers-degenerate unoccupied bands above $\varepsilon_F$ per primitive cell, justifying an effective spin-$1/2$ description.
The substantial \ce{O} contribution to these bands reflects strong Cu--O covalency, as discussed in the main text. 

The bare GGA will be gapped out by the onsite electron-electron interaction $U$. In Fig.~\ref{fig:gga_bands} we also show the GGA+U density of states, using the Liechtenstein approach~\cite{liechtenstein1995}, with \ce{Cu} $3d$ onsite values of $U=8\,\text{eV}$ and $J=0.7\,\text{eV}$. As expected a gap opens up, with upper and lower Hubbard bands formed from the \ce{Cu} $d$ states. In between them we see an island of ligand bands resembling what is found in other copper minerals like herbertsmithite~\cite{PustogowPRB2017}.

\subsection{Structural relaxations}

For optimizing the crystal structure, we performed a full relaxation (atomic positions, unit cell shape, unit cell volume), using the PBE functional, with a kinetic energy cutoff of $600\,\text{eV}$, and a force convergence criterion of $10^{-3}\,\text{eV/\AA}$. The full crystallographic results after relaxation are shown in Table~\ref{tab:relaxed_crystal_data} in standard conventional settings. The primitive cell lattice vectors $\mathbf{a}$, $\mathbf{b}$, $\mathbf{c}$, which we have adopted in this work, are related to the crystallographic conventional unit cell $\mathbf{a}^c$, $\mathbf{b}^c$, $\mathbf{c}^c$ reported in the table, by the transformation
\begin{equation}
    \begin{pmatrix} \mathbf{a} & \mathbf{b} & \mathbf{c} \end{pmatrix}= 
    \begin{pmatrix} \mathbf{a}^c & \mathbf{b}^c & \mathbf{c}^c \end{pmatrix}
    \begin{pmatrix} 1/2 & 0 & 0 \\
                    1/2 & 1 & 0 \\
                      0 & 0 & 1
    \end{pmatrix}
\end{equation} 

Additionally, we have performed the relaxations using the revised PBE implementation for solids (PBEsol)~\cite{PBEsol}. A comparison of PBE and PBEsol is reported in Table~\ref{tab:comp_functionals}. Empirically, PBE tends to overestimate volumes, while PBEsol tends to underestimate them. We find a $6\%$ shrinkage in volume using PBEsol, resulting in all copper bonds having slightly shorter length. Nevertheless, the overall shape and geometry is not significantly changed. This is seen already in the monoclinic angle $\beta$ which only changes by $0.02^\circ$ between the functionals. More importantly for the superexchange paths, the copper-oxygen bond angles are minimally affected.

\subsection{Magnetocrystalline anisotropy}

Anisotropic magnetic terms manifest from the effects of spin-orbit coupling (SOC), resulting in preferred lower energy magnetic orientations. In order to investigate the magnitude of magnetic anisotropy in bobkingite, we performed GGA+U+SOC non-self-consistent calculations restarted from collinear converged GGA+U calculations, which is the recommended way in VASP to study magnetocrystalline anisotropy. 

For the collinear seeds we use a FM starting point with all copper sites in the up state, and a configuration of $3$ up on \ce{Cu}($1$) and \ce{Cu}($2$) and $2$ down on \ce{Cu}($3$) ($3\text{u}2\text{d}$). 
Checking the total energy for several directions, including $\mathbf{a}$, $\mathbf{b}$, $\mathbf{b}^c$, $\mathbf{c}^*\!=\!\mathbf{a}\!\times\!\mathbf{b}$, and $\mathbf{c}$, we find a difference of $0.041$ and $0.056$ meV per primitive cell, for the FM and $3\text{u}2\text{d}$ collinear seed, respectively. 
This energy difference is exceedingly small, confirming that anisotropic spin interactions are weak in bobkingite. 
The scale of $10^{-2}$meV energy difference is below the accuracy of our DFT simulations, therefore we cannot reliably determine the preferred orientation (but it likely is the $\mathbf{b}$ direction). 

The extremely small energy differences indicate that anisotropic spin interactions must be much smaller in magnitude. This would be consistent with other copper minerals like herbertsmithite \cite{HerbAniso2008,HerbAniso2012,HerbAniso2025}. As a first order analysis we neglect them, and proceed with isotropic Heisenberg exchanges (see the next section for details of their determination).

\section{Total energy mapping analysis for several \texorpdfstring{$U$}{U}}
The isotropic spin Hamiltonian parameters are extracted by the total energy mapping analysis (TEMA) method~\cite{GlasbrennerNature2015,razpopov2023j,garcia2026microscopic}. The estimation using TEMA is a two step procedure, which we present below.

First, we perform spin-polarized GGA+U calculations for $76$ different magnetic spin configurations. Calculations are performed in the VASP framework using the GGA+U exchange-correlation functional for the \ce{Cu} $3d$ orbitals, where we apply the Liechtenstein approach~\cite{liechtenstein1995}.
We vary the Hubbard parameter $U$ and fix the Hund's coupling $J_{\rm H} = 0.7$ eV which is a realistic value for Cu.
Calculations are carried out within the conventional unit cell, which comprises $10$ magnetic Cu atoms. The BZ is sampled using a $5 \times 6 \times 6$ $\mathbf{k}$-point grid, which was tested to ensure total energy convergence.

In the second step, we fit the GGA total energies of the different magnetic configurations to an effective spin Hamiltonian via the method of least squares. The complete results for several values of $U$ are listed in Table~\ref{tab:heis_mod_all_data}.
\begin{figure}[t]
	\centering
    \begin{overpic}[width=1.0\columnwidth,percent,grid=false,tics=5,angle=0]{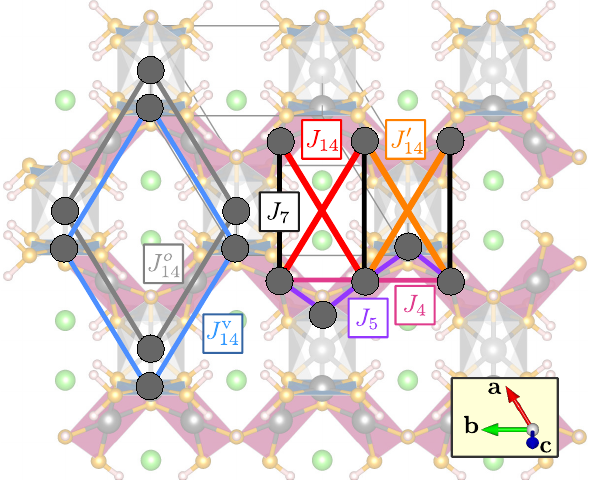}
    \end{overpic} 
    \caption{Top view of bobkingite, looking down the normal to the $\mathbf{ab}$ plane, with the sawtooth running along $\mathbf{b}$. The second interchain base-to-base coupling $J_{14}$, of length $6.227$\AA, is shown, along with the first interchain base-to-base coupling $J_7$. The additional symmetry inequivalent $14$\textsuperscript{th} neighbor bonds $J_{14}^\prime$, $J_{14}^\mathrm{v}$, and $J_{14}^o$, are also shown for completeness. The sawtooth bonds $J_4$ and $J_5$ serve as a reference point. }\label{fig:additionalbonds}
\end{figure}

While the main focus is the first few bonds which form the nearest neighbor sawtooth chain with dominant exchanges, $J_7$ has been kept and discussed in the main text as the first contribution to directly couple the base of the chains to neighboring chains. The second contribution comes from the $14$\textsuperscript{th} neighbor by length at $6.227$\AA. In Fig.~\ref{fig:additionalbonds} we present $J_{14}$ and we note the nuance that there are symmetry inequivalent bonds at exactly the same length, which couple other parts like the dangling sites (Wyckoff position $1$) and the vertices. Given the TEMA setup described above, without enlarging the calculations, we can use the same data set to test the fit on a $J_1 - J_7 + J_{14}$ model, omitting $J_8 - J_{13}$ and the symmetry inequivalent $J_{14}^{\mathrm{v}}$, $J_{14}^{\prime}$, and  $J_{14}^{o}$. We perform this extraction for $U=8\text{ eV}$ and the results are compared against the original model
\begin{equation}
\begin{array}{c*{2}{d{3.5}}}
\multicolumn{1}{c}{ } &
\multicolumn{1}{c}{\text{Original }J_1-J_7} & 
\multicolumn{1}{c}{\text{Adding }J_{14}} \\
\multicolumn{1}{c}{a} &
\multicolumn{1}{c}{J_a\,\mathrm{(meV)}} & 
\multicolumn{1}{c}{J_a\,\mathrm{(meV)}} \\
\hline
 1 &   -6.79 &   -6.89 \\
 2 &    8.83 &    8.84 \\
 3 &   -0.49 &   -0.48 \\
 4 &   11.13 &    11.04 \\
 5 &   18.12 &    18.13 \\
 6 &   -0.08 &   -0.07 \\
 7 &    0.50 &    0.23 \\
14 &         &    0.71 \\
\end{array}
\end{equation}
Most importantly, the $J_1-J_6$ are minimally affected, meaning the resulting model is stable. $J_{14}$ is small, and importantly AFM, meaning there is competition between $J_{14}$ and $J_{7}$, which can reduce ordering tendencies as mentioned in the main text. The extended TEMA fit finds that it can do a better job by reweighting $J_7$ and placing the weight into $J_{14}$. If we keep the intermediate omitted $J_8-J_{13}$ we would have found a different redistribution of the small weights between all of them, even more so over the omitted $J_{14}^\prime$, $J_{14}^\mathrm{v}$, and $J_{14}^o$. This is a known limitation of TEMA, however, the first few bonds are stable indicating the trustworthy nature of the results overall. Given the stability of the existing results we do not perform more extended additional bond model TEMA fitting.

\section{Microscopic estimate of the sawtooth exchanges}

The magnetocrystalline anisotropy energy difference is already indicative that any anisotropic exchanges will be small, however, it is difficult to deduce the magnitude of these exchanges directly. To this end, we carry out a simplified microscopic analysis to estimate the approximate value of these anisotropic interactions. In particular, we aim to estimate the anisotropic exchanges on the sawtooth chain bonds, for which we already found the largest Heisenberg couplings $J_4$ (base-to-base) and $J_5$ (base-to-vertex).

\subsection{Symmetry allowed exchanges}
\begin{figure*}[t]
	\centering
    \begin{overpic}[width=1.0\textwidth,percent,grid=false,tics=2,angle=0]{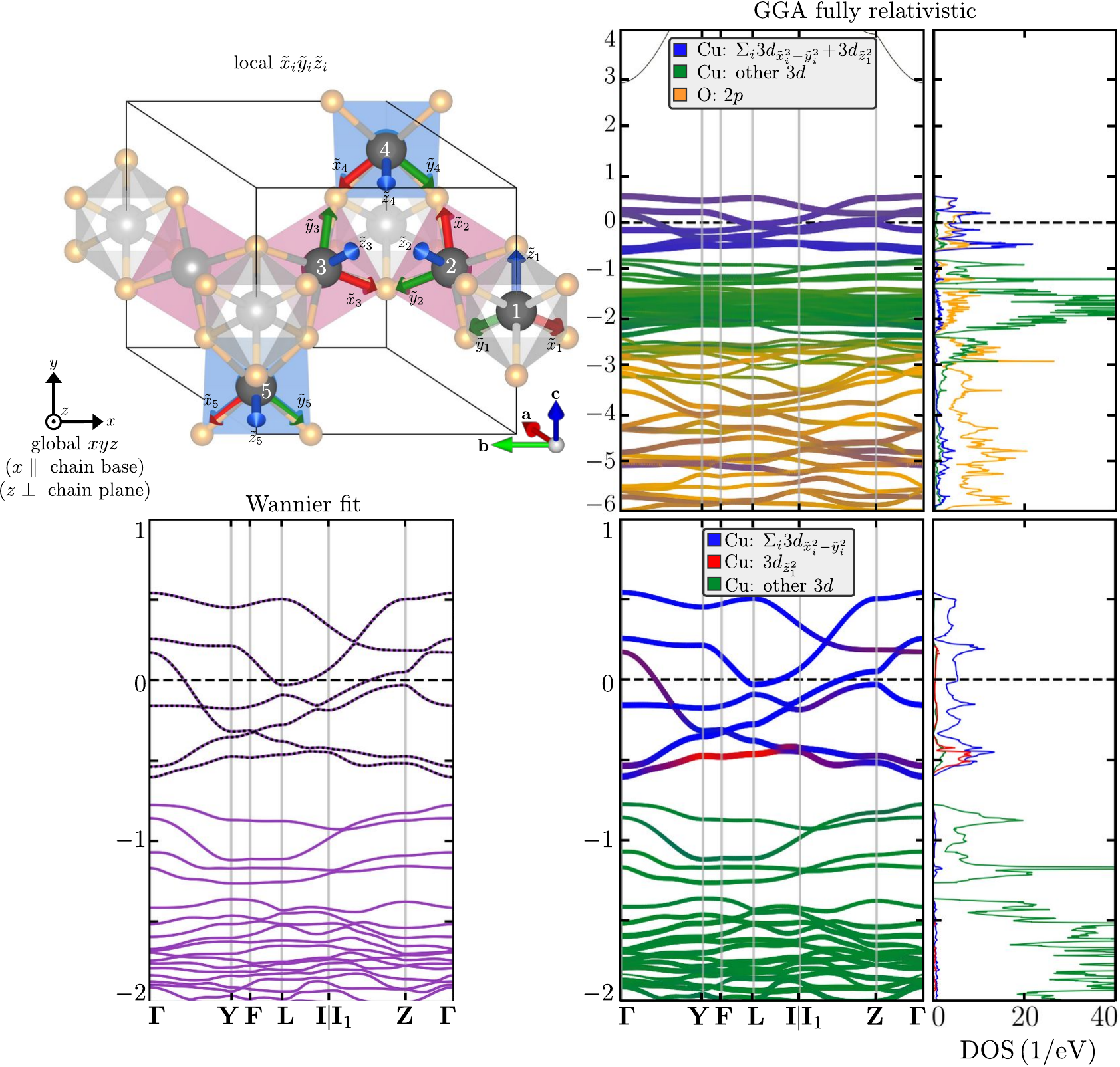}
        \put(2,93){(a)}
        \put(51,93){(b)}
        \put(8,47){(d)}
        \put(51,47){(c)}
    \end{overpic} 
    \caption{(a) The primitive unit cell, with the five \ce{Cu} sites and their corresponding local basis $\tilde{x}\tilde{y}\tilde{z}$ (red, green, blue) following from the local crystal environment. The global basis $xyz$ which is such that $z$ is perpendicular to the sawtooth plane is also shown. (b) The unpolarized fully relativistic PBE bands and DOS found from FPLO, resolved in the dominant metallic Fermi level contribution $\sum\nolimits_{i=1}^{5} d_{\tilde{x}_i^2-\tilde{y}_i^2} + d_{\tilde{z}_1^2}$ (blue), hybridizing oxygen (orange), and the rest of the $d$-orbitals (green). (c) Examination of the $d_{\tilde{z}_1^2}$ (red) mixture into the Fermi level bands, mainly located at the bottom of the manifold but mixing into the other Fermi level bands. (d) The Wannier extracted bands (black dashed), plotted over the DFT bands (purple), showing excellent agreement.}\label{fig:microscopics_struct_bands_dos}
\end{figure*}

Before proceeding to the microscopics, the symmetry of the bonds under the $C2/m$ space group reveals the symmetry reduced form exchanges should have in principle. We chose one of the bonds to symmetry reduce, since the remaining bonds are generated by symmetry. For $J_4$ we chose the bond between \ce{Cu} site $2$ to site $3$, with copper site enumeration as shown in Fig.~\ref{fig:microscopics_struct_bands_dos}(a). The point group operations of  $C2/m$, centered at the primitive unit cell origin, are $C_{2h}=\{E, C_{2\mathbf{b}}, I, m_{\mathbf{b}}\}$, where $E$ is the identity operator, $C_{2\mathbf{b}}$ is a two-fold rotation around $\mathbf{b}$, $I$ is spatial inversion, and $m_{\mathbf{b}}$ is the mirror through the plane perpendicular to $\mathbf{b}$. The self mapping operations that leave bond $(2\rightarrow 3)$ invariant, up to lattice translations, is $\{E,m_{\mathbf{b}}^T\}\cong C_s$. Thus bond $ (2\rightarrow 3)$ is invariant under $C_s$. For $J_5$ we chose the bond between \ce{Cu} site $2$ to site $4$, which has no self-mapping operations, and is trivially invariant under $C_1$.

To write down anisotropic exchanges, we chose a right-handed global coordinate system such that the global $z$-axis is perpendicular to the plane containing the sawtooth chain, and the global $x$-axis is along the $-\mathbf{b}$ direction, in other words along the chain base, see Fig.~\ref{fig:microscopics_struct_bands_dos}(a). The chosen $J_4$ bond $(2\rightarrow 3)$ will symmetry reduce the $\bm{\mathcal{J}}_{4,(2\rightarrow 3)}$ exchange matrix to
\begin{equation}\label{eq:symJ4}
\begin{array}{c}
 H_{4,(2\rightarrow 3)} = \mathbf{S}_2\cdot \bm{\mathcal{J}}_{4,(2\rightarrow 3)} \cdot \mathbf{S}_3,\\[10pt]
\bm{\mathcal{J}}_{4,(2\rightarrow 3)} = J_4 \bm{1}_{3\!\times\!3} + 
\left(\begin{array}{ccc}
               \phantom{-}K^x_{4} &\phantom{-}D^z_{4} & -D^y_{4} \\[4pt]
               -D^z_{4} &               \phantom{-}K^y_{4} & \phantom{-}\Gamma^x_{4} \\[4pt]
               \phantom{-}D^y_{4} & \phantom{-}\Gamma^x_{4}  &                \phantom{-}K^z_{4} \\[4pt]
\end{array}\right),
\end{array}
\end{equation}
where $K^\gamma$ are the spin component selective diagonal exchanges of the form $K^\gamma S^{\gamma}_i S^{\gamma}_j$ where $\gamma=x,y,z$ spin components, $\mathbf{D}^{(2\rightarrow 3)}=[0,D^y_4,D^z_4]$ are components of the Dzyaloshinskii-Moriya (DM) vector describing the antisymmetric DM interaction~\cite{DzyaloshinskyJPCS1958,MoriyaPRL1960} of the form $\mathbf{D}^{i,j}\cdot(\mathbf{S}_i\times\mathbf{S}_j)$, and $\Gamma^{\gamma}$ are the symmetric off-diagonal interactions of the form $\Gamma^\gamma (S^{\alpha}_i S^{\beta}_j + S^{\beta}_i S^{\alpha}_j)$ where $(\gamma,\alpha,\beta)$ are cyclic permutations of the $(x,y,z)$ spin components. The DM interaction is fully consistent with the DM rules~\cite{MoriyaPR1960} under the single mirror symmetry. While not necessary,  we will maintain the separation of $J_4$ from $K^\mu_4$ to better facilitate tracking the anisotropic exchanges on top of the existing isotropic Heisenberg prediction. Since the $J_5$ bond $(2\rightarrow 4)$ has no symmetry, the  $\bm{\mathcal{J}}_{5,(2\rightarrow 4)}$ exchange matrix remains fully general
\begin{equation}\label{eq:symJ5}
\begin{array}{c}
 H_{5,(2\rightarrow 4)} = \mathbf{S}_2\cdot \bm{\mathcal{J}}_{5,(2\rightarrow 4)} \cdot \mathbf{S}_4,\\[10pt]
\multicolumn{1}{l}{\bm{\mathcal{J}}_{5,(2\rightarrow 4)} = J_{5} \bm{1}_{3\!\times\!3} }\\[8pt]
\ \ \ \ \ \ \ \ \ \ \ \ \ + 
\left(\begin{array}{ccc}
               K^x_{5} & \Gamma^z_{5} + D^z_{5} & \Gamma^y_{5} - D^y_{5} \\[4pt]
\Gamma^z_{5} - D^z_{5} &                K^y_{5} & \Gamma^x_{5} + D^x_{5} \\[4pt]
\Gamma^y_{5} + D^y_{5} & \Gamma^x_{5} - D^x_{5} &                K^z_{5} \\[4pt]
\end{array}\right).
\end{array}
\end{equation}

\subsection{Simplified microscopic approach}
As discussed in the main text, and further elucidated in the bare GGA DOS Fig.~\ref{fig:gga_bands}, the \ce{O} square-like and octahedral environment of the \ce{Cu} sites will leave `one hole' $3d$ states as the leading contributions at the Fermi level, along with some \ce{O} $2p$ hybridization. The spin exchanges will then be set in the canonical strong-coupling expansion sense, where the kinetic hopping term $t_{dp}$ connecting \ce{Cu} and \ce{O} is treated perturbatively on top of a dominant $e-e$ interaction $U$ on the \ce{Cu} sites, which is the physical microscopic picture of the gapping out the GGA+U DOS in Fig.~\ref{fig:gga_bands}. To estimate these exchanges we need to estimate the kinetic term between the metal and ligand $t_{dp}$, and the direct metal to metal $t_{dd}$ (which are canonically much smaller than the metal to ligand), along with the full crystal field split levels, and resum a plethora of virtual hopping processes carrying out a strong-coupling expansion up to at least $4$\textsuperscript{th} order in perturbation~\cite{Fazekas1999}. Alternatively we could use the estimated kinetic terms $t_{dp}$, $t_{dd}$, and crystal field levels, and use numerical methods like projED~\cite{WinterPRB2016,RiedlPRB2016,KoniecznaNPJQM2025} which effectively reproduces the perturbative expansion in a numerical diagonalization and projection scheme. Such methods will be expensive to implement and carry out. 

Given that the material has not yet been synthesized and characterized under laboratory conditions, and since the microscopic details will of course change with more accurate measurement of ligand positions, we opt to perform a more simplified effective analysis, which will at least provide an order of magnitude prediction of the anisotropic exchanges. We start with the effective metal to metal hoppings $t^{\text{eff}}$, and schematically write $t^{\text{eff}} = t_{dd} + t^2_{dp}/\Delta_{pd}$, with a downfolded correction of the metal-ligand-metal process $t^2_{dp}/\Delta_{dp}$ added to the direct metal-metal $t_{dd}$, where $\Delta_{dp}$ is the atomic energy difference between the metal and ligand orbitals. Then the needed $4$\textsuperscript{th} order perturbative expansion using the full kinetic terms is effectively captured in a $2$\textsuperscript{nd} order expansion, and for example the isotropic $\mathcal{J}_{\text{iso}}$ exchange, like the Heisenberg exchange, is estimated as $\mathcal{J}_{\text{iso}} \sim (t^{\text{eff}})^2/U$. To get the anisotropic exchanges we still need to have information of SOC somehow backed into the expansion, since anisotropic interactions arise as a result of entangling spin space and real space. For this, we in principle have to again go higher in the effective perturbation, to allow for the SOC interaction in the virtual processes to occur. However we can once again downfold the effect of SOC into the hoppings if we can find the effective spin resolved $t^{\text{eff}}_{\sigma\sigma'}$ hoppings, with anisotropic exchanges scaling as an appropriate sum of spin dependent hoppings $\mathcal{J}_{\text{aniso}} \sim (t^{\text{eff}}_{\sigma\sigma'})^2/U$. This can be achieved in DFT given the appropriate setup which we now present.

\subsection{Local basis definitions}

The $d_{\tilde{x}^2-\tilde{y}^2}$ character of the copper one hole states makes sense in an appropriate local $\tilde{x}\tilde{y}\tilde{z}$ basis definition. In Fig.~\ref{fig:microscopics_struct_bands_dos}(a) the local co-ordinates for each of the five copper sites of the primitive unit cell are shown. In terms of the global $xyz$ basis, we have the local to global site dependent $\widetilde{\mathbf{R}}_i$ transformations, which can be most easily constructed by writing down the active rotation matrix $\mathbf{R}_i$ that rotates the global axis vectors into the local axis vectors, and taking the transpose 
\begin{equation}\label{eq:basischange}
\begin{array}{rcl}
\widetilde{\mathbf{R}}_i &=& \mathbf{R}_i^T, \\[4pt]
\mathbf{R}_1 &=& \bm{\mathcal{R}}(\psi_1,\mathbf{x})\bm{\mathcal{R}}(\pi/4,\mathbf{z}),\\[4pt]
\mathbf{R}_2 &=& \bm{\mathcal{R}}(\delta\xi,\mathbf{d}) \bm{\mathcal{R}}(2\pi/3,\mathbf{z}) \mathbf{R}_1,\\[4pt]
\mathbf{R}_3 &=& \bm{\mathcal{R}}(\delta\xi,\mathbf{M}_x\mathbf{d}) \bm{\mathcal{R}}(-2\pi/3,\mathbf{z}) \mathbf{R}_1,\\[4pt]
\mathbf{R}_4 &=& \bm{\mathcal{R}}(\psi_4,\mathbf{x})\bm{\mathcal{R}}(-3\pi/4,\mathbf{z}), \\[4pt]
\mathbf{R}_5 &=& \mathbf{R}_4,\\[8pt]
\multicolumn{3}{c}{\psi_1 =  -113.802^\circ,\ \psi_4 =  37.138^\circ,}\\
\multicolumn{3}{c}{\delta\xi = 1.874^\circ,\ \mathbf{d} = [0.253,0.936,0.246],}
\end{array}
\end{equation}
where $\bm{\mathcal{R}}(\theta,\mathbf{n})$ are the active right-hand rotation matrices~\cite{Rodrigues1840} of angle $\theta$ around the unit axis $\mathbf{n}$, and $\mathbf{M}_x = -\bm{\mathcal{R}}(\pi,\mathbf{x})$ is the mirror matrix operation with a $yz$ mirror plane. The particular choice of rotations is such that some special features are exemplified: \textit{i)} both site $1$ and $4$ are a very simple rotation since the local ligand planes for both these sites is just a rotation about the $x$ axis. \textit{ii)} sites $2$ and $3$ are nearly $C_3$ rotations away from site $4$, with the remaining correction rotation $\delta \xi$ being a small angle.

\subsection{Extracting the kinetic terms}

In order to estimate all the $t^{\text{eff}}_{\sigma\sigma'}$ hoppings we use the Wannierization~\cite{fploWF} as implemented in the DFT software \textit{ Full-Potential Local-Orbital code} (FPLO)~\cite{fplocite1,fplocite2}, using the fully relativistic implementation~\cite{fploRel}, and keeping the spin moment fixed to zero. The PBE version of the GGA is used as the exchange-correlation functional~\cite{perdew1996}.
The BZ sampling is performed on a $12\!\times\!12\!\times\!8$  primitive $\mathbf{k}$-grid, and a energy convergence criteria of $10^{-8}$ eV is used.
The resulting bands and DOS of a zero magnetization fully relativistic FPLO run are shown in Fig.~\ref{fig:microscopics_struct_bands_dos}. They are in excellent agreement with the VASP bands Fig.~\ref{fig:gga_bands}, both in the band dispersion around the Fermi level, and the general orbital  resolved DOS. This is indicative of a system with a small SOC.

 We now confirm the local basis orbital picture in the bands and DOS plot Fig.~\ref{fig:microscopics_struct_bands_dos}(b), where the local $\sum\nolimits_{i=1}^5 d_{\tilde{x}_i^2-\tilde{y}_i^2}$ dominate the Fermi level, along with oxygen orbitals. The oxygen hybretization is larger at the Fermi level than in the rest of the lower lying $d$-orbitals, consistent with the expected significant overlap of $d_{\tilde{x}_i^2-\tilde{y}_i^2}$ with the oxygen $p$. However, one notices an additional band, which is found to be of $d_{\tilde{z}_1^2}$ character, coming only from site $1$.
 Since \ce{Cu} site $1$ is in an octahedral environment, the multiplet $ \tilde{e}_{g,1} = \{ d_{\tilde{x}_1^2-\tilde{y}_1^2}, d_{\tilde{z}_1^2} \}$ split off together and remain close in energy, enough to mix both orbitals at the Fermi level. 
 This is more carefully examined in Fig.~\ref{fig:microscopics_struct_bands_dos}(c) where the one $d_{\tilde{z}_1^2}$ is projected separately (red) from the $\sum\nolimits_{i=1}^5 d_{\tilde{x}_i^2-\tilde{y}_i^2}$, showing that $d_{\tilde{z}_1^2}$  is mainly weighted at the bottom of the manifold but of course mixes in throughout the Fermi level bands. Therefore we cannot trivially disentangle it. This does not pose any hurdle to the estimate we are after, it simply informs us that we need to Wannierize by also including it. We proceed with a $\sum\nolimits_{i=1}^5 d_{\tilde{x}_i^2-\tilde{y}_i^2} + d_{\tilde{z}_1^2}$ Wannierization, and the resulting Wannier bands in in Fig.~\ref{fig:microscopics_struct_bands_dos}(d) are in excellent agreement with the DFT bands. 
 
 Having established the orbital character and found a good Wannier fit, we can now read off the effective spin resolved $t^{\text{eff}}_{4,\sigma\sigma'}$  and $t^{\text{eff}}_{5,\sigma\sigma'}$ hoppings with confidence from the fully relativistic calculation, where the effective downfolded ligand and SOC corrections are numerically incorporated in the numerical Wannierization projection. The resulting $2\times2$ spin-resolved hopping matrices need to be at least time reversal invariant, which is of the general form $\bm{t}^{\text{eff}} = t^o \bm{1}_{2\times 2} + i t^\gamma \bm{\sigma}^\gamma$ where $\bm{\sigma}^\gamma$ are the three Pauli matrices for $\gamma=x,y,z$. We find that for the two particular bonds we are examining, $(2\rightarrow 3)$ and $(2\rightarrow 4)$, we have the hopping matrices
 \begin{equation}\label{eq:spinreshopnum4}
\begin{array}{l}
 \bm{t}^{\text{eff}}_{4,(2\rightarrow 3)} =  \\[8pt]
\begin{pmatrix}
 -77.98-101.25 i & 49.10 + 49.10 i \\
 -49.10 + 49.10 i & -77.98+101.25 i
 \end{pmatrix}\text{ meV}\\[20pt]
 \Rightarrow(t^o_4,\ t^x_4,\ t^y_4,\ t^z_4)=\\[4pt]
 \phantom{\Rightarrow}\ (-77.98,\ 49.10,\ 49.10,\ -101.25)\text{ meV},
\end{array}
 \end{equation}
 \begin{equation}\label{eq:spinreshopnum5}
\begin{array}{l}
 \bm{t}^{\text{eff}}_{4,(2\rightarrow 3)} =  \\[8pt]
\begin{pmatrix}{cc}
 -80.53+108.72 i & -54.21 - 50.77 i \\
  54.21 - 50.77 i & -80.53-108.72 i
 \end{pmatrix}\text{ meV}\\[20pt]
 \Rightarrow(t^o_5,\ t^x_5,\ t^y_5,\ t^z_5)=\\[4pt]
 \phantom{\Rightarrow}\ (-80.53,\ -50.77,\ -54.21,\ 108.72)\text{ meV}.
\end{array}
 \end{equation}
Both perfectly agree with the time reversal symmetric form expected. The  $\bm{t}^{\text{eff}}_{5,(2\rightarrow 4)}$ matrix has all $t^\mu_5$ independent which is expected given the bond has no symmetry. The $\bm{t}^{\text{eff}}_{4,(2\rightarrow 3)}$ has an extra $t^x_4=t^y_4$ relation, reflecting the bond mirror symmetry. We note that it should not be confusing why $x$ and $y$ are getting related: while the mirror plane is $yz$ in the global axis, the Wannier hoppings are defined and extracted in the local basis, twisting the axes, and transforming the global picture into the local one.

\subsection{Numerical estimates of the sawtooth exchanges}
Having performed the fully relativistic calculation, understood the local orbital picture, and derived spin resolved hoppings $t^\mu_w$ in the local frames, we can now extract exchanges in the strong-coupling expansion, using existing $2$\textsuperscript{nd} order time reversal symmetric hopping formulas \cite{CarterPRB2012}, with our numerically estimated hoppings Eq.~\ref{eq:spinreshopnum4}, \ref{eq:spinreshopnum5} and $U=8\text{ eV}$. The resulting exchange matrix $\widetilde{\bm{\mathcal{J}}}$ is in the local basis definitions, in which the Wannier functions were extracted. Using the basis transformations Eq.~\ref{eq:basischange}, the final bond exchange in the global $xyz$ basis is numerically found for the base-to-base sawtooth bond
\begin{equation}
\begin{array}{l}
\bm{\mathcal{J}}_{4,(2\rightarrow 3)} = \widetilde{\mathbf{R}}_2^T\widetilde{\bm{\mathcal{J}}}_{4,(2\rightarrow 3)}\widetilde{\mathbf{R}}_3 = \\[12pt]
10.490 \bm{1}_{3\!\times\!3} + 
\left(\begin{array}{rrr}
-0.043 & -0.722 & -1.488\\
\phantom{-}0.722 & \phantom{-}0.062 & -0.051\\
\phantom{-}1.488 & -0.051 & -0.019
\end{array}\right)
\end{array}
\end{equation}
in units of meV, from where the symmetry expected exchange couplings of Eq.~\ref{eq:symJ4} are trivially extracted $J_4=10.490\text{ meV}$, $\mathbf{D}_4=[0,1.488,-0.722]\text{ meV}$, $\Gamma^x_4=-0.051\text{ meV}$ and $K^\mu_4=(-0.043,0.062,-0.019)\text{ meV}$. Similarly for the base-to-vertex sawtooth bond we find
\begin{equation}
\begin{array}{l}
\bm{\mathcal{J}}_{5,(2\rightarrow 4)} = \widetilde{\mathbf{R}}_2^T\widetilde{\bm{\mathcal{J}}}_{5,(2\rightarrow 4)}\widetilde{\mathbf{R}}_4 = \\[12pt]
11.862 \bm{1}_{3\!\times\!3} + \left(\begin{array}{rrr}
 0.014 & 0.749 & -0.500\\
-0.707 & -0.012 & 0.959\\
0.558 & -0.926 & -0.002
\end{array}\right)
\end{array}
\end{equation}
in units of meV, from where the symmetry expected exchange couplings of Eq.~\ref{eq:symJ5} are extracted by appropriate symmetric and antisymmetric combinations of the matrix elements, and are $J_5=11.862\text{ meV}$, $\mathbf{D}_5=[0.943,0.529,0.728]\text{ meV}$, $\Gamma^\mu_5=(0.016,0.029,0.021)\text{ meV}$, and $K^\mu_5=(0.014,-0.012,-0.002)\text{ meV}$. 

The numerical exchanges are found to perfectly respect the expected symmetry reduced form. The largest exchange are found to be the Heisenberg $J_4$ and $J_5$. In particular $J_4$ is nearly identical to the one extracted from TEMA. For $J_5$, while it is found to be slightly larger than $J_4$, following the expected TEMA trend, it is nearly identical to $J_4$ while TEMA predicts a much larger deviation. Given the reduced method we applied, not all ligand contributions are being incorporated, and this deviation is expected. The next thing to observe is that the leading order anisotropic exchange is the DM vector $\mathbf{D}_4$ and $\mathbf{D}_5$, which are one order of magnitude smaller than the dominant Heisenberg exchanges. As predicted, the result is quite similar to the herbertsmithite case where DM exchanges are about an order of magnitude smaller that the Heisenberg exchange ~\cite{HerbAniso2008,HerbAniso2012,HerbAniso2025}. 

Finally we comment on the energy hierarchy of the exchanges we found from the microscopics. Anisotropic exchanges, which includes $\mathbf{D}$, $K$, and $\Gamma$, scale with the SOC parameter $\lambda$, however, the DM vector $\mathbf{D}$ scales with $\lambda \delta t$, where $\delta t$ is the inversion symmetry breaking hopping contribution, while $K$ and $\Gamma$ scale with $\lambda^2$~\cite{YildirimPRB1995}. For systems with a small $\lambda$, the symmetric anisotropic interactions $K$ and $\Gamma$ remain small, while $\mathbf{D}$ can be appreciable if the inversion breaking ($\delta t$) is large enough, and the expected hierarchy is first $J$, then $\mathbf{D}$, then $K$ and $\Gamma$. This intuition manifestly appears in the numerics, with the DM vector $\mathbf{D}$ being an order of magnitude smaller than $J$, and $K$ and $\Gamma$ being yet another order of magnitude smaller.

\section{Classical $\bm{T=0}$ Minimization}

We employ parallel tempering Monte Carlo methods~\cite{SwendsenPRL1986,HukushimaJPSJ1996} to minimize the classical spin model and find the $T=0$ ground state. In parallel tempering, we simulate $240$ replicas, with the replica temperatures $k_BT$ spread logarithmically between $3$ and $10^{-5}$ in order to sample equally across all orders in the range. We perform a minimum of $5\cdot10^5$ temperature updates between replicas. In between every temperature update, each replica performs $500$ Monte-Carlo sweeps, with each sweep consisting of system size Metropolis-Hastings trial updates. 

At the end of the simulation, thermal noise on the order of the lowest simulated temperature still persists in the system. To refine the state towards $T=0$, we perform torque updates. Demanding that the ground state is a stationary state against time evolution equates to setting the Landau–Lifshitz–Gilbert equation~\cite{LakshmananLLGeq2011} to zero, 
$\mathrm{d}S^{\alpha}_i/\mathrm{d}t= \tau_i^{\alpha}  = 0$, 
where 
$\tau_i^{\alpha} = -\varepsilon^{\alpha\beta\gamma} S^{\beta}_i B_i^{\gamma}$
is the local torque, and 
$B_i^{\gamma} = \partial\mathcal{H}/\partial S^{\gamma}_i $
the local field, on site $i$. To achieve zero torque per site in the system, we sweep through many times and perform the update 
$S^{\alpha}_i \rightarrow - B_i^{\alpha}/\|\mathbf{B}_i\|$. 
Convergence is declared once the maximum remaining torque $\|\bm{\tau}_i\|$ for all sites is smaller than $10^{-14}$.

\subsection{Additional details on the classical state}

The classical $T=0$, as described in the main and presented in main Fig.3, forms a collection of coplanar sawtooth states, with a rotational freedom between them and an emergent $2$D degeneracy $SO(2)^{N_a N_c}$. This comes about as a result of dominant $J_4$ (base-to-base) and $J_5$ (base-to-vertex) exchanges. The role of the other exchanges is also covered in the main, and here we expand on that in more detail.

The sawtooth vertices are all ordered in the ground state in the same direction $\mathbf{S}_{\text{v}}$. This comes out as a result of $J_2$ and $J_1$. In particular $J_2$ couples vertices between adjacent chains along $\mathbf{a}$ by passing through the dangling site (Wyckoff position $1$). The $J_3$ and $J_6$ which also couple to the dangling site are very small, and as a result the AFM $J_2$ results into an effective FM coupling between interchain vertices by anti-aligning the dangling site $\mathbf{S}_{\text{o}}=-\mathbf{S}_{\text{v}}$. In this sense $J_2$ is unfrustrated. Vertically along $\mathbf{c}$, the FM $J_1$ also aligns the vertices, and in total all vertex spins are ordered.

The much smaller $J_3$ and $J_6$ are too small to cause qualitative differences, but they explicitly enter in the angle relation between base sites $\alpha$.
This can be understood by the network these exchanges form. First we remind our serves that the isolated sawtooth chain $J_4+J_5$ would result in $\cos\alpha = -J_5/(2J_4)$ for the reasons covered in the End Matter. The $J_6$ exchange forms a triangle by coupling two adjacent base sites to the vertex of a neighboring site, see main Fig.1(c). If we extend this we can easily see this form a sawtooth like connectivity, where $J_4$ is still the base-to-base, and $J_6$ plays the role of base-to-vertex. Since all vertices are ordered and aligned, this results in adding a $-J_6/(2J_4)$ factor to $\cos\alpha$. Similarly, it is clear that $J_3+J_4$ is forming a sawtooth like network, however, it does this by treating the dangling site as the vertex. Since the dangling sites are all anti-aligned to the vertex (as described above) it has exactly the opposite sing contribution $+J_3/(2J_4)$, and in total the angle is set by $\cos\alpha = (-J_5+J_3-J_6)/(2J_4)$, leading to the expression for $\alpha$ covered in the main text.

\section{Order by disorder}

\begin{figure}[t]
	\centering
    \begin{overpic}[width=1\columnwidth,percent,angle=0]{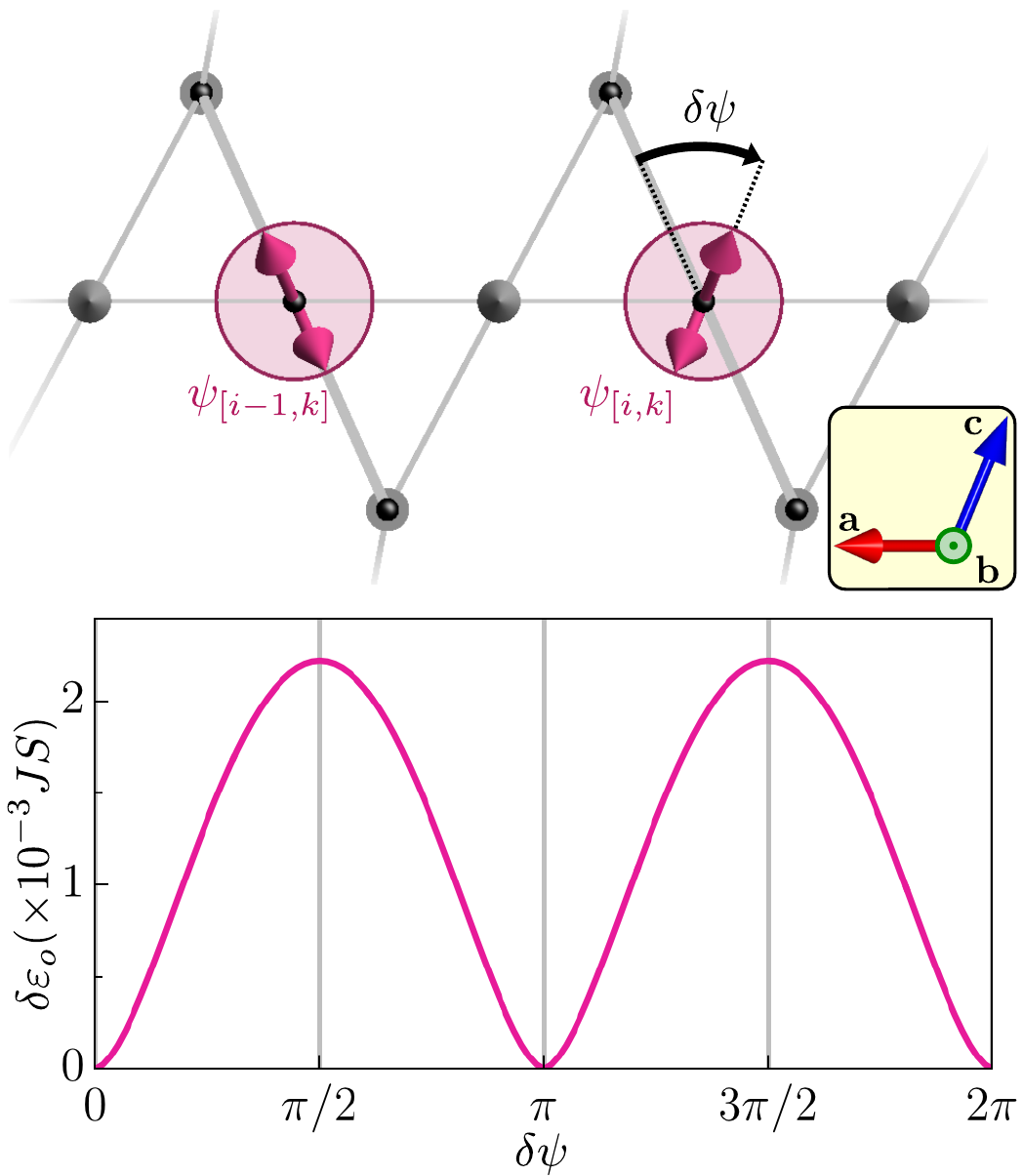}
        \put(1,96){(a)}
        \put(1,45){(b)}
    \end{overpic} 
    \caption{(a) The definition of the relative angle 
    $\delta\psi=\psi_{\left[i-1,k\right]}-\psi_{\left[i,k\right]}$. (b) The zero point energy correction $\delta \varepsilon_o = \Delta \varepsilon_o(\delta\psi) - \text{Min}\left[\Delta \varepsilon_o(\delta\psi)\right]$.
     The quantum fluctuations lift the $SO(2)$ degeneracy and select aligned and anti-aligned $\mathbf{S}_A$, leaving an Ising degree of freedom. }\label{fig:obdplot}
\end{figure}

The classically degenerate states discussed in the main may acquire different zero-point energy corrections $\varepsilon = \varepsilon_{\text{cl}} + \Delta\varepsilon_o(\psi)$, giving rise to quantum order-by-disorder selection. Restricting to the coplanar family of states, as leading-order spin-wave theory is known to favor coplanar over non-coplanar orders, we compute the zero-point energy as a function of the relative angle $\delta\psi\!=\!\psi_{[i-1,k]}-\psi_{[i,k]}$ between neighboring chains, with results in Fig.~\ref{fig:obdplot}. The minimum is achieved at $\psi_{[i-1,k]}\!=\!\pm\psi_{[i,k]}$, corresponding to $\mathbf{S}_{A(B)_{[i-1,k]}}\!=\!\mathbf{S}_{B(A)_{[i,k]}}$, i.e., full coplanarity. However, a $\pm 1$ ambiguity persists for every chain in the system, so that the low-energy effective description within the classical degenerate subspace retains an Ising degree of freedom per chain.

\section{Classical Thermal Sampling\label{sec:ClassicalThermalSampling}}

We employ parallel tempering Monte Carlo methods~\cite{SwendsenPRL1986,HukushimaJPSJ1996} to estimate thermal averages in the $J_1 - J_6$ spin model, on symmetric $ L\times L\times L$ periodic clusters, with $L=6,7,8,9,10$. In parallel tempering we simulate 580 replicas, logarithmically spaced across $k_BT=10^{-3}J_5$ and $k_BT=10 J_5$. Finding in trials that $k_BT=1.5 J_5$ is already pretty high in terms of the models features, we split the simulations into separate high  ($k_BT=[1.5,10]J_5$) and low ($k_BT=[10^{-3},1.5]J_5$) temperature ranges, with 240 replicas each, to sample as well as possible the low temperatures, and also verify the large temperature limit, but run them in parallel.

For each simulation we perform $10^7$ temperature updates between replicas, while doing 8 Monte-Carlo sweeps between each temperature update, and an additional 2 energy conserving microcanonical overrelaxation sweeps~\cite{CreutzPRD1987,KankiPRD2005} per Monte-Carlo sweep.
 To avoid local minima trapping of the particular low temperature state, 
 we prepare 240 random arrangements of the $T=0$ state discussed in the main, 
 and distribute them to the replicas, which are then further left to run upsampled for
  half the simulation steps ($0.5 \cdot 10^7$) in order to guarantee a good equilibrium
   of the temperature ladder.
  Monitoring the rate of temperature swaps, for the largest system size simulated ($L=10$), we find that at the lowest temperatures ($10^{-3}$) we have a quite small $12\%$ swapping between neighbour replicas, while on the high end ($1.5$) it is about $60\%$, which makes the low temperature data somewhat noisy but of acceptable quality. For the second half of the simulation, we sample in regular intervals but sparsely, to avoid correlated sample gathering, collecting $2.5 \cdot 10^5$ samples.

\subsection{Sampled observables}

\begin{figure}
	\centering
    \begin{overpic}[width=1.0\columnwidth,percent,grid=false,tics=2,angle=0]{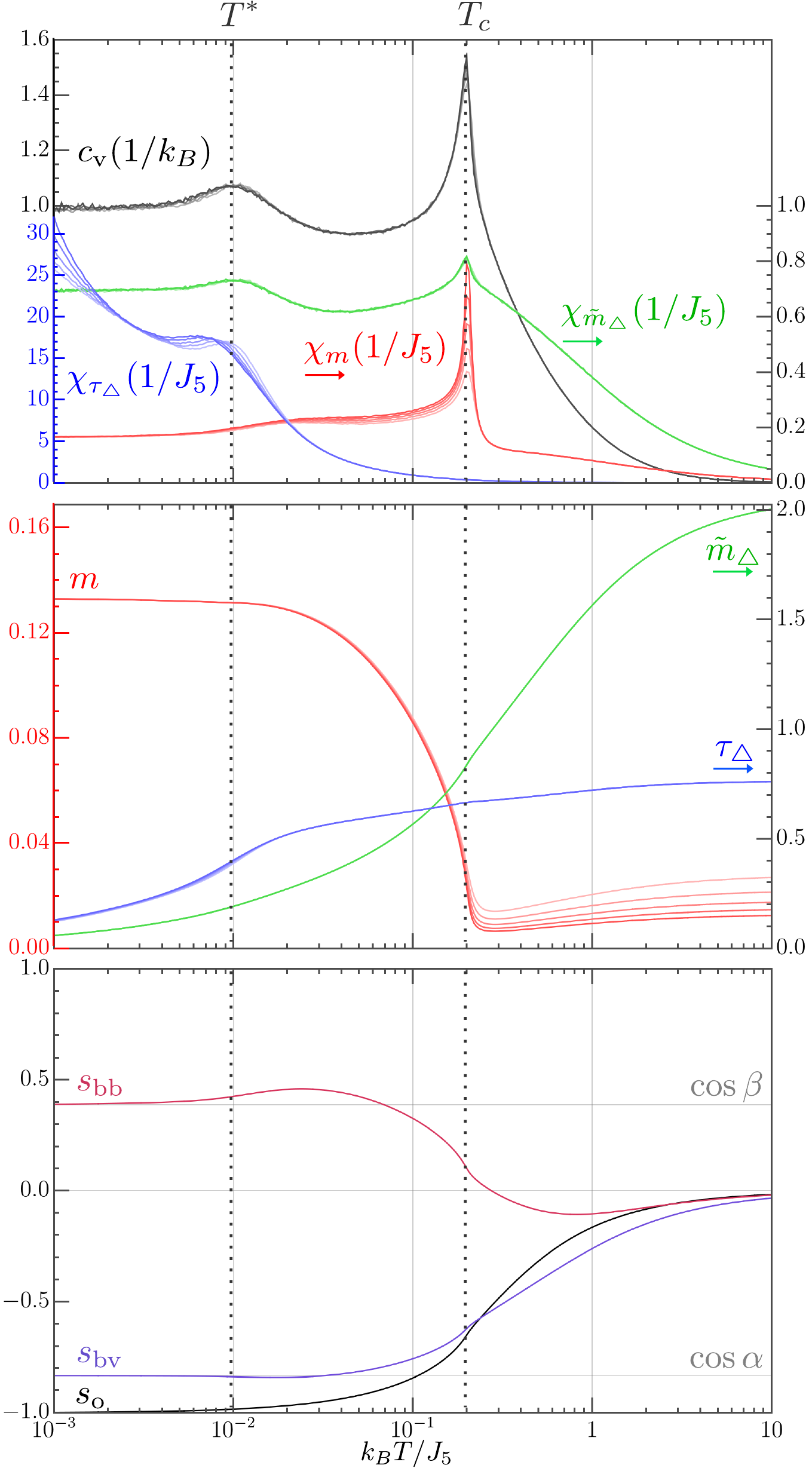}
	\put(4.5,95.5){(a)}
	\put(4.5,64){(b)}
	\put(4.5,32.5){(c)}
    \end{overpic} 
    \caption{Result from thermal averages in Monte-Carlo simulations. The smaller $L=6$ to larger $L=10$ system results are plotted from light to dark colors, against a logarithmic temperature scale. Panel (a) summarizes the heat capacity $c_{\mathrm{v}}$ and the susceptibilities $\chi_{m}$, $\chi_{\tilde{m}_{\triangle}}$, and $\chi_{\tau_{\triangle}}$. The corresponding observables $m$,$\tilde{m}_{\triangle}$, and $\tau_{\triangle}$ are plotted in panel (b). The base-to-base, base-to-vertex, and vertex-to-dangling (Wyckoff position $1$) correlators are plotted in panel (c). The special $T_c$ and $T^*$ temperatures are also indicated with dotted lines.\label{fig:MC_thermal}}
\end{figure}

In the simulations we need to decide on observables to sample. The obvious simple ones to sample are heat capacity, magnetization magnitude, and magnetic susceptibility
\begin{equation}
\begin{array}{rcl}
c_{\mathrm{v}}/k_B & = & N(\langle \varepsilon^2\rangle - \langle \varepsilon\rangle^2)/(k_{B}T)^2,  \\[12pt]
m & = & \dfrac{1}{N}\left\Vert\displaystyle\sum_{l\in\text{sites}} \!\!\! \mathbf{S}_l   \right\Vert,  \\[20pt]
\chi_{m} & = & N(\langle m^2\rangle - \langle m\rangle^2)/(k_{B}T),
\end{array}
\end{equation}
where $\varepsilon$ is the energy per site, and $N=5L^3$ the number of sites in the system. In particular we measure the per site definitions above, to collapse the plots over system sizes.

We can recover the expected $m(T=0)$ magnetization if we find our selves in the $T=0$ classical ground state described in the main text. A primitive unit cell contains $5$ sites: $2$ vertex sites (Wyckoff position $3$), $2$ base sites (Wyckoff position $2$), and $1$ dangling site (Wyckoff position $1$). The dangling site $\mathbf{S}_{\mathrm{o}}=-\mathbf{S}_{\mathrm{v}}$ and one vertex $\mathbf{S}_{\mathrm{v}}$ cancel out, thus the magnetization per site is $m(T=0) = \|\mathbf{S}_{\mathrm{v}} + \mathbf{S}_{\mathrm{A}} + \mathbf{S}_{\mathrm{B}}\|/5 = (\sqrt{2+2\cos\beta}-1)/5 = 0.1332$

An important feature of the $J_1 - J_6$ spin model is that at $T=0$ the generalized magnetization $\widetilde{\mathbf{M}}_{\triangle,l}$, as defined in the main text, needs to be zero on every sawtooth triangle. We additionally measure the average generalized triangle magnetization magnitude in the system, and its susceptibility,
\begin{equation}
\begin{array}{rcl}
\tilde{m}_{\triangle} & = &\dfrac{1}{N_{\triangle}}\displaystyle\sum_{l\in\triangle} \Vert \widetilde{\mathbf{M}}_{\triangle,l}\Vert, \\[18pt]
\chi_{\tilde{m}_{\triangle}} & = & N_{\triangle}(\langle \tilde{m}_{\triangle}^2\rangle - \langle \tilde{m}_{\triangle}\rangle^2)/(k_B T),  
\end{array}
\end{equation}
where $N_{\triangle}=2L^3$ since there are 2 sawtooth triangles in every primitive cell. The $\tilde{m}_{\triangle}(T\rightarrow\infty)$ behavior is recovered by integrating $\Vert \widetilde{\mathbf{M}}_{\triangle,l}\Vert$ over all directions for one triangle, meaning three sites. We instead perform the numerical estimate by uniformly sampling the sphere $10^6$ times for three sites, finding that $\tilde{m}_{\triangle}(T\rightarrow\infty) = 2.06614\pm0.0005$.

In the isolated sawtooth chain, we saw in the End Matter how multiplanar states exist, where each triangle defines a single plane, which can be arbitrarily rotated to its neighbor triangles. To capture such potential features, we further define now the triangle torsion
\begin{equation}
\begin{array}{rcl}
\mathbf{T}_{l,l^\prime} & = & \mathbf{n}_{l} \times \mathbf{n}_{l^\prime}, \\
\mathbf{n}_{l} & = & \text{Normalize} \left[ \mathbf{S}_{l,1}\!\times\mathbf{S}_{l,2} + \mathbf{S}_{l,2}\!\times\mathbf{S}_{l,3}+ \mathbf{S}_{l,3}\!\times\mathbf{S}_{l,1} \right] 
\end{array}
\end{equation}
where $\mathbf{S}_{l,w}$ are the three spins in the $l^{\text{th}}$ triangle, with $w=1,2,3$ ordered clockwise for every triangle. For three coplanar spins, $\mathbf{n}_{l}$ is the unit vector normal to the spin plane. In this way the torsion $\mathbf{T}_{l,l^\prime}$ is a measure of how much neighboring triangles have their spin planes rotated relative to each other. In the coplanar sawtooth state $\mathbf{T}_{l,l^\prime} = \mathbf{0}$, but it is finite for multiplanar sawtooth states. We measure the average NN triangle torsion magnitude in the system, and its susceptibility,
\begin{equation}
\begin{array}{rcl}
\tau_{\triangle} &=& \dfrac{1}{N_{\triangle}}\displaystyle\sum_{(l,l^\prime)\in(\text{NN  $\triangle$})} \!\!\!\!\!\!\!\!\! \Vert \mathbf{T}_{l,l^\prime} \Vert, \\[18pt]
 \chi_{\tau_{\triangle}} &=& N_{\triangle}(\langle \tau_{\triangle}^2\rangle - \langle \tau_{\triangle}\rangle^2)/(k_B T),
\end{array}
\end{equation}
to diagnose the onset of the coplanar order or multiplanar fluctuations. The $\tau_{\triangle}(T\rightarrow\infty)$ behavior is recovered by integrating $\Vert \mathbf{T}_{l,l^\prime} \Vert$ over all directions for two triangles, meaning six sites. We instead perform the numerical estimate by uniformly sampling the sphere $10^6$ times for six sites, finding that $\tau_{\triangle}(T\rightarrow\infty) = 0.785212\pm0.00005$.

Finally it is instructive to measure some correlators. The base-to-base and base-to-vertex are obvious good choices. Additionally we measure the vertex-to-dangling intermediate site (Wyckoff position $1$). We specifically measure the averages
\begin{equation}
\begin{array}{c}
s_{\text{bb}} = \dfrac{1}{N_{\triangle}}\displaystyle\sum\limits_{(b,b^\prime)\in(\text{b to b})} \!\!\!\!\!\!\!\!\! \mathbf{S}_b\cdot \mathbf{S}_{b^\prime}  \\[18pt]
s_{\text{bv}} = \dfrac{1}{2 N_{\triangle}}\displaystyle\sum\limits_{(b,v)\in(\text{b to v})} \!\!\!\!\!\!\!\!\! \mathbf{S}_b\cdot \mathbf{S}_{v} \\[18pt]
s_{\text{o}} = \dfrac{1}{N_{\triangle}}\displaystyle\sum\limits_{(v,l)\in(\text{v to dangling})}\!\!\!\!\!\!\!\!\!\!\!\!\!\!\!\!\! \mathbf{S}_v\cdot \mathbf{S}_{l}
\end{array}
\end{equation}

\begin{figure}
	\centering
    \begin{overpic}[width=1.0\columnwidth,percent,grid=false,tics=5,angle=0]{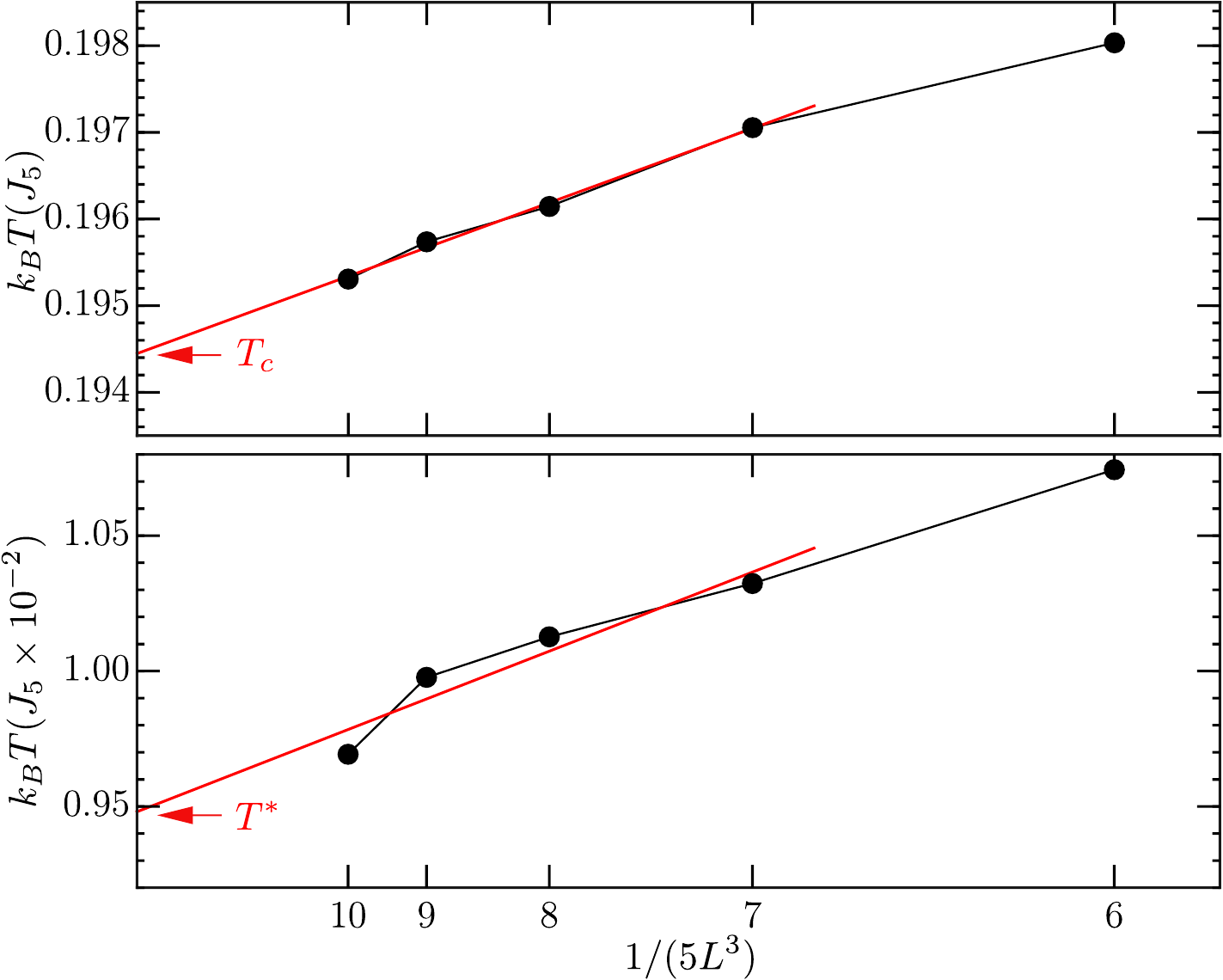}
	\put(13,76){(a)}
	\put(13,39){(b)}
    \end{overpic} 
    \caption{Heat capacity peaks are extracted by fitting a 4\textsuperscript{th} order polynomial in the temperature range $[0.17,0.22]$ and $[0.4,1.6]\times10^{-2}$ and finding the maxima, with results plotted (black points) as a function of $1/(5L^3)$. Omitting the smallest system $L=6$, a linear fit is extrapolated down to zero ($L\rightarrow\infty$), to identify $k_BT_c = 0.194 J_5$ and $k_BT^* = 0.0095 J_5$.}\label{fig:sysscalTextr}
\end{figure}

\subsection{Results}

The results of the simulations, for all systems, are plotted in Fig.~\ref{fig:MC_thermal}, where the temperature axis is plotted logarithmically. Looking first at the heat capacity $c_{\mathrm{v}}$ in panel (a) there are two features to notice, and we extract two temperatures by fitting the peaks as a function of system size in Fig.~\ref{fig:sysscalTextr}. Coming in from high temperatures, there is a sharp peak around $k_BT_c = 0.194 J_5$, which becomes more sharp with increasing system size. This is characteristic of a second order phase transition. Lowering the temperature further there is a soft bump feature around $k_BT^* = 0.0095 J_5$, which is mostly unaffected by system size scaling. Both these features are also seen in the magnetic susceptibility $\chi_m$, with the $T_c$ being pronounced and becoming sharper with system size, characteristic of a magnetic second order transition, however, the low $T^*$ feature is quite weak and more like a shoulder which becomes even weaker with increasing system size. The magnetization magnitude $m$, plotted in panel (b), also shows a very canonical second order phase transition development around $T_c$, saturating to the expected $T\rightarrow 0$ value $0.1332$, but crosses through $T^*$ without any signatures. This is already indicative that some order onsets at $T_c$ but something more subtle is happening at $T^*$.

Looking at the generalized triangle magnetization susceptibility $\chi_{\tilde{m}_{\triangle}}$ and torsion susceptibility $\chi_{\tau_{\triangle}}$ in panel (a) we see a very interesting distinction. The $\chi_{\tilde{m}_{\triangle}}$  susceptibility picks up both $T_c$ as a sharper feature and $T^*$ as a broader bump. However, $\chi_{\tau_{\triangle}}$ crosses through $T_c$ featureless, and blows up at $T\rightarrow 0$ with a shoulder feature at $T^*$. This is consistent with the raw observables in panel (b). The generalized triangle magnetization $\tilde{m}_{\triangle}$ initiates at the expected $T\rightarrow \infty$ value of $2.0661$ (verified by uniformly sampling the sphere) and remains large until $T_c$ where it drastically falls and starts tending to zero as $T\rightarrow 0$. This indicates that at the magnetic $T_c$ transition, the local $\tilde{m}_{\triangle}$ starts to settle into their expected $T=0$ behaviour covered in the main text. On the other hand the torsion $\tau_{\triangle}$ initiates at the expected $T\rightarrow \infty$ value of $0.78521$ (verified by uniformly sampling the sphere), but remains large across $T_c$, taking a drastic dip much later around $T^*$, after which it tends to zero, consistent with the coplanar state. 

The above results paint a particular picture. The system undergoes a magnetic second order phase transition at $T_c$, but the system only partially orders, with the local triangle magnetizations approaching their ground state value of zero, in other words local correlators start becoming well defined. However the system is not a coplanar state variant, because the torsion shows that multiplanar fluctuations are still prominent in the system. At the much lower $T^*$ the system refines into its coplanar character expected at $T=0$ with the torsion drastically dropping off. This leaves an intermediate temperature regime with multiplanar fluctuations prevalent in the system. This is further corroborated by the correlators plotted in panel (c). The vertex and the dangling sites become anti-aligned as expected, with $s_o$ tending to -1 below $T_c$, and there is partial ordering in the system. However, the base-to-base ($s_\mathrm{bb}$) and base-to-vertex ($s_\mathrm{bv}$) correlations are dramatically reconstructed at $T_c$ indicating local correlations developing, allowing local ordering in the triangle. Nevertheless, they only settle into their expected $\cos\alpha$ and $\cos\beta$ values much later around $T^*$.

\section{Exact diagonalization of sawtooth chains}

Quantum solvers applied to three-dimensional models are a herculean task. Instead, we treat quantum mechanically the isolated sawtooth chain, with the chain model as defined in the End Matter. 
We perform spin $S=1/2$ exact diagonalization (ED) on an isolated 28-site sawtooth chain with periodic boundary conditions, setting $J_{\mathrm{bb}}=J_4$ and $J_{\mathrm{bv}}=J_5$. 
The quantum ground state is obtained using a custom ED code interfaced with the iterative ARPACK eigensolver~\cite{lehoucq1998arpack} with seven Arnoldi basis vectors and a relative convergence tolerance of $10^{-12}$.

\begin{figure}[t]
	\centering
    \begin{overpic}[width=1.0\columnwidth,percent,grid=false,tics=2,angle=0]{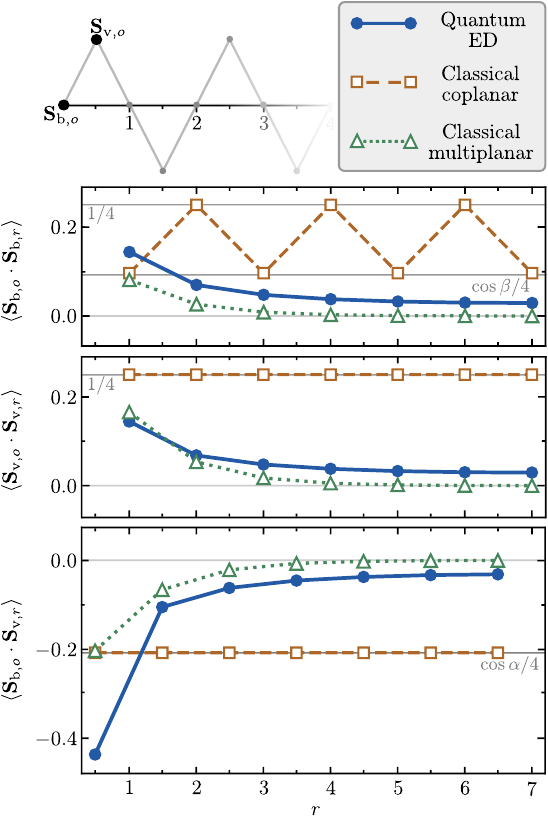}
	\put(0,75){(a)}
	\put(0,53.5){(b)}
	\put(0,32){(c)}
    \end{overpic} 
    \caption{Quantum correlators for base-to-base (a), vertex-to-vertex (b), and base-to-vertex (c) as a function of separation distance $r$ from ED (blue). The classical coplanar (brown) as well as the averaged multiplanar (green) are also plotted.
    The gray horizontal lines indicate certain reference values of the coplanar state, such as $\cos \alpha/4$ and $\cos\beta/4$ (definitions in the End Matter). 
    The ED results are shown for $N=28$; similar qualitative correlation patterns are also found for $N=16$ and $24$.}\label{fig:EDres}
\end{figure}

The resulting base-to-base (a), vertex-to-vertex (b), and base-to-vertex (c) quantum correlators (blue), as a function of separation distance $r$, are plotted in Fig.~\ref{fig:EDres}. The plots go up to half the system size, as the correlators will repeat due to the periodic boundary conditions. As a reference we also plot the classical correlators, scaled by 1/4 for comparable spin length, for the coplanar state (brown), as well as for averaging over the classically degenerate multiplanar ground states of the isolated chain (green).  While the coplanar classical correlators are finite and repeating, the average multiplanar correlators decay with distance. This follows from the fact that in the multiplanar ensemble, the random rotation allowed between adjacent spin planes as described in the End Matter will eventually decorrelate the spins at a large enough distance.

Most interestingly though, we find a qualitative similarity in both  the signs and the distance dependence between the quantum and the averaged multiplanar correlators, as seen in all three panels of the figure. This suggests that there is some relation between the quantum state and the multiplanar degeneracy, akin to classical-to-quantum correspondences found in other frustrated magnets with macroscopically degenerate classical manifolds~\cite{Samarakoon2017PRB,Samarakoon2018PRB}. 
This finding is quite intriguing in light of the classical thermal Monte Carlo results for the full three-dimensional system, discussed in Supplemental Sec. \ref{sec:ClassicalThermalSampling}, which shows that there is a finite-temperature window where multiplanar correlations are present. This indicates that some features of the quantum sawtooth chain may survive and be present in the three-dimensional model~\cite{LakeNatPhys2005,Banerjee2016,Rousochatzakis2019}.

\ifdefined\suppresssupbibliography
\else
\input{main_reduced_ref.bbl}
\fi

\end{document}